\PassOptionsToPackage{table,svgnames,dvipsnames}{xcolor}
\PassOptionsToPackage{scaled=.8}{beramono}

\documentclass[fleqn, 5p,times,letterpaper]{elsarticle} 
\usepackage[protrusion=true,expansion=true]{microtype}
\usepackage[switch]{lineno}

\usepackage[greek,english]{babel}
\usepackage{amssymb}
\usepackage{amsmath}
\usepackage{centernot}
\usepackage{bbm}

\usepackage[utf8]{inputenc}
\usepackage[T1]{fontenc}
\usepackage[scaled=.8]{beramono}
\usepackage{subcaption}
\usepackage{makecell} \usepackage{wrapfig}
\usepackage{url}
\usepackage[
   pdftex, final, breaklinks, hidelinks,
   colorlinks = false, bookmarks = false, bookmarksnumbered, plainpages = false,
   pdfpagelabels, pdfborder = 0 0 0, hyperindex
]{hyperref}
\usepackage{hyphenat}
\usepackage{pifont}
\usepackage{paralist}
\usepackage{booktabs}
\usepackage{color, colortbl}
\usepackage{changepage}
\usepackage{bbding}
\usepackage{multirow}
\usepackage{multicol}

\usepackage{setspace}
\usepackage{orcidlink}

\setdefaultenum{i)}{(a)}{i.}{A.}

\usepackage{tabularx}
\usepackage{adjustbox}
\usepackage[most]{tcolorbox}
\tcbset{
    mybox/.style={
      enhanced,
      attach boxed title to top center={yshift=-3mm,yshifttext=-1mm},
      colback=blue!10!white,
      colframe=blue!50!white,
      colbacktitle=red!30!white,
      coltitle=black,
      title=#1,
      fonttitle=\bfseries,
      boxed title style={size=small,colframe=red!50!black}
  },
}
\newtcolorbox{quack}{
  breakable,
  enhanced,
  frame hidden,
  interior hidden,
  size=minimal,
  left skip=16pt,
  right skip=16pt,
  borderline west={2pt}{-8pt}{lightgray},
}

\newtcbtheorem{mydefinition}{Highlight}{myboxstyle=yellow}{def}

\usepackage[frozen]{showcode}

\usepackage{titlesec}
\titleformat{\paragraph}[runin]{\normalfont\fontsize{10}{1.2}\bfseries}{\theparagraph}{.5em}{}
\titlespacing*{\paragraph}{15pt}{0ex plus 1ex minus .2ex}{1em}

\journal{The Journal of Systems \& Software}

\tikzstyle{every picture}+=[remember picture]
\tikzstyle{na} = [baseline=-2.5pt]
\usetikzlibrary{shapes.arrows}

\input{fix_star}

\usepackage{xspace}
\newcommand{\toolname}[1]{\textsf{#1}\xspace} 
\newcommand{\fontrq}[1]{\mathcal{#1}}
\newcommand{\fontc}[1]{\mathbfcal{#1}}
\newcommand{\ourstar}{\ding{79}}

\DeclareMathAlphabet\mathbfcal{OMS}{cmsy}{b}{n}

    \usepackage{fancyhdr}
    \usepackage{tcolorbox}
\begin{document}
\begin{frontmatter}

	\title{
		\textbf{Code Less to Code More}\tnoteref{t1} \\
		\large Streamlining Language Server Protocol and Type System Development for Language Families
	}

	\tnotetext[t1]{This work was partly supported by the MUR project ``T-LADIES'' (PRIN 2020TL3X8X).}

	\author[unimi]{Federico Bruzzone\,\orcidlink{0009-0004-6086-8810}} 
	\ead{federico.bruzzone@unimi.it}
	\author[unimi]{Walter Cazzola\,\orcidlink{0000-0002-4652-8113}\corref{cor1}} 
	\ead{cazzola@di.unimi.it}
	\author[unimi]{Luca Favalli\,\orcidlink{0000-0001-7452-2440}} 
	\ead{favalli@di.unimi.it}
	\affiliation[unimi]{
		organization={Università degli Studi di Milano, Computer Science Department},
		city={Milan},
		country={Italy}
	}

	\cortext[cor1]{Corresponding author.}

	\begin{abstract}
        Developing editing support for \(\fontc{L}\) languages in \(\fontc{E}\) editors is complex and time-consuming.
Some languages do not provide dedicated editors, while others offer a single native editor.
The \textit{language server protocol} (LSP) reduces the language-editor combinations \(\fontc{L} \times \fontc{E}\) to \(\fontc{L} + \fontc{E}\), where a single language server communicates with editors via LSP plugins.
However, overlapping implementations of linguistic components remain an issue. Existing language workbenches struggle with modularity, reusability, and leveraging type systems for language server generation.
In this work, we propose: \begin{inparaenum}
    \item Typelang, a family of domain-specific languages for modular, composable, and reusable type system implementation,
    \item a modular language server generation process, producing servers for languages built in a modular workbench,
    \item the variant-oriented programming paradigm and a cross-artifact coordination layer to manage interdependent software variants, and
    \item an LSP plugin generator, reducing \(\fontc{E}\) to \(\mathbf{1}\) by automating plugin creation for multiple editors.
\end{inparaenum}
To simplify editing support for language families, each language artifact integrates its own Typelang variant, used to generate language servers.
This reduces combinations to \(\fontc{T} \times \mathbf{1}\), where \(\fontc{T} = \fontc{L}\) represents the number of type systems.
Further reuse of language artifacts across languages lowers this to \(\fontc{N} \times \mathbf{1}\), where \(\fontc{N} << \fontc{T}\), representing unique type systems. We implement Typelang in Neverlang, generating language servers for each artifact and LSP plugins for three editors. Empirical evaluation shows a 93.48\% reduction in characters needed for type system implementation and 100\% automation of LSP plugin generation, significantly lowering effort for editing support in language families, especially when artifacts are reused.

	\end{abstract}

	\begin{keyword}
		Software product lines \sep{} Feature modularity \sep{} Language Server Protocol \sep{} Integrated development environments \sep{}  Software systems architectures \sep{}  Extensible languages \sep{}  Domain-specific languages \sep{} Neverlang
	\end{keyword}
\end{frontmatter}

\thispagestyle{fancy}

\section{Introduction}\label{sec:introduction}

\vspace{0.5em}\paragraph*{Context.}
Programming languages require editing support for a proficient use~\cite{Barros22}.
This applies to both \textit{general-purpose} (GPL) and \textit{domain-specific languages} (DSL).
Modern \textit{integrated development environments} (IDE) and \textit{source-code editors}\footnote{For sake of brevity, from here onwards, when not otherwise specified, we will use the term editor meaning both IDE and code editor.} provide editing support capabilities---e.g., highlighting, code completion, and hovering---but the development of such support is complex and time-consuming~\cite{Rodriguez-Echeverria18}.
Supporting \(\fontc{L}\) languages across \(\fontc{E}\) editors requires to implement editing features for each language-editor combination, resulting in a total of \(\fontc{L} \times \fontc{E}\) implementations.
Language development effort is therefore burdened by the editing support for each targeted editor, whereas editing features can overlap across editors~\citet{Rask21}: the risk is introducing inconsistencies and useless implementation overhead.
Among many results achieved over the years to improve upon this aspect, we mention type systems implementations~\cite{Bettini16b, Bettini19}, \textit{language workbenches}~\cite{Erdweg13b} and modular language development~\cite{Basten15, Mernik05, Parr09, Cazzola15c}.
\citet{Bettini16} demonstrated that type systems are key components for language editing support. For instance, type inference rules can be used to provide inlining hints or error messages and suggestions thereof.

\begin{figure}[t]
	\centering
	\includegraphics[width=1\linewidth]{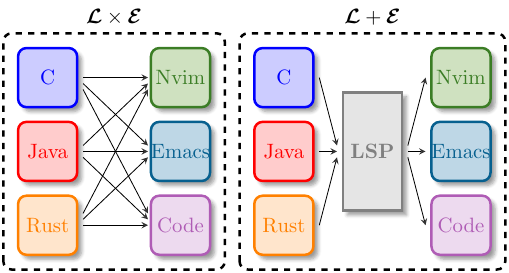}
	\caption{Traditional vs LSP approach to language editing support}%
	\label{fig:introduction:lsp_combination}
\end{figure}

\vspace{0.5em}\paragraph*{LSP as a solution.}
In 2016, Microsoft, along with RedHat and Codenvy, proposed the LSP\footnote{https://microsoft.github.io/language-server-protocol}~\cite{Gunasinghe21} as to limit the editing support implementation effort. The LSP model reduces the number of combinations to be implemented from \(\fontc{L} \times \fontc{E}\) to \(\fontc{L} + \fontc{E}\) by decoupling the implementation of the language editing support from the editor itself. It is based on a component dubbed \textit{language server} which communicates with the editor through an editor-dependent \textit{LSP plugin} (as shown in Fig.~\ref{fig:introduction:lsp_combination}), both adhering to the LSP specification.
The LSP achieves reuse of language services across different editors, but the overlap among different languages remains unaddressed: the number of language servers and the LSP plugins remain \(\fontc{L}\) and \(\fontc{E}\) respectively.
Type systems development for the language server and LSP plugin generation is complex and usually carried out manually and monolithically in a top-down fashion~\cite{Bettini16, Bettini19}.
Recently, researchers explored \textit{software product lines} (SPL) and their impact on editors~\cite{Cazzola19,Cazzola23d}, which could enhance the modularity and reusability of LSPs and their type systems.
SPLs applied to programming languages lead to the creation of \textit{language product lines} (LPL)~\cite{Cazzola15f}---\emph{i.e.} SPLs in which each product is a language variant. LPLs have been used in both GPLs~\cite{Cazzola16, Cazzola16i, Cazzola15f, Cazzola24d} and DSLs~\cite{Cazzola10b, Haugen08, Liebig13, Cazzola14e, Wende09, White09, Cazzola20} development.

\vspace{0.5em}\paragraph*{Modularization and Reusability Issues.}
Notwithstanding this, the editing support is still coupled to the editor ecosystem and can not be reused across different editors---for example, support implemented for Eclipse cannot be reused in IntelliJ IDEA\@. Most language workbenches are integrated with their native editor and most of them do not provide editing support with modularization in mind. Neverlang~\cite{Cazzola19} is the first language workbench to leverage LPLs towards the reusability of the editing support, yet limited to the Eclipse IDE~\cite{Cazzola19}. \toolname{Xtext} is the only language workbench that generates the language server for the LSP support~\cite{Bunder19a, Bunder19} through \toolname{XTypeS}~\cite{Bettini11}, and its successor \toolname{XSemantics} for the type system definition~\cite{Bettini13, Bettini16, Bettini16b, Bettini19}. Yet, \toolname{Xtext} does not support modular language development and the language server implementation cannot be spread across language artifacts. The main challenge, as \citet{Cazzola23b} reported, lies in the overlap between the implementations of similar linguistic components, which stems from the \textit{monolithic} design of language servers. It hinders their extensibility and reusability in other languages. Maintaining consistency between the language server and the language itself demands time and effort. While \(\fontc{L} + \fontc{E}\) is a linear number of combinations, the number of editor \(\fontc{E}\) still remains significant and cannot be overlooked. As shown in Table~\ref{ttab:problem-statement:lw-comparison}, none of the analyzed language workbenches provides support for LSP plugin generation, nor for modular language servers.

\vspace{0.5em}\paragraph*{Contribution.}
In this work, we propose \toolname{Typelang}, a family of domain-specific languages for the implementation of type systems in a modular, composable and reusable way.
We demonstrate how the language server implementation can be modularized and reused by automatically generating it for each language artifact with a specific \toolname{Typelang} variant.
The \toolname{Typelang} family follows the \textit{variant-oriented programming} paradigm and manages a set of interdependent software components through the \textit{cross-artifact coordination} layer. This approach is particularly suited to complex software systems, which pose numerous challenges~\cite{Holl12, Krueger06}.
The former defines properties for the integration of product variants within a software system as first-class citizens. The latter provides properties for integrating different variants across various artifacts---self-contained software components that may belong to a product line---to address issues of modularity and reusability.
Finally, we demonstrate that the proposed approach can reduce the number of combinations required to provide editing support for \(\fontc{L}\) languages in \(\fontc{E}\) editors. Initially, the combinations decrease from \(\fontc{L} + \fontc{E}\) to \(\fontc{T}\times\mathbf{1}\), where \(\fontc{T}\) represents the set of type systems associated with the languages. This is further reduced to \(\fontc{N}\times\mathbf{1}\) where \(\fontc{N} << \fontc{T}\)\footnote{
	The notation \(N << T\) means that the number of type systems \(N\) is \textit{much smaller} than the number of languages \(T\).
} and \(\fontc{N}\) is the number of type systems without overlaps, achieved by reusing the language artifacts---and their type system---across different languages.
The number of editors is reduced to \(\mathbf{1}\) because the LSP client generator can automatically generate the LSP plugin for any editor.

\newcolumntype{H}{>{\setbox0=\hbox\bgroup}c<{\egroup}@{}} 

\begin{table*}[t]
    \centering
    \resizebox{.7\linewidth}{!}{%
    \rowcolors{1}{white}{gray!20}
    \renewcommand{\arraystretch}{1.1}
        \begin{tabular}{lccHcccc}
            \toprule
            \multicolumn{1}{>{\bfseries}p{1.8cm}}{\centering Language Workbench} &
            \multicolumn{1}{>{\bfseries}c}{\multirow{2}{*}{Modularity}} &
            \multicolumn{1}{>{\bfseries}p{2.2cm}}{\centering Separate Compilation} & &
            \multicolumn{1}{>{\bfseries}c}{\multirow{2}{*}{IDE}} &
            \multicolumn{1}{>{\bfseries}p{2.0cm}}{\centering LS Generation} &
            \multicolumn{1}{>{\bfseries}p{2.0cm}}{\centering LS Modularity} &
            \multicolumn{1}{>{\bfseries}p{1.8cm}}{\centering LSP Plugin Generation} \\
            \midrule
            JustAdd~\cite{Ekman07b}       & \LEFTcircle     & \Circle     & \Circle        & \Circle           & \Circle        & \Circle   & \Circle   \\
            Melange~\cite{Degueule15}     & \(\circledwedge\) & \Circle     & \ourstar       & \makecell{3rd p.} & \ourstar       & \ourstar  & \ourstar  \\
            MontiCore~\cite{Gronninger08} & \LEFTcircle     & \LEFTcircle & \Circle        & \CIRCLE           & \Circle        & \Circle   & \Circle   \\
            MPS~\cite{Volter11}           & \(\circledwedge\) & \Circle     & \ourstar       & \CIRCLE           & \ourstar       & \ourstar  & \ourstar  \\
            Rascal~\cite{Klint09}         & \Circle         & \Circle     & \Circle        & \CIRCLE           & \Circle        & \Circle   & \Circle   \\
            Spoofax~\cite{Visser10}       & \(\circledwedge\) & \LEFTcircle & \ourstar       & \CIRCLE           & \ourstar       & \ourstar  & \ourstar  \\
            Xtext~\cite{Bettini13b}       & \Circle         & \LEFTcircle & \CIRCLE        & \CIRCLE           & \CIRCLE        & \Circle   & \Circle   \\
            Neverlang~\cite{Cazzola14c}   & \(\circledvee\)   & \CIRCLE     & \ourstar       & \makecell{3rd p.} & \ourstar       & \ourstar  & \ourstar  \\
            \bottomrule
        \end{tabular}
        }
        \caption{Comparison of language workbenches in terms of modular development support, separate compilation support, IDE support, language server generation, language server modular development, and LSP plugin generation. The \(\CIRCLE\) symbol indicates full support, \(\Circle\) no support, \(\LEFTcircle\) limited support, \(\circledvee\) fine-grained modularization,  \(\circledwedge\) coarse-grained modularization, and \ourstar\@ indicates no support; however, based on our motivation, it should theoretically be achievable.}%
        \label{ttab:problem-statement:lw-comparison}
\end{table*}

\vspace{0.5em}\paragraph*{Evaluation.}
To evaluate our approach, we extended Neverlang to support the \toolname{Typelang} family of DSLs and the LSP plugin generation.
Using \toolname{Typelang}, we modularly defined the type system for \toolname{SimpleLanguage}\footnote{https://github.com/graalvm/simplelanguage}---a general-purpose language that is part of the GraalVM project~\cite{Wurthinger13}---and Neverlang~\cite{Cazzola15c, Cazzola14c} itself. Then, we generated the plugins for three editors (\toolname{Visual Studio Code}, \toolname{NeoVim}, and \toolname{Vim}).
We analyzed the impact of \toolname{Typelang} on the language server generation process, which is entirely based on \toolname{Typelang}: we compared the amount of code needed to implement a type system in Neverlang before and after the introduction of \toolname{Typelang} by calculating the percentage reduction in \textit{lines of code} (LoC) and in the \textit{number of characters} (NoC). We also calculated LoC and NoC needed to develop LSP plugins for the three considered editors to estimate the effort needed to support a new editor.
The degree of reusability of type systems across different languages is measured with two metrics: \textit{Normalized Absolute Reuse Degree} and \textit{Operator Conditional Reuse Degree}.

This work is validated by answering the research questions:\smallskip
\begin{compactdesc}
	\item[\(\fontrq{RQ}\)\textit{\textsubscript{1}}] \textit{To what degree is it possible to streamline by associating variants to language artifacts?}
	\item[\(\fontrq{RQ}\)\textit{\textsubscript{2}}] \textit{To what degree is it possible to automate the generation of LSP clients, lowering \(\fontc{E}\) to \(\mathbf{1}\)?}
	\item[\(\fontrq{RQ}\)\textit{\textsubscript{3}}] \textit{Can the language server be automatically generated starting from the \toolname{Typelang} variants lowering \(\fontc{L}\) to \(\fontc{N}\)?}\smallskip
\end{compactdesc}

\vspace{0.5em}\paragraph*{Structure.}
The rest of the paper is organized as follows. Sect.~\ref{sec:background} provides foundation, terminology, and background information. Sect.~\ref{sec:theoretical-overview} introduces the \textit{variant-oriented programming} paradigm and \textit{cross-artifact coordination} layer, providing a formal definition of the concepts. Sect.~\ref{sec:typelang} presents the \toolname{Typelang} family of DSLs and how it can be used to implement type systems in a modular, composable, and reusable way.
Sect.~\ref{sec:case-study} presents the case study and provides the answers to the research questions.
Sect.~\ref{sec:lsp} shows how the language server generation can be modularized and reused by automatically generating it for each artifact part of a language variant using its specific \toolname{Typelang} variant. Sect.~\ref{sec:related-work} discusses related work. Finally, Sect.~\ref{sec:conclusion} concludes the paper and outlines future work.

\begin{figure}[t]
    \centering
    \includegraphics[width=1\linewidth]{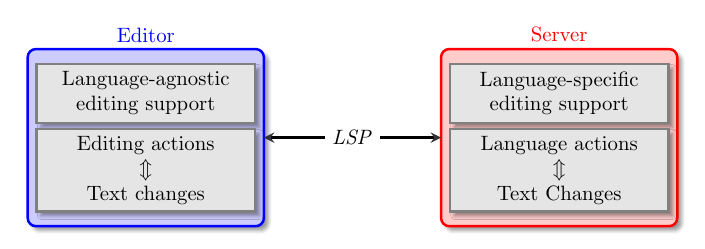}
    \caption{LSP approach to language support. Borrowed from~\cite{Rodriguez-Echeverria18b}}%
    \label{fig:background:lsp_diagram}
\end{figure}

\section{Background}\label{sec:background}
We briefly introduce the main concepts and technologies relevant to our work.

\subsection{Language Server Protocol}
In 2016 Microsoft, RedHat and Codenvy defined the LSP\@. It is a JSON-RPC based protocol that describes the communication between a language server code editor using an LSP plugin. The LSP permits to decouple a language-agnostic editor from the language-specific features, as shown in Fig.~\ref{fig:background:lsp_diagram}. The language server provides language-specific analysis, such as go to definition, semantic highlighting, and error checking, while the editor focuses on providing a user-friendly interface and managing the overall development environment. This separation of concerns allows developers to use their preferred editor while benefiting from language-specific features provided by the language server. Consequently, the LSP reduces the number of combinations to implement the language server and editing support from \(\fontc{L}\times\fontc{E}\) to \(\fontc{L}+\fontc{E}\), where \(\fontc{L}\) is the number of languages and \(\fontc{E}\) is the number of editors (see Fig.~\ref{fig:introduction:lsp_combination})~\cite{Rodriguez-Echeverria18b, Rask21}.
This reduction is achieved by decoupling the implementation of the editing support from the editor. Since its introduction, the LSP has gained significant traction within the development community. Many popular programming languages\footnote{https://microsoft.github.io/language-server-protocol/implementors/servers} now have language servers, and various editors support the protocol, making it a \textit{de facto} standard for language support.

The set of commands that a language server can handle is defined by the LSP specification\footnote{https://microsoft.github.io/language-server-protocol/specifications/lsp/3.17/specification}.
The specification defines four categories of commands: \textit{language features}, \textit{text document synchronization}, \textit{workspace features}, and \textit{window features}.
The LSP client usually triggers the commands in response to user actions, such as opening a file, typing, or interacting with the editor's UI\@.
The \textit{language features} respond to requests related to the language semantics, such as code completion, hover information, and go to definition.
These requests are sent as a tuple of \texttt{TextDocument} and \texttt{Position} objects, where the \texttt{TextDocument} represents the content of the file being edited, and the \texttt{Position} represents a specific location in the file.
The \textit{text document synchronization} category is responsible for keeping the language server up-to-date with the editor's content via notifications, such as \texttt{textDocument/didOpen} to notify the server that a document has been opened, \texttt{textDocument/didChange} to notify the server that a document has been modified, and \texttt{textDocument/didClose} to notify the server that a document has been closed.
The \textit{workspace features} category provides commands to interact with the workspace. A workspace is a collection of files and folders that are opened in the editor. The commands in this category allow the language server to interact with the workspace, such as \texttt{workspace/symbol} to search for symbols in the workspace, and \texttt{workspace/configuration} to retrieve the workspace configuration.
The \textit{window features} category provides commands to interact with the editor's UI, such as \texttt{window/showMessage} to display a message to the user, \texttt{window/showMessageRequest} to display a message with a set of actions, and \texttt{window/logMessage} to log a message to the editor's console.

\begin{Listing}[t]
   \setminted[neverlang]{escapeinside=``}
   \showneverlang*{Backup.nl}\vskip -10pt%
   \caption{Syntax and semantics for the backup task.}\label{lst:backup}
   \newcommand{\nt}[1]{%
      \node[draw, fill=none, shape=circle, inner sep=0pt, minimum width=.375cm] (c\i) at (s\i) {};%
      \coordinate (d\i0p) at ($(d\i0)+(-1pt,2pt)$);
      \coordinate (d\i1p) at ($(d\i1)+(1pt,-3pt)$);
      \node[draw, thick, shape=rectangle, name=d\i, inner sep=1pt, fit=(d\i0p) (d\i1p)] {};
   }
   \begin{tikzpicture}[overlay, thick, BloodRed]
      \coordinate (s0) at ($(s0)+(2pt,0pt)$) ;
      \foreach \i in {0, 1, 2} {%
         \nt{\i}
         \draw[-stealth, rounded corners=7pt] (c\i) -- (d\i);
      }
   \end{tikzpicture}
\end{Listing}

\subsection{Neverlang in a Nutshell}
The \textit{Neverlang}~\cite{Cazzola12c, Cazzola13e, Cazzola15c} language workbench promotes code reusability and separation of concerns in the implementation of programming languages, based on the \textit{language-feature} concept.
The basic development unit is the \textbf{module}, as shown in line 1 of Listing~\ref{lst:backup}.
A module may contain a \textbf{reference syntax} and could have zero or multiple \textbf{role}s. A role, used to define the semantics, is a composition unit that defines actions that should be executed when some syntax is recognized, as defined by \textit{syntax-directed translation}~\cite{Aho86}.
Syntax definitions are defined using \textit{Backus-Naur form} (BNF) grammars, represented as sets of \textit{productions} and \textit{terminals}.
Syntax definitions and semantic \inlineneverlang{role}s are tied together using \inlineneverlang{slice}s.
Listing~\ref{lst:backup} shows a simple example of a Neverlang module implementing a backup task of the \texttt{LogLang} LPL\@. Reference syntax is defined in lines 2--6; the \textit{categories} (line 5) are used to generate the syntax highlighting for the IDEs.
Semantic actions may be attached to a non-terminal using the production's label as a reference, or using the position of the non-terminal in the grammar, as shown in line 8, numbering start with 0 from the top left to the bottom right.
The two \texttt{String} non-terminals on the right-hand side of the \texttt{Backup} production are referenced using 1 and 2, respectively.
Each \inlineneverlang{role} is a compilation phase that can be executed in a specific order, as shown in line 24.
In contrast, the \texttt{BackupSlice} (lines 14--18) reveals how the syntax and semantics are tied together; choosing the \inlineneverlang{concrete syntax} from the Backup module (line 15), and two \inlineneverlang{role}s from two different modules (lines 16--17).
Finally, the \inlineneverlang{language} can be created by composing multiple \inlineneverlang{slices} (line 20).
The composition in Neverlang is twofold: between modules and between slices. Thus, the grammars are merged to generate the complete language parser. On the other hand, the semantic actions are composed in a pipeline, and each \inlineneverlang{role} traverses the syntax tree in the order specified in the \inlineneverlang{roles} clause (line 24).

\subsection{Software and Language Product Lines}
Variability-rich software systems development leverages principles from product line engineering, commonly referred to as \textit{feature-oriented programming}~\cite{Prehofer97} and \textit{software product line}~\cite{Clements01} engineering. An SPL consists of a family of software products, where their similarities and differences are characterized by their features. A feature is a unit that provides a piece of functionality that satisfies a requirement, represents a design decision, or corresponds to a stakeholder's interest.
A key task in SPL engineering is feature modeling, which involves creating and maintaining a \textit{feature model}. The concept of feature model was first introduced by \citet{Kang90} in the FODA method and serves to represent the variability of a system through its features and their interdependencies. In SPLs, the feature model formalism is essential for configuring software products by defining valid feature sets, known as \textit{configurations}. A feature is considered \textit{active} if it belongs to the selected subset of features in a configuration, while all other features are deemed \textit{inactive}. The structure of a feature model implicitly captures feature dependencies by specifying mandatory, optional, alternative, and grouped features, alongside parent-child relationships, where a feature can only be \textit{active} if its parent features are also \textit{active}.

On the other hand, the development of families of programming languages and DSLs has gained popularity among researchers and practitioners~\cite{Ng11, Kuehn14, Crane05, Zschaler09}. Similar to other software, DSL interpreters and compilers can be designed around the concept of product line. When a SPL is applied to the implementation of a programming language, each product corresponds to a language variant taking the name of \textit{language product lines}~\cite{Cazzola15f}. LPLs, widely used~\cite{Cazzola23d, Cazzola21b, Cazzola20}, have been successfully used in both GPLs~\cite{Cazzola16, Cazzola16i, Cazzola15f} and DSLs~\cite{Haugen08, Cazzola13g, Cazzola14e, White09}. Feature-oriented programming~\cite{Apel13, Czarnecki04, Prehofer01} embraces the idea of modularizing software systems into feature modules, which encapsulate specific functionality and can be composed with other feature modules to form a software system; similar to an aspect module that encapsulates a crosscutting concern in \textit{aspect-oriented programming}~\cite{Kiczales01, Kiczales97, Laddad03}. Using feature-oriented programming in language development, a family of languages~\cite{Liebig13} can be defined by composing feature modules~\cite{Wende09}, and a language can be seen as a product of the family.
A special case of a product line is the \textit{multi product line}~\cite{Rosenmuller10, Rosenmuller08, Rosenmuller11}, where multiple product lines are integrated into another software product line.

\section{Foundational Concepts Overview}\label{sec:theoretical-overview}
As noted by~\citet{Holl12}, ``\textit{managing a set of interdependent software product lines involves numerous challenges}'', \emph{e.g.} ensuring coexistence among product variants within the same system.
Poorly coordinated variant management can increase system complexity and hinder maintainability.
To mitigate this, it is essential to adopt programming and architectural approaches that explicitly support per-artifact variant integration, thereby simplifying the process and minimizing the need for source code modifications to either the system or the variants.
To this end, we propose the \textit{variant-oriented programming} paradigm and the \textit{cross-artifact coordination} layer as two components that enable the integration of product variants into a complex system.

\subsection{Variant-Oriented Programming Paradigm}\label{sec:variant-oriented-programming}

\begin{figure*}[t]
\centering
\includegraphics[width=.9\linewidth, keepaspectratio]{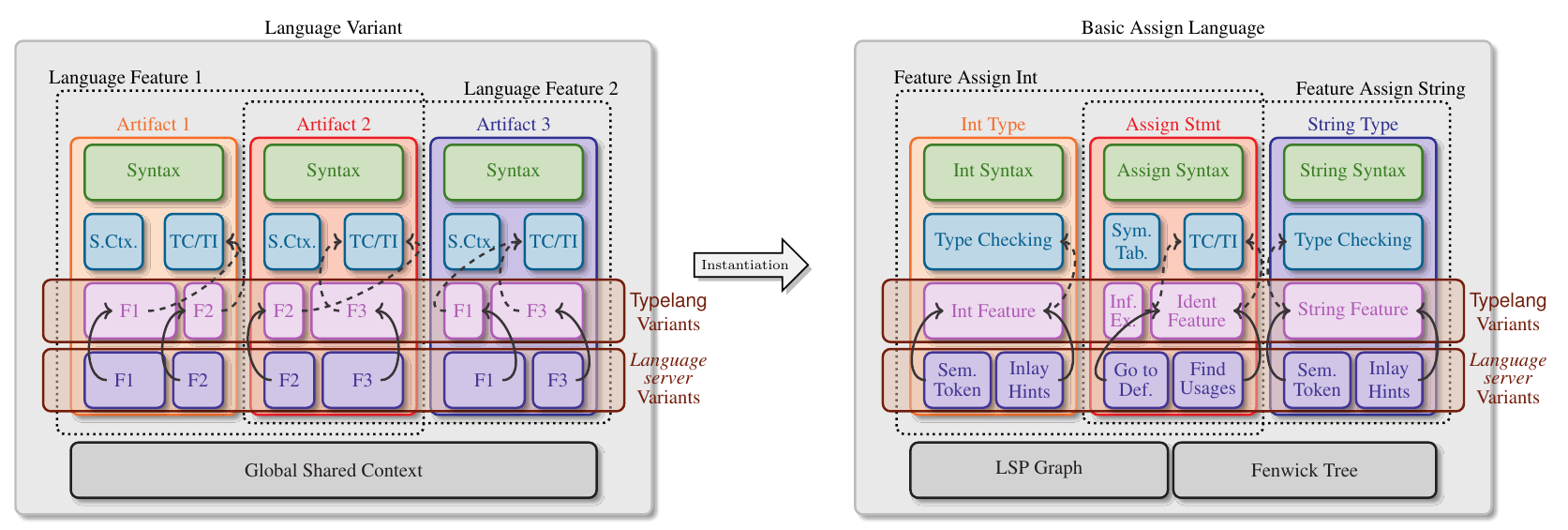}
\caption{The diagram on the left illustrates the two dimension of variability: (1) \textit{vertical dimension}: each artifact has its own syntax and semantics defined through a grammar and set of rules. The \texttt{S.\,Ctx.} represents the artifact's shared context. Additional semantic for type checking (TC) and type inferencing (TI) are specified using the \toolname{Typelang} DSL\@. The arrows pointing to the TC/CI rules indicate that the \toolname{Typelang} features (F1, F2, etc.) used by the artifact can be inferred from those rules. Finally, the LSP features can also be derived from the \toolname{Typelang} features employed by the artifact. (2) \textit{Horizontal dimension}: this dimension pertains to the \textit{cross-artifact coordination} layer. The \toolname{Typelang} features used by each artifact represent a specific variant within the \toolname{Typelang} family, and the same applies to the corresponding LSP features. The \textit{global shared context} is depicted by the gray box at the bottom of the diagram.\\
On the right, a concrete instantiation of the abstract diagram is shown to clarify the concept. This example models a simple \textit{assignment} language with three artifact: \texttt{Int Type}, \texttt{String Type}, and the \texttt{Assign Stmt}. The \texttt{Sym.\,Tab.} is the \textit{shared context} concretely defined by the \texttt{Assign Stmt} artifact.
}%
\label{fig:lsp:module_with_lsp}
\end{figure*}

\paragraph*{Overview.}
We define the \textit{variant-oriented programming} as a programming paradigm in which product variants are treated as first-class entities within a unique software system, referred to as \textit{variant-oriented software}.
\begin{quack}
  Fig.~\ref{fig:lsp:module_with_lsp} illustrates, both abstractly and concretely, the \textit{variant-oriented programming} paradigm, in which a (modular) language variant---referred to as \textit{variant-oriented software}---is defined as a subset of artifacts associated with one or more features for the \toolname{Typelang} and the LSP variant.
These variants are defined by the \toolname{Typelang} features and LSP languages, respectively.
\end{quack}
\noindent This approach aims to improve reusability, modularity, and maintainability of software systems. A \textit{variant-oriented software} is designed to support multiple product variants by providing \textit{shared contexts} that drive the interoperability among variants.
Variants interact with the shared contexts to perform their tasks, and the shared contexts can either be global or local.
Thus, the paradigm revolves around two concepts: \textit{variants} and \textit{shared contexts}.
\begin{quack}
  In Fig.~\ref{fig:lsp:module_with_lsp}, the \textit{shared contexts} are depicted as the artifact's \texttt{S.\,Ctx.} and \texttt{Sym.\,Tab.}, which represents its \textit{local} contexts. The \textit{global shared contexts} is illustrated by the gray box at the bottom of the diagram, while the \textit{local shared contexts} are shown as gray boxes that do not overlap the vertical division.
\end{quack}
\noindent \textit{Variants} are self-contained, independent entities with a feature-based behavior and identity~\cite{Booch07}, and may also have an internal state. The \textit{Variants} can either be explicitly defined or derived from a set of features, though the latter approach does not guarantee the validity of the derived variants.\ \textit{Shared contexts} are abstract data types that provide operations enabling the interaction between variants and defining the conditions under which they can interact. These contexts can either have a global scope, making them accessible by all variants, or a local scope, restricting access to a subset of variants.
\begin{quack}
    The dashed lines in Fig.~\ref{fig:lsp:module_with_lsp} (from the \toolname{Typelang} features to the \texttt{TC/CI} rules) indicate the possibility of deriving the \toolname{Typelang} features from those used in the \texttt{TC/CI} rules. Similarly, the dashed lines from the LSP features to the \toolname{Typelang} features suggest that LSP features can be derived from the corresponding \toolname{Typelang} features.
\end{quack}

\vspace{0.5em}\paragraph*{Conceptual Example.}
Let us take into consideration a \textit{type‑checker} product line, which is a family of tools that includes two distinct variants of type checking. For instance, consider a \textit{type checker} with two variants:
\begin{inparaenum}
  \item \textit{assignment‑statement checker}, and
  \item \textit{expression checker}.
\end{inparaenum}
Both variants traverse their portions of the AST to enforce type safety, each focusing on a distinct node.
Now, consider a \textit{compiler pipeline}---representing the \textit{variant‑oriented software}---which programmatically\footnote{We assume the \textit{compiler} is configured at setup time rather than at runtime.} selects the \textit{assignment‑statement} variant for modules where the correctness of variable bindings is critical, and the \textit{expression} variant for modules where complex expression evaluation requires more thorough analysis.
Assuming the AST---serving as the \textit{shared context}---can be checked by both variants, the compiler must be designed to support both strategies and define a clear dispatching policy for assigning each AST node to the appropriate checker. For instance, during the front‑end pass the compiler might:
\begin{compactitem}
  \item apply the \textit{assignment‑statement checker} to all \texttt{assignment} nodes (ensuring that the left‑ and right‑hand sides are type-compatible);
  \item apply the \textit{expression checker} to nodes such as \texttt{BinaryOp}, \texttt{FunctionCall}, and other expression constructs (verifying operand types and return types).
\end{compactitem}
This strategy interweaves detection of mis-assigned variables with validation of expression correctness across the AST\@.

\vspace{0.5em}\paragraph*{Integration of Variants.}
The \textit{variant-oriented programming} paradigm deliberately leaves the definitions of \textit{variants} and \textit{shared contexts} open. This design choice enables the creation of programming languages that embody the paradigm, referred to as \textit{variant-oriented programming languages}. For example, such a language might automatically handle shared contexts and allow variants to be defined using a syntax like:\begin{center}\texttt{foo = \textit{variant} X \textbf{of} Foo {\ldots}},\end{center} where \texttt{Foo} denotes a family of related products.
\begin{quack}
  The \toolname{Typelang} language variant can be defined either explicitly, using a syntax such as \begin{center}\texttt{foo = \textbf{variant} X \textbf{of} Typelang {\ldots}},\end{center} or implicitly, as illustrated in Fig.~\ref{fig:lsp:module_with_lsp}, through a set of features that uniquely characterize the variant. The same approach applies to the LSP language variant.
\end{quack}

The primary goal of this paradigm is to provide strong encapsulation of concerns while promoting interoperability among product variants. By reducing coupling between variants, it enhances the cohesion of the software. At the source code level, \textit{separation of concerns}~\cite{Videira-Lopes95} is achieved through shared contexts, which facilitate the integration of independent variants within a \textit{variant-oriented software}.
In \textit{variant-oriented programming}, encapsulation is supported by the design of self-contained variants. These variants can be integrated into a \textit{variant-oriented software} system without requiring changes to the existing codebase or the variants themselves. The paradigm directly addresses the challenge of integrating product variants in complex software systems~\cite{Damiani19, VanOmmering01}.
When a \textit{variant-oriented software} system serves as a product line, the \textit{variant-oriented programming} paradigm can be seen as an instance of the \textit{multi product line} model~\cite{Rosenmuller10, Rosenmuller08, Rosenmuller11}, enabling interoperability of variants in complex systems.

However, not all product line variants can be seamlessly integrated into a \textit{variant-oriented software}, and not all software systems can support multiple variants of a given product line. Therefore, the concept of \textit{integration} is central to the \textit{variant-oriented programming} paradigm, and certain properties must be satisfied to ensure successful integration.

\vspace{0.5em}\paragraph*{Integration Properties.}
Given a \textit{variant-oriented software} implementation \(\mathcal{S}\), a set of \textit{shared contexts} \(C_{\mathcal{S}} = \{\Gamma_1, \Gamma_2, \ldots, \Gamma_n\}\) (where each \(\Gamma_i\) is independent), and the variants \(v_i\) and \(v_j\) of a product family \(P\), we define the \textit{contextual compatibility} relation \(v_i \rightleftharpoons_{\Gamma}^{\mathcal{S}} v_j\). This relation holds between \(v_i\) and \(v_j\) in \(\mathcal{S}\) wrt.\ a shared context \(\Gamma\in C_{\mathcal{S}}\) if and only if:
\begin{compactenum}
   \item \(v_i\) and \(v_j\) are \textit{independent};
   \item they can \textit{coexist} simultaneously in \(\mathcal{S}\);
   \item they are \textit{semantically interoperable} in \(\mathcal{S}\); and
   \item they can \textit{cooperate} on \(\Gamma\) simultaneously.
\end{compactenum}

\noindent A variant-oriented software system must ensure the \textit{coexistence} of variants in \(\mathcal{S}\) during the feature selection phase of the product line engineering process~\cite{Metzger14}.
If two variants cannot coexist, they cannot be integrated into \(\mathcal{S}\) and it should be reported as a constraint violation. Constraints must be explicitly declared so that verification tools can detect such violations.
\textit{Independence} of variants must be guaranteed by \(P\) or \(\mathcal{S}\), especially if \(\mathcal{S}\) generates the variants of \(P\).
When \(P\) is responsible for generating the variants, it could be designed in such a way that the variants are inherently independent or Constraints can be defined to ensure independence.
On the other hand, if \(\mathcal{S}\) is responsible for generating the variants, it must enforce independence by design. It could be achieved through design-time analysis (e.g., static verification of configurations) or runtime isolation (e.g., sandboxing, namespace separation).
As \citet{Briand99} pointed out for object-oriented systems, \textit{independence} is crucial for maintaining low coupling between objects. Also, the \textit{independence} of variants is essential for safe coexistence in \(\mathcal{S}\)---\emph{i.e.}, if two variants are not independent but must coexist in the same system, they may introduce conflicts (\emph{e.g.}, conflicting dependencies, logical contradictions, or runtime errors).
Similarly, \textit{independence} and \textit{coexistence} aim to minimize coupling between variants \(v_i\) and \(v_j\).
\textit{Semantic interoperability} ensures that the variants of \(P\) can be used interchangeably in \(\mathcal{S}\) without any semantic loss and they interact seamlessly, thereby increasing the \textit{cohesion} of \(\mathcal{S}\).
Finally, the \textit{cooperation} of variants on \(\Gamma\) must be determined by both \(\Gamma\) (e.g., enforcing mutual exclusion) and \(\mathcal{S}\) (\emph{e.g.}, preventing deadlocks and race conditions).
Thus, the set of variants of a product family \(P\) can be integrated in \(\mathcal{S}\), denoted as \(P\Rrightarrow\mathcal{S}\), if and only if \(\forall v_i, v_j \in P \mid \exists\,\Gamma\in C_{\mathcal{S}}:v_i \rightleftharpoons_{\Gamma}^{\mathcal{S}} v_j\).

\begin{quack}
    In Fig.~\ref{fig:lsp:module_with_lsp}, each artifact provides its own variant of \toolname{Typelang}. These variants are designed to be independent, meaning that they can be used in isolation without relying on other artifacts. For instance, the \texttt{Int Type} artifact can function independently of the \texttt{String Type} artifact, and vice versa.
    This independence enables multiple variants to coexist within the same language variant, allowing developers to select only the specific artifacts relevant to their projects.
    \textit{Semantic interoperability} among variants is ensured by the locality of the artifact's \textit{shared context}, represented by \texttt{S.\,Ctx.} and \texttt{Sym.\,Tab}.
    For instance, the \texttt{Assign Stmt} artifact expects \texttt{Sym.\,Tab.} to be in a specific state during execution.\ \textit{Semantic interoperability} ensures that switching the \toolname{Typelang} variant used in \texttt{Assign Stmt} does not affect the \texttt{Sym.\,Tab.} state.
    \textit{Cooperation} among variants is supported by the \textit{global shared context}---the gray box at the bottom of the diagram.
\end{quack}

\subsection{Cross-Artifact Coordination Layer}\label{sec:cross-artifact-modularization}

\vspace{0.5em}\paragraph*{Overview.}
We define the \textit{cross-artifact coordination} layer as a layer that integrates independent product variants into artifacts of an artifact-based \textit{variant-oriented software} without modifying the code of either the software system or the variants.
The \textit{cross-artifact coordination} layer has as assumptions all the properties of the \textit{variant-oriented programming} paradigm.
A \textit{variant-oriented software} in the \textit{variant-oriented programming} paradigm definition could not be modular, and it could not have artifacts.
The \textit{cross-artifact coordination} layer, built on top of the \textit{variant-oriented programming} paradigm, requires that the software system is artifact-based.
This layer helps managing the complexity of integrating product variants across artifacts, particularly when a global shared context allows all product variants in a family to interact with a set of common resources.
\begin{quack}
    Fig.~\ref{fig:lsp:module_with_lsp} illustrates in the \textit{horizontal dimension} the \textit{cross-artifact coordination} layer, where the \toolname{Typelang} features used by each artifact form the associated variant of the \toolname{Typelang} family, and the same for the LSP features.
\end{quack}

\vspace{0.5em}\paragraph*{Conceptual Example.}
Let us bring back the \textit{type‑checker} example to illustrate the concept of \textit{cross‑artifact coordination layer}. Suppose that each \textit{module} (the artifact) in the \textit{compiler pipeline}---the artifact‑based \textit{variant‑oriented software}---has a variant of the \textit{type‑checker} product. Each module contains a finite number of \textit{type environments} (the shared contexts) that can be managed by the type‑checker variants. Note that type environments can be seen as shared contexts because they are resources that can be accessed by multiple variants, if a module admits multiple type‑checking strategies.
Now, consider the \textit{global program environment}---the \textit{global shared context}---with several type environments which determine the typing rules of the entire program. The global environment is connected to each module via interfaces. Each type environment is governed by the type checker in the nearest module, which can be either the \textit{assignment‑statement checker} or the \textit{expression checker} variant. An issue arises: how can the type‑checkers coordinate to enforce consistent typing across module boundaries? To avoid conflicting type conclusions in the global environment, a coordination layer is needed, managing the interaction between the type‑checkers and the global type environments. This layer must ensure that the variants can interoperate to maintain a coherent type system across the program.

\begin{quack}
    Fig.~\ref{fig:lsp:module_with_lsp} illustrates the \textit{LSP Graph} and the \textit{Fenwick Tree} as examples of \textit{global shared contexts}. Each variant populates these structures by adding the features it supports. For example, the assign statement artifact adds a node to the \textit{LSP Graph} representing the assigned variable. Additional edges---from variable's uses to its declaration---are added to the \textit{LSP Graph} by other artifacts.
\end{quack}

\vspace{0.5em}\paragraph*{Integration Properties.}
Using the notation from Sect.~\ref{sec:variant-oriented-programming}, let \(\mathcal{S}\) be an artifact-based \textit{variant-oriented software} and let \(A_{\mathcal{S}} = \{a_1, a_2, \ldots, a_n\}\) denote the set of artifacts in \(\mathcal{S}\).
Consider \(P\), a family of products such that \(P\Rrightarrow\mathcal{S}\).
The family \(P\) can leverage the \textit{cross-artifact coordination} layer if and only if, for every \(v_i\in P\) and \(a\in A_{\mathcal{S}}\), the following conditions are met:\smallskip
\begin{compactenum}
   \item \(a\) is defined in terms of a \(v_i\);
   \item \(a\) provides its specific shared context \(\Gamma\in C_{\mathcal{S}}\) with which \(v_i\) can interact; and
   \item a \textit{global shared context} of \(\mathcal{S}\), denoted as \(\Gamma_{\mathcal{S}}\), must exist to enable semantic interaction among all variants of \(P\).\smallskip
\end{compactenum}
The first property states that a part of an artifact \(a\), or its behavior, can be defined by a variant \(v_i\). Here, a part refers to an essential component of an artifact---one without which the artifact cannot be defined---while behavior denotes the rules governing the artifact's interaction with other artifacts.
The second property states that each artifact \(a\) must maintain a state with which the variant \(v_i\) can interact.
This interaction between \(v_i\) and the shared context \(\Gamma\) ensures that the variant has access to the resources it needs to perform its tasks within the artifact.
If \(v_i\) does not require interaction with \(a\)'s state, \(a\) a may still be defined by \(v_i\), though in such cases, the shared context may be empty or entirely omitted.
Lastly, the third property is crucial for integrating product variants into artifact-based \textit{variant-oriented software}. It guarantees that all variants within a product family can interact through a global shared context, enabling consistent and coordinated behavior across the system.

Note that,
\begin{inparaenum}
\item \(\mathcal{S}\) can include multiple families of products, each of which can be integrated into \(\mathcal{S}\) using the \textit{cross-artifact coordination} layer, provided the above properties are met,
  \item an artifact \(a\) cannot be defined in terms of more than one variant \(v_i\),
  \item the global shared context \(\Gamma_{\mathcal{S}}\) is not simply the union of all shared contexts of the artifacts in \(\mathcal{S}\), and
  \item \(\mathcal{S}\) can provide multiple global shared contexts \(\Gamma_{\mathcal{S}}\).
\end{inparaenum}

\begin{quack}
    In Fig.~\ref{fig:lsp:module_with_lsp}, as previously mentioned, the \textit{type checking} and \textit{type inferencing} behavior of the artifacts is defined by the variants in the horizontal dimension---i.e., the artifacts are defined in terms of the \toolname{Typelang} and LSP variants.
    The \texttt{S.\,Ctx.}\ and the concretely defined \texttt{Sym.\,Tab.}\ are the shared contexts of the artifacts, which are defined by the variants in the horizontal dimension.
    The language variant---the artifact-base \textit{variant-oriented software}---provides two global shared contexts: the \textit{LSP Graph} and the \textit{Fenwick Tree}.
\end{quack}

\subsubsection{Considerations}

The introduction of \textit{cross-artifact coordination} in artifact-based \textit{variant-oriented software} improves the management of complex product lines. This approach extends variability management from individual artifacts to interactions across multiple artifacts, fostering a comprehensive understanding of modular layers and their relationships across features. It enables seamless integration of variants within a product family and ensures consistent interoperability between artifacts across features and the entire product line. By enhancing modularization, \textit{cross-artifact coordination} layer provides coherence and structure in \textit{variant-oriented software} in spite of the potential high number of feature-artifact combinations.


\section{\toolname{Typelang}: Towards Type Systems Composition}\label{sec:typelang}
\citet{Bettini16} demonstrated that type systems are crucial for language editing support.
A type system is a set of rules that assigns a type to language constructs, ensuring program's correctness and the proper applications of operations.
Additionally, the type system can be used to infer the type of a construct and verify its compatibility with another  construct's type.
The overlap between linguistic components extends to their type systems. In this section, we introduce \toolname{Typelang}, a family of DSLs for defining type systems in a modular, composable, and extensible way.\ \toolname{Typelang} aims to enhance the reusability of type system definitions by leveraging the \textit{variant-oriented programming} paradigm and a \textit{cross-artifact coordination} layer.\ \toolname{Typelang} ensures the reusability of type system definitions by coupling \toolname{Typelang} variants to language artifacts.

\subsection{Overview.}
\toolname{Typelang} is a family of DSLs hosted by a language workbench for language syntax and semantics definition. It provides rules for
\begin{inparaenum}
	\item type definition,
	\item type checking,
	\item type inferencing, and
	\item error catching.
\end{inparaenum}

Its goal is to decouple type system definitions from a specific technological space, enabling artifacts to be used across different language workbenches. A specific \toolname{Typelang} variant can define the type system for a whole language variant or of a number of its language artifacts. Therefore, multiple variants of \toolname{Typelang} can be used to target different language artifacts within the same language workbench\@ to define their type systems. The following example illustrates a snippet of \toolname{Typelang} code.
\showtl{TypelangExample.tl}
\noindent It is used to infer the type to an identifier \inlinetl{<id>} with respect to the expression \inlinetl{<expr>}. Then, it checks if the inferred type is invariant with the type of the expression.
\begin{Listing}[t]
	\subfloat[core module\label{lst:typelang:typelang_grammar_fixed}]{\showbnf{typelang1.ebnf}}
	\vspace{1em}
	\subfloat[mutable hooks\label{lst:typelang:typelang_grammar_variable}]{\showbnf{typelang2.ebnf}}
	\vspace{1em}
	\caption{EBNF Grammar for \toolname{Typelang}}%
	\label{lst:typelang:typelang_grammar}
\end{Listing}
In \toolname{Typelang}, the grammar (Listing~\ref{lst:typelang:typelang_grammar}) consists of an immutable core providing essential functionalities and mutable hooks that allow adaptation to language-specific requirements. By integrating language-specific features with the immutable core, \toolname{Typelang} can be configured into a tailored variant within its product line.

\subsection*{Typelang as a Family of Domain-Specific Languages.}

Typelang is best understood not as a single DSL, but as a \textit{family} of DSLs, organized according to a \textit{language product line} architecture.
Its design enables the systematic derivation of language variants for different domains. For example, in some domain-specific languages, the ability to declare functions may be unnecessary. In such cases, the corresponding Typelang variant can be instantiated without the function declaration feature. Conversely, certain application domains may require features that were not originally anticipated during the design of Typelang. The modular nature of the language allows for the extension of the feature set to accommodate such unforeseen requirements.

Modeling Typelang as a family of DSLs rather than a single monolithic DSL serves two primary purposes:\smallskip
\begin{compactdesc}
	\item \textbf{language restriction}---it enables the creation of minimal, domain-specific variants by omitting unnecessary language features, thereby simplifying the language surface;
	\item \textbf{language extension}---it supports the introduction of new, potentially unconventional or unforeseen features, enhancing Typelang’s ability to express a wide variety of type systems.
\end{compactdesc}

\subsubsection{Core Module.}
The fixed part of the grammar (Listing~\ref{lst:typelang:typelang_grammar_fixed}) is designed to support language server generation.
In \toolname{Typelang}, the entry point of the DSL is \inlinebnf{<program>}, which consists of a sequence of statements. New type bindings are defined using the \inlinebnf{"define"} keyword, which must be followed by two elements: a type---either a \inlinebnf{<lw type>} or a \inlinebnf{<type>}---and a \inlinebnf{<lw token>} to associate the type with a token. An optional \inlinebnf{<callback>} can be specified using the \inlinebnf{"then"} keyword, allowing the definition of a callback function that is triggered when the respective binding occurs.

\begin{figure*}[t!]
	\centering
	\includegraphics[width=\linewidth]{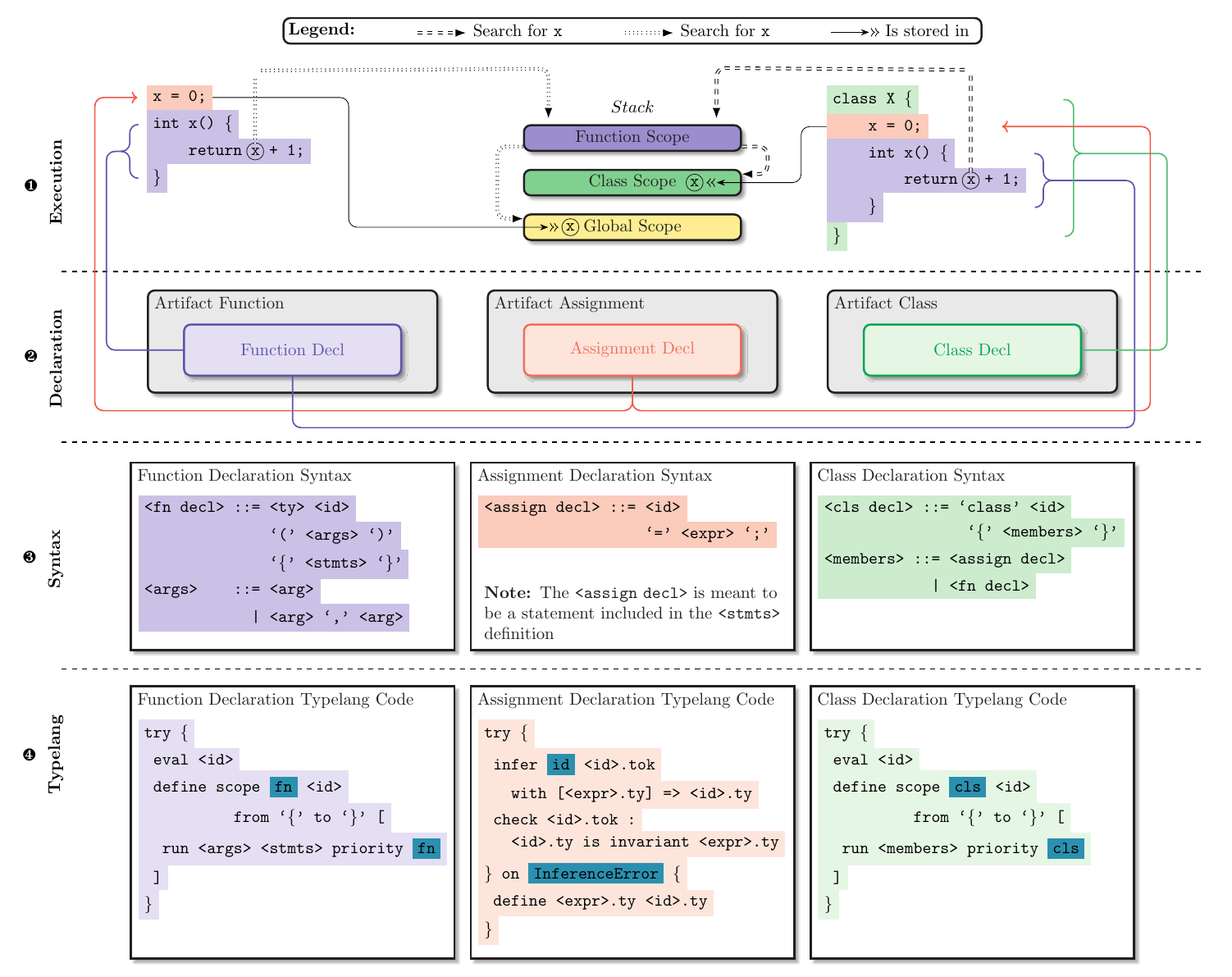}
	\caption{The diagram shows the four layers of \toolname{Typelang} in action.
		\textit{Execution Layer} \ding{182}: Divided into three vertical sections. The left shows a \textit{function scope} created within the \textit{global scope}; the right shows a \textit{function scope} within a \textit{class scope}. The center displays the contents of each scope and their hierarchical relationships.
		\textit{Declaration Layer} \ding{183}: Includes three artifacts—\textit{function}, \textit{class}, and \textit{assignment}. Each artifact contains the same information shown in Figure~\ref{fig:lsp:module_with_lsp}. Scopes are defined modularly, and the grammar supports nesting.
		\textit{Syntax Layer} \ding{184}: Defines the artifacts’ syntax modularly. From this grammar, the language workbench generates the \textit{hooks} used in the \toolname{Typelang} DSL\@.
		\toolname{Typelang} \textit{Layer} \ding{185}: Contains the \toolname{Typelang} code defining type checking and inference rules for the artifacts (denoted in Fig.~\ref{fig:lsp:module_with_lsp} by \texttt{TC/TI} boxes).\ \toolname{Typelang} features—called \textit{hooks}—are highlighted in teal.}%
	\label{fig:typelang:modular_type_system}
\end{figure*}

\vspace{0.5em}\paragraph*{Type Inference.}
Type inference rules in \toolname{Typelang} are defined using the \inlinebnf{"infer"} keyword, which depends on the presence of \inlinebnf{<signature>}s. The \inlinebnf{"infer"} keyword is followed by a \inlinebnf{<signature>} and a \inlinebnf{<lw token>} to associate the inferred type with a token. The \inlinebnf{<signature>}s are designed to support both \textit{structural type systems}~\cite{Cardelli88, Cook90} and \textit{nominal type systems}~\cite{Cooper22, Pierce02}. If the type is known in advance, the programmer can help the inference process specifying the type through the \inlinebnf{"=>"} operator, followed by a \inlinebnf{<lw type>} to indicate the type to be inferred. If the type constructor is not nullary\footnote{A nullary type constructor is one that does not take any arguments.}, the \inlinebnf{"with"} keyword can be used to specify its argument types. A type constructor can represent various types, including a \textit{product type}, a \textit{sum type}, a \textit{function type}, or a \textit{parametric polymorphic type}. As an example, Fig.~\ref{fig:typelang:modular_type_system} layer \ding{185} presents the \toolname{Typelang} code used to define the type inference rules for the assignment statement defined in layer \ding{183}. Likewise, Section~\ref{sec:case-study} shows our implementation of \toolname{Typelang} with the type inference rule for the assignment statement in Neverlang.
For example, the following snippet defines a type inference rule for a hypothetical assignment statement that infers the type of the identifier \inlinetl{<id>} from the type of the expression \inlinetl{<expr>}.
\showtl{TypeInference.tl}

\vspace{0.5em}\paragraph*{Type Checking.}
Type checking in \toolname{Typelang} is defined by using the \inlinebnf{"check"} keyword. This keyword is followed by a \inlinebnf{<lw token>} and two types, separated by the variance type. The \inlinebnf{<lw token>} associates the type checking rule with a token. The developer can explicitly define the variance type between the two types using the \inlinebnf{"covariant"}, \inlinebnf{"contravariant"}, or \inlinebnf{"invariant"} keywords. Similar to type inference rules (shown in boxes denoted with \texttt{TC/TI} in Fig~\ref{fig:lsp:module_with_lsp}), Fig.~\ref{fig:typelang:modular_type_system} layer \ding{185} presents the \toolname{Typelang} code defining the type checking rules for the assignment statement, while Section~\ref{sec:case-study} showcases our Neverlang implementation.
For example, the following snippet defines a type checking rule for a hypothetical assignment statement that ensures the left-hand side and the right-hand side share the same type.
\showtl{TypeChecking.tl}

\vspace{0.5em}\paragraph*{Scope Management.}
As~\citet{Cooper22} explained, scopes are crucial for defining the visibility and lifetime of bindings.\ \toolname{Typelang} enables the programmer to define scopes at the artifact level using the \inlinebnf{"define scope"} keyword. The \inlinebnf{<scope>} represents the name of the scope, and the \inlinebnf{<lw token>} associates the scope with a token. Two elements can be specified: \inlinebnf{<range>} and \inlinebnf{<priority>}. The former uses the \inlinebnf{"from"} and \inlinebnf{"to"} keywords followed by two terminals (e.g., the \texttt{\{} and \texttt{\}} tokens) to set the folding range for the language server. The latter specifies the order in which scopes are executed; each \inlinebnf{<scope>} automatically defines a priority level, but the programmer can specify the scope’s priority within square brackets using the \inlinebnf{"run"} keyword, followed by a sequence of non-terminals, the priority level, and an optional \inlinebnf{<callback>}. This configuration indicates that all the non-terminals should be executed with the specified priority level. Note that the language workbench must provide a way for the programmer to define a total order between priority levels. The \inlinebnf{"enter"} and \inlinebnf{"exit"} keywords, followed by the \inlinebnf{<scope>}, are used to define the scope's entry and exit points. Typically, this involves pushing and popping the scope from a stack.
For example, the following snippet defines a scope named \inlinetl{module} with a priority level with the same name, which is executed upon entering the \inlinetl{module} scope
\showtl{ScopeManagement.tl}

\vspace{0.5em}\paragraph*{Error Handling.}
Some operations may fail during the type checking and type inference phases. The \inlinebnf{"try"} and \inlinebnf{"on"} keywords are used to define error-handling rules. If an error occurs during the \inlinebnf{"try"} block, the \inlinebnf{"on"} block is executed. If the \inlinebnf{"on"} block is not specified, the error is automatically caught and passed to the \textit{compilation helper} (see Sect.~\ref{sec:typelang:collection-assembling}).

\vspace{0.5em}\paragraph*{Root of the Scope Hierarchy.}
Finally, the \inlinebnf{"init"} keyword is used to define the root of the scope hierarchy tree, followed by the \inlinebnf{<scope>} to specify the primary scope. Note that if the \inlinebnf{<scope>} used in the \inlinebnf{"init"} keyword should have the highest priority level because it is the first scope to be executed during the type checking and type inference phases.

\subsubsection{Mutable Hooks.}
The variable part of the grammar (Listing~\ref{lst:typelang:typelang_grammar_variable}) entails the concepts of \inlinebnf{<scope>}, \inlinebnf{<type>}, \inlinebnf{<signature>}, and \inlinebnf{<callback>}. For example, an implementation of the \texttt{int} type can be hooked to \inlinebnf{<type>} to create a \toolname{Typelang} variant with definition, inference, and integers type checking capabilities. This enables two main reuse opportunities:
\begin{inparaenum}
	\item the resulting \toolname{Typelang} variant can be leveraged to define multiple type systems involving similar types; and
	\item The \texttt{int} type can be reused in different \toolname{Typelang} configurations.
\end{inparaenum}
The concrete implementations of \inlinebnf{<scope>}, \inlinebnf{<signature>}, \inlinebnf{<type>}, and \inlinebnf{<callback>} hooks can be developed independently, allowing them to be reused across various \toolname{Typelang} variants.

\vspace{0.5em}\paragraph*{\toolname{Typelang} Configuration.}
Configuring a \toolname{Typelang} DSL and implementing \toolname{Typelang}-based type systems primarily focus on the \inlinebnf{<scope>} and \inlinebnf{<type>} concepts. For instance, if a language artifact defines two scopes (\textit{function} and \textit{class}) and two types (\textit{int} and \textit{float}), then the corresponding \toolname{Typelang} variant grammar will include the following definitions:
\showebnf{es-variant.ebnf}
\noindent Such types are then involved in the definition of \inlinebnf{<signature>}s, to provide a way to specify the expected types of an expression.
The \inlinebnf{<signature>}s hook can be used to decorate untyped parse trees or abstract syntax trees with type definitions and inference rules.
For example, consider the function call \inlinerust{let res = add(1, 2);} in \toolname{Rust}, where \inlinerust{add} is a \textit{parametric polymorphic} function over the type \inlinerust{T}, defined as:
\showrust{add.rs}
\noindent Although the monomorphization phase has not yet been performed,\footnote{Monomorphization typically occurs after the type checking and inference phases.} the \inlinebnf{<signature>} \inlinerust{add: (T, T) -> T} can be used to infer the type of the \inlinerust{res} identifier based on the argument types of the function call.

The type system can be further refined through the \inlinebnf{<callback>} hook, where each callback serves as a unique identifier for a function invoked in response to a specific event.
Each of these core concepts is abstract, and the corresponding \toolname{Typelang} rule is flexible, enabling programmers to define their own \inlinebnf{<scope>}, \inlinebnf{<type>}, \inlinebnf{<signature>}, and \inlinebnf{<callback>} elements. The semantics of the variable parts of \toolname{Typelang} variants are not predetermined; their implementation is left to the language workbench\@.
In Listing~\ref{lst:typelang:typelang_grammar_variable}, such cases are indicated by the phrase \textit{``defined by the language workbench\@'',} meaning the left-hand side of the production rule is fixed, while the right-hand side must be provided by the language workbench.\footnote{We assume that the fixed part of the \toolname{Typelang} grammar is in a language workbench-compatible format; otherwise, it must be adapted accordingly.}
For example, \inlinebnf{<t>} and \inlinebnf{<nt>} serve as placeholders for terminal and non-terminal symbols associated with syntax tree nodes, as defined by the language workbench\@. Similarly, \inlinebnf{<lw type>} and \inlinebnf{<lw token>} fulfill analogous roles, referencing syntax tree nodes. However, this distinction highlights whether nodes are accessed directly or used to retrieve their type (\inlinebnf{<lw type>}) or source code (\inlinebnf{<lw token>}).
The \inlinebnf{<exception>} construct is designed to be modular and can be raised during the type checking and inference phases.
The language server can catch these exceptions and provide feedback to the editor.

\subsection{Collection and Assembling Phases}\label{sec:typelang:collection-assembling}

\vspace{0.5em}\paragraph*{Overview.}
Inspired by the \textit{application engineering} phase of \textit{product line engineering}~\cite{Pohl05, VanDerLinden07}, the \toolname{Typelang} DSL relies on: the \textit{collection} phase and the \textit{assembling} phase.
During the \textit{collection} phase, the language workbench gathers all features specified through \textit{product configurations}. These configurations can be explicitly defined by the programmer or implicitly derived from the semantics of the language artifacts, following the principles of the \textit{variant-oriented programming} paradigm. This phase determines how the \toolname{Typelang} variant should be generated. The \textit{assembling} phase combines the concepts of \textit{product derivation}~\cite{Rabiser10,Deelstra05,Hotz06,Griss00,Rabiser11} and \textit{product validation}. In this phase, the language workbench creates the \toolname{Typelang} variant for a given language artifact based on the features collected during the \textit{collection} phase. By assembling the \toolname{Typelang} variant before the compilation phase, the necessary language features are made available during the compilation of the language artifacts.

\vspace{0.5em}\paragraph*{Collection Phase.}
In our proposal, the \textit{collection} phase gathers all definitions to generate the \toolname{Typelang} variant for a given language artifact. The variant generation is demanded to the \textit{assembling} phase. Any definition that should be used during \textit{assembling} phase but not collected by the \textit{collection} phase will cause the \textit{assembling} phase to fail. Therefore, the \toolname{Typelang} DSL grammar is not fixed: it depends on the language artifact requirements and it is defined before the artifact is processed.
We define a \textit{feature box} as a concrete implementation of a feature. A feature box contains all the necessary information to define the variable part of the \toolname{Typelang} DSL grammar. A feature box combines concepts from \textit{feature modules}~\cite{Batory04, Kastner11} in feature-oriented programming and \textit{delta modules} in delta-oriented programming~\cite{Damiani14, Koscielny14}.
From feature-oriented programming, we inherit the concept of a \textit{feature module} as a modular unit that encapsulates a feature definition~\cite{Apel08}. From delta-oriented programming, we inherit the concept of a \textit{delta module} as a unit that specifies changes to be applied to the core module to implement further products by adding, modifying, and removing code~\cite{Schaefer10}.

\vspace{0.5em}\paragraph*{Feature Box.}
Each variable definition must be encapsulated within its own feature box, such as a language workbench module or a Java class. Feature boxes should be independent but composable when needed. A feature box acts as a provider of a definition and contains all relevant LSP information associated with that definition.
For example, a feature box can define a \inlinebnf{<type>} and specify whether it serves as an LSP \textit{semantic token}---a token enriched with semantic information used to provide language-aware editor features such as semantic highlighting and code navigation.
A concrete example of a feature box is shown in Listing~\ref{lsp:type_function}.
To facilitate development, language workbenches should provide default feature boxes for common definitions while allowing developers to define new ones modularly. This approach suggests that an increasing number of feature boxes will likely result in more \toolname{Typelang} variants.\footnote{Note that adding new feature boxes may not necessarily lead to new variants.} All feature boxes are designed to be reusable across multiple language artifacts to form distinct \toolname{Typelang} variants. These variants can be applied across different languages and, by extension, across various language variants. However, a feature box may depend on---or rather, require---the existence of another feature box to work correctly. Each feature box can be viewed as a provider of a definition, and dependencies between them can be considered \textit{requirements} necessary to utilize the definitions provided by other feature boxes. This brings forth to the need for a \textit{composer}. The \textit{composer}, e.g., a language workbench language compilation unit, is responsible for collecting all the necessary feature boxes to generate \toolname{Typelang} variants.

\vspace{0.5em}\paragraph*{Assembling Phase.}
The \textit{composer} is responsible for distributing all necessary information to the internal components of the language workbench to support the type checking and type inferencing compilation phases using a specific \textit{assembled} \toolname{Typelang} variant. A proper \toolname{Typelang} variant can be automatically generated for each language artifact, as the \textit{collection} phase solely collects the features actually used in the semantics related to type checking and type inferencing rules. This approach avoids the need for an explicit, verbose form---such as \texttt{ty\_1 = \textbf{variant} \{F1, F2, \dots\} \textbf{of} ty}---favouring a desugaring mechanism that generates the variant based on the features used in the semantics of the language artifact. With this approach, the \textit{product configuration}, \textit{product derivation}, and \textit{product validation} phases of the \textit{application engineering} phase can be automated. Figure~\ref{fig:lsp:module_with_lsp} shows the \textit{assembling} phase for a given language artifact (the dashed gray arrows).
For instance, the type checking semantics of Artifact 1 uses the \toolname{Typelang} feature F1, while the type inferencing semantics relies on feature F2. The \textit{composer} collects the feature boxes required by F1 and F2 to generate the appropriate \toolname{Typelang} variant for Artifact 1. The teal-colored layer (Fig.~\ref{fig:typelang:modular_type_system} \ding{185}) highlights the \toolname{Typelang} features used by different artifacts, introduced as \textit{hooks} in the \toolname{Typelang} grammar.
Indeed, the other \toolname{Typelang} code shown in Fig.~\ref{fig:typelang:modular_type_system} is the \textit{core} of the \toolname{Typelang} variant.
Each variant only includes the features needed to define the type checking and type inference rules for the corresponding artifact and remains unaware of features used by other artifacts. The type checking and type inference semantic actions assume that the required features are present in the \toolname{Typelang} LPL and that the \textit{composer} can successfully collect them. A number of issues may arise during the \textit{collection} and \textit{assembling} phases. For instance, if a feature box declaration---such as \texttt{fn}, \texttt{id}, and \texttt{cls} in layer \ding{185} of Fig.~\ref{fig:typelang:modular_type_system}, or the \texttt{identifier} used in Listing~\ref{lst:typelang:AssignStatement}---is not found during the \textit{collection} phase, it should be treated as a parsing error.

\subsection{\toolname{Typelang} Integration in Modular language workbenches}\label{sec:typelang:integration}

\begin{Listing}[t]
	\centering
	\showjava*[\linewidth]{SymbolTableEntry.java}
	\caption{The \texttt{SymbolTableEntry} interface.}%
	\label{lst:typelang:SymbolTableEntry}
\end{Listing}

Using the notation introduced in Sect.~\ref{sec:theoretical-overview}, let the \textit{variant-oriented software} \(\mathcal{S}\) represent a modular language workbench, and let \(P\) denote the \toolname{Typelang} LPL\@. The \textit{shared contexts} \(\Gamma_i\in C_\mathcal{S}\) are the \textit{typing environments}. A \textit{typing environment} is a map from identifiers to symbol table entries, which contain information such as the identifier type and additional metadata (details in Sect.~\ref{sec:lsp}).
As shown in Listing~\ref{lst:typelang:SymbolTableEntry}, a symbol table entry is a data structure that contains information about the type associated to the entry, the location in the source code, and other relevant information.

\begin{figure}[t]
	\centering
	\includegraphics[width=\linewidth]{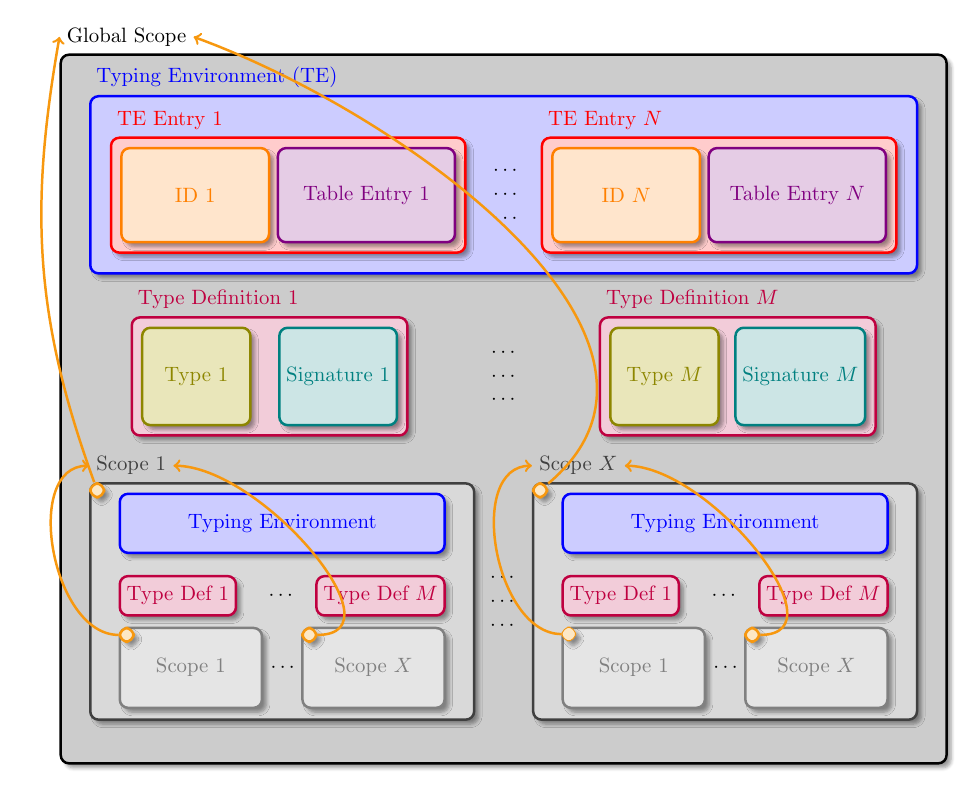}
	\caption{The modularity and extensibility of \toolname{Typelang}, illustrating the independent definition and composition of scopes, types, and signatures as reusable building blocks that can be assembled into language-specific variants at compile-time. The diagram demonstrates a simplified version of \toolname{Typelang} scopes structure, with each scope associated with a typing environment that contains the unordered pairs of identifiers and table entries and types.}%
	\label{fig:typelang:type_system}
\end{figure}

\vspace{0.5em}\paragraph*{Variant-oriented Programming.}
The \textit{composer} generates \toolname{Typelang} variants in accordance with the principles of \textit{variant-oriented programming}. It enforces the first property, that is, that each variant is generated for its language artifact irrespective of the other variant generations. Moreover, each variant is also unaware of the other variants existence. The \textit{coexistence} property is satisfied because \(\mathcal{S}\), as a modular language workbench, perceives \(v_i\) and \(v_j\) as different configurations of the same DSL, which are already used in modularized language artifacts. The language workbench enforces constraints on the \toolname{Typelang} variants through internal APIs, which serve as an abstraction layer to interact with the variants. The language workbench does not need to be aware of the specific \toolname{Typelang} variant used in a given language artifact; instead, it only needs to know how to interface with the variants and enable their cooperation within the same shared context \(\Gamma\). As a result, all \toolname{Typelang} variants are considered \textit{semantically interoperable} within \(\mathcal{S}\), as they can be used interchangeably. The definition of a scope involves the definition of a \textit{typing environment} \(\Gamma\). It is the responsibility of \(\mathcal{S}\) to ensure correct access to \(\Gamma\) by the \toolname{Typelang} variants. This may require mechanisms such as mutual exclusion to regulate access, handling race conditions to prevent data corruption, and addressing potential issues related to data consistency.

\vspace{0.5em}\paragraph*{Modular Scopes.}
Fig.~\ref{fig:typelang:modular_type_system}, layer \ding{185}, illustrates the \toolname{Typelang} code, where the \textit{artifact function} and \textit{artifact class} define their own scopes using distinct \toolname{Typelang} variants. The eligibility for scope nesting is directly derived from the grammar (layer \ding{184}) defined at the artifact level. In this example, both the \textit{artifact class} and the \textit{global scope} grammars permit function scope definitions. Consequently, the relationships between scopes do not need to be predetermined; instead, they can be established at compile time based on the grammar of the language artifact used in the language variant. Similarly, the \textit{assignment declaration} is valid within all scopes, including the \textit{global scope}, \textit{function scope}, and \textit{class scope}. In the left-side of layer \ding{182}, the variable \(\texttt{x}\) is declared within the \textit{global scope}, whereas in the right-side, it is declared within the \textit{class scope}. When \(\texttt{x}\) is used, its value is retrieved from the nearest enclosing \textit{nested lexical scope}.\footnote{According to~\citet{Cooper22}, ``scopes that nest in the order they are encountered in the program are called lexical scopes''.} In the first case, the value is retrieved from the \textit{global scope}, while in the second case, it is retrieved from the \textit{class scope}. According to~\citet{Cooper22}, the \textit{execution layer} can be entirely generated at compile time based on the \toolname{Typelang} scopes defined within the artifact, even if the scopes are modularly defined. Proper data structures that dynamically track the most recent binding of a free variable\footnote{A \textit{free variable} is a name defined outside the current scope.} must be generated. Being the artifacts self-contained ensures that all required feature boxes are included and foster their reuse across different language implementations. Fig.\ref{fig:typelang:type_system} expands and abstracts the stack shown in the center of Fig.~\ref{fig:typelang:modular_type_system}, illustrating the content of scopes and their hierarchical relationships. This basic representation could be further detailed with elements such as \textit{callbacks} and \textit{exceptions}, although this is beyond the scope of this work.

\vspace{0.5em}\paragraph*{Cross-artifact Coordination.}
Once \toolname{Typelang} variants are generated and the type systems are defined, the language workbench can use them like a conventional type system for type checking and type inference phases. 
Scopes in \toolname{Typelang} are defined at the artifact level, each providing multiple type definitions along with an associated \textit{typing environment} \(\Gamma\). A type definition consists of the type itself and its associated signature. With blue boxes shows in Fig.~\ref{fig:typelang:type_system}, \(\Gamma\) is represented as a map linking identifiers to symbol table entries. The global scope typing environment is initialized after the invocation of \inlinebnf{"init"} \inlinebnf{<scope>}, so that the context \(\Gamma\) and type definitions are defined at the artifact level. Each artifact \(a \in A_{\mathcal{S}}\) provides its own interoperability context \(\Gamma\in C_{\mathcal{S}}\), where variant \(v_i\) can operate, thus satisfying the first property of the \textit{cross-artifact coordination} layer. As shown by Fig.~\ref{fig:typelang:modular_type_system}, layer~\ding{185}, it is possible for an artifact \(a\) to omit scope definitions.
The omission of scope definitions occur for several reasons:
\begin{compactitem}
	\item a language does not require scopes---e.g., a simple calculator language;
	\item not all artifacts want to define namespaces---e.g., an assignment statement;
	\item an artifact depends, after the \textit{composition} phase, on the scopes defined by other artifacts---e.g., a function call.
\end{compactitem}
Moreover, each artifact \(a\) can be defined in terms of a variant \(v_i\), given that its type checking and type inferencing semantics are derived from \(v_i\). Consequently, as long as the behaviors of \(a\) are defined with \toolname{Typelang}, the artifacts are indirectly defined by the \toolname{Typelang} variants. Although not depicted in Fig.~\ref{fig:typelang:type_system}, the global shared context \(\Gamma_{\mathcal{S}}\) acts as a repository for common resources accessible to all \toolname{Typelang} variants for seamless interaction. This context could be implemented as a priority queue containing \textit{compilation unit tasks} (further details in Sec.\ref{sec:lsp}), which are executed lazily and in increasing priority order. Variants interact by adding their tasks to the queue using the \inlinebnf{"run"} keyword.

\begin{Listing}[t]
   \showjava*{TypeFunction.java}\vskip -5pt
   \caption{The \inlinejava{TypeFunction} feature box.}%
   \label{lsp:type_function}
\end{Listing}

\section{Generative Modular LSP Design}\label{sec:lsp}

\vspace{0.5em}\paragraph*{Overview.}
The LSP defines a common API for language servers, enabling a single implementation to be used across multiple editors (clients). It operates on a \textit{client-server} model, communicating over \textit{pipes} or \textit{sockets}.
Language server implementations are created manually in a \textit{top-down} fashion, with significant time invested in complex data structures and algorithms~\cite{Gunasinghe21}. In contrast, we propose a \textit{bottom-up}~\cite{Cazzola16i}, modular approach to generate language servers. This method leverages the variant-oriented programming and cross-artifact coordination principles. LSP capabilities are conceptualized as separate \textit{feature boxes}, composed at the artifact level for a more flexible and reusable implementation.

\vspace{0.5em}\paragraph*{LSP Features and Feature Boxes.}
The features of the language server variant---represented as solid gray arrows in Fig.~\ref{fig:lsp:module_with_lsp}---are derived from the \toolname{Typelang} feature. Since a \toolname{Typelang} feature box could require multiple LSP features, the language server variant is composed of the union of the features provided by the feature boxes used by the artifact. Thus, the language server variant is tailored to the artifact's needs, without unnecessary features and \textit{feature boxes} contain the necessary information to provide LSP capabilities. The language server implementation emerges through a generative process involving the definition of feature boxes, their \toolname{Typelang} representations, and their composition.
Fully modular feature boxes---\textit{i.e.}, without endogenous bindings---extend their modularization and reusability to the LSP capability implementations, so that type system components can be reused across languages.
Finally, this approach (Fig.~\ref{fig:background:lsp_diagram}) takes advantage of the LSP’s language-agnostic nature~\cite{Niephaus20, Rodriguez-Echeverria18}. Since the protocol imposes no restrictions on language server implementation, each LSP implementation needs to comply with the protocol specification, with no concern for the other implementations.

\subsection{Language Server Generation Process}\label{lsp:building_blocks}
In the LSP context, the implementation of a language server can be seen as a \textit{generative} process based on \toolname{Typelang} feature boxes. As explained in Sect.~\ref{sec:typelang}, the feature boxes encapsulate all the needed information to generate the language server implementation.

\vspace{0.5em}\paragraph*{Feature Box Definition.}
For example, Listing~\ref{lsp:type_function} shows the \inlinejava{TypeFunction} feature box which implements the LSP capabilities for \textit{semantic tokens}, \textit{document symbols}, and \textit{inlay hints}. In our implementation, the \textit{collector} mechanism is realized as an annotation processor. Specifically, the \inlinejava{@TypeAnnotation} annotation and the value \inlinejava{Type.FN} (line~5) tell the annotation processor to bind the \inlinejava{TypeFunction} Java class to a feature box of kind function type. The \inlinejava{Type.FN} identifier in the annotation corresponds to the kind of feature box and becomes part of the variable section of \toolname{Typelang}'s grammar whenever the feature box is employed. The \inlinejava{@DocumentSymbol} (line~9) and \inlinejava{@InlayHint} (line~22) annotations mark specific methods, enabling the client to collect document symbols and provide inlay hints for the function type. At line 15, the \inlinejava{@SemanticToken} annotation is followed by a list of \inlinejava{SemanticTokenTypes}---a class provided by LSP4J---that specifies the token types, allowing the client to semantically highlight source code based on these types. Overall, Listing~\ref{lsp:type_function} consolidates all LSP capabilities related to the function type without assumptions on other language features: the feature box is then reusable across any language with functions, regardless of its broader structure.

\vspace{0.5em}\paragraph*{Feature Box Collection.}
All annotated classes are collected only if the feature box associated with the specified kind is actually used by \toolname{Typelang} in a semantic action. Consequently, LSP-related annotations are processed only when the corresponding feature box is employed in the artifact, and the code for the respective language server features is generated only under these conditions. This process is repeated for all feature boxes used by \toolname{Typelang} within an artifact and subsequently for all artifacts associated with a language implemented in the language workbench\@. The union of all capabilities specified by the feature boxes constitutes the active LSP features for the language server variant of the artifact. Other modular language workbenches can implement the \textit{collector} mechanism and similar methods using alternative strategies, such as visitor patterns~\cite{Gamma95}.

\subsection{Building Blocks for the Language Server}
A language server implementation requires in-memory runtime data structures to properly work, providing all the necessary information to the LSP client.
The language workbench is in charge of providing the necessary data structures to generate the language server implementation.
Among these data structures, some are automatically populated by \toolname{Typelang} during the execution of the tasks, while others are still automatically populated, but are used by the language server to provide correct responses to the LSP client requests.
This section provides an overview of the data structures that the language workbench must provide to generate the language server implementation.

\vspace{0.5em}\paragraph*{Table Entry.}
The table entry is at the core of compiler design and type systems.
Out interface for the \textit{Table Entry} is shown in Listing~\ref{lst:typelang:SymbolTableEntry}.
It must contain at least the following fields to provide the information to the LSP client:
\begin{compactitem}
    \item the \textit{entry type}, which represents a symbol---ideally a lexical token---in the source code;
    \item the \textit{entry kind}, which represents the kind of the usage of the symbol in the source code---e.g., a definition or a reference to a symbol;
    \item the \textit{entry location}, which represents the location of the symbol in the source code;
    \item the \textit{entry range}, which represents the range of the symbol in the source code.
\end{compactitem}
Some feature box provided capabilities need a \textit{Table Entry} to compute a query and to generate the language server implementation (see Listing~\ref{lsp:type_function} at line 10, 15 and 23).
In addition to the fields, the implementation includes routines for computing information required by the LSP client. For example, the \textit{isAssignableFrom} routine determines whether the right-hand side of an assignment is compatible with the left-hand side, enabling the client to deliver precise diagnostics.
Further, behind the method invocation at line 24 in Listing~\ref{lsp:type_function}, the \textit{getSignature} routine is called to provide the signature of the symbol used in the inlay hint.
Hovering capabilities are provided by the \textit{getHover} routine, which is used to provide relevant information upon hovering over a symbol in the code.
The LSP provides a wide range of capabilities that can be implemented by the language server, for more details see the documentation for either the LSP or LSP4J.
All the pieces of information regarding table entries are collected within the \textit{typing environment} (shown in Fig.~\ref{fig:typelang:type_system}); the typing environment is a data structure that maps the relevant symbols in the source code to their respective table entry.
The language workbench generates the code to populate the table entry at runtime. Thus, the language server only needs to access the data structure and provide the information to the client.

\vspace{0.5em}\paragraph*{Compilation Unit.}
For each scope defined in \toolname{Typelang}, a \textit{Compilation Unit} is created and associated to a \textit{Compilation Unit Task}. The compilation unit is a data structure that contains:
\begin{compactitem}
    \item the \textit{scope}, which contains the typing environment \(\Gamma\) associated with it;
    \item a reference to the \textit{stack} of scopes shown in Fig.~\ref{fig:typelang:modular_type_system};
    \item the \textit{type inference strategy}~\cite{Gamma95}, used by the language workbench to infer the types of the expressions in the scope;
    \item the associated \textit{compilation unit task} with the associated priority of the task;
    \item the \textit{parent compilation unit}, which is used to ensure the tasks are executed in the correct order.
\end{compactitem}
The language workbench generates the code to populate the compilation unit at runtime.
The compilation unit serves as a container for the tasks that are executed to bring the compilation unit to a fully typed state via the type inference.
Without a fully typed compilation unit, the language server cannot provide correct information to the client.
Combining compilation units and incremental parsers with error recovery~\cite{Ammann78, Graham79, Richter85}, a small modification made to the source code triggers the execution of only the tasks related to the compilation unit and its scope, as well as any of its dependencies. Other tasks do not need to be repeated.

\begin{figure*}[t]
    \centering
    \includegraphics[width=.7\linewidth, keepaspectratio]{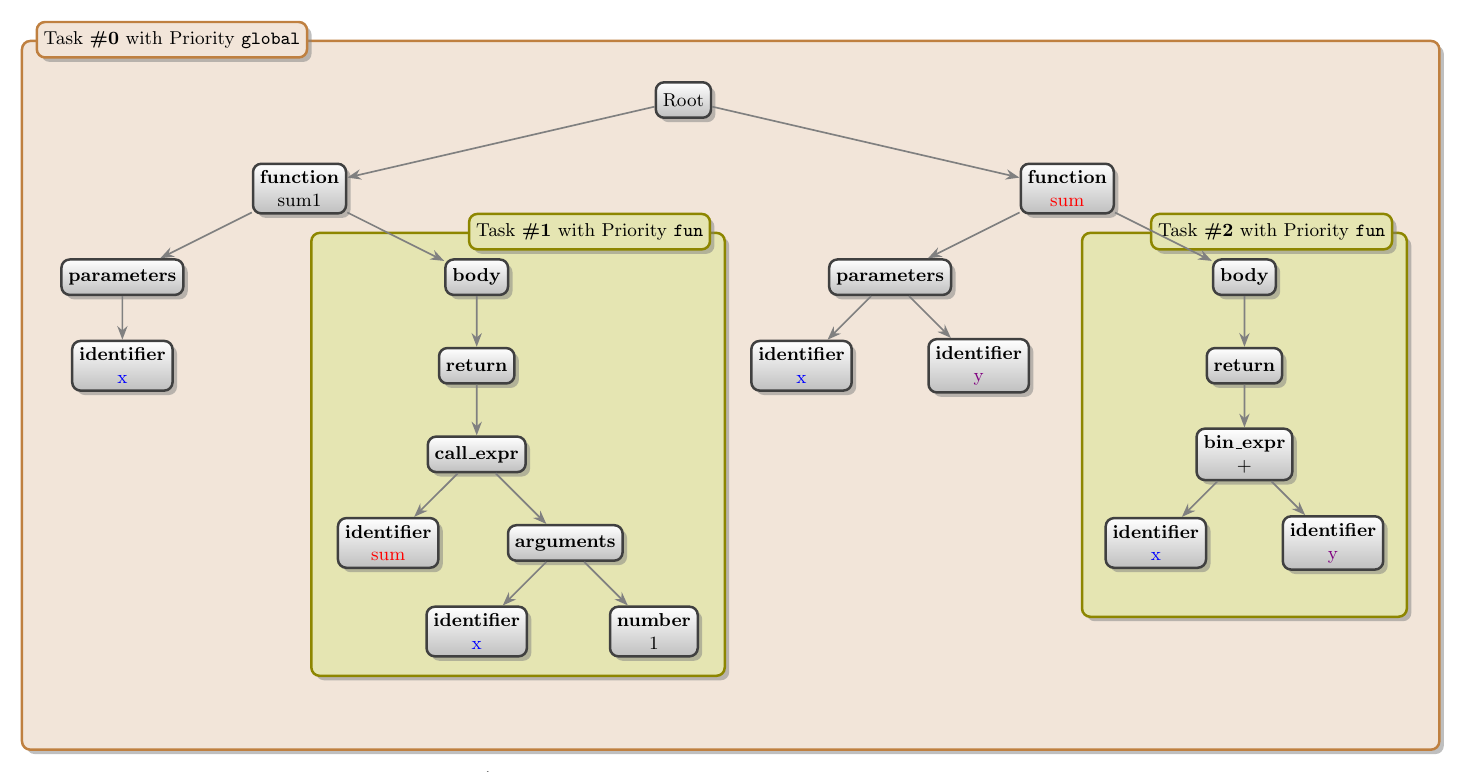}
    \caption{The diagram shows an AST with the \textit{compilation unit tasks} associated.}%
    \label{fig:typelang:ast_with_priority}
\end{figure*}

\vspace{0.5em}\paragraph*{Compilation Unit Task.}
The \textit{compilation unit tasks} are needed to solve the following limitation: in any given moment, a compilation unit may not be fully typed---\emph{e.g.}, because the type of a symbol is not yet defined or was not inferred yet. Such an example is shown in Fig.~\ref{fig:typelang:ast_with_priority}:
the \texttt{sum} identifier in Task \#1 (on the left side) with priority \texttt{fun} is highlighted in red because it is not typed yet.
The language server cannot provide correct information to the client if the compilation unit is not fully typed.
The compilation unit task component contains the following fields:
\begin{compactitem}
    \item the reference to the \textit{language workbench context}, used to access a mutable reference of the AST\@;
    \item a procedure,\footnote{A procedure is a function that does not return a value.} which accept a \textit{language workbench context} and performs some operations on it, such as modifying the AST\@;
    \item the \textit{priority} of the task.
\end{compactitem}
As discussed in Sect.~\ref{sec:typelang}, the priority is used to order the tasks in the language workbench\@. To ensure modularity, the feature boxes are unaware of the existence of any other priorities except their own.
However, the language workbench and the language server must have a global knowledge of all existing priorities to order the tasks correctly, thus bringing the compilation unit to a fully typed state.
This dependency should be declared exogenously, using a language construct that specifies the evaluation order of tasks without affecting the individual feature boxes.
This solution embraces the \textit{Hollywood principle}~\cite{Sweet85, Fowler05c} design pattern\footnote{``\textit{Don't call us, we'll call you}'' is what the Hollywood Principle is all about---known also as Hollywood's Law or Inversion of Control. It is a design pattern that allows low-level components to hook themselves into a system, but the high-level components determine when they are needed, and how.} guaranteeing that the separation of concerns and the reusability are preserved.
To provide a visual representation of the tasks and their priorities, Fig.~\ref{fig:typelang:ast_with_priority} presents an AST with associated \textit{compilation unit tasks} denoted as boxes around the AST nodes. By assuming a left-to-right depth-first AST traversal, a task associated with the global scope is added to \(\Gamma_{\mathcal{S}}\) with the highest priority.
The function \texttt{sum1} is processed before the function \texttt{sum}. Since the body of \texttt{sum1} calls \texttt{sum} with the formal parameter \texttt{x} and an integer literal \texttt{1} as actual parameters, this may result in an incomplete AST and related errors.
Each variant \(v_i\) is responsible for creating and adding its corresponding \textit{compilation unit task} to the priority queue \(\Gamma_{\mathcal{S}}\) with the appropriate priority level as detailed in Sect.~\ref{sec:lsp}.

\vspace{0.5em}\paragraph*{Compilation Unit Executor.}
The \textit{Compilation Unit Executor} is a component that is used by the language workbench to execute the tasks in the correct order. It contains:
\begin{compactitem}
    \item the \textit{execution listener}, which is used to listen to the execution of the tasks and report the errors to the \textit{Compilation Helper} (see the last paragraph);
    \item the \textit{executor}, which accepts a configuration and executes the tasks concurrently according to the \textit{executor service};
    \item the \textit{executor service}, which provides routines to manage termination and methods that can produce a \textit{Future} for tracking the progress of asynchronous tasks;
    \item the \textit{task queue}---a min-heap priority queue---in which the task dependencies are lazily registered and resolved during the execution of the tasks;
    \item the \textit{task provider} which is used to provide the tasks to the \textit{task queue}.
\end{compactitem}
The compilation unit executor is used to run the tasks in the correct order, and it is essential for the language server to bring the compilation units to a fully typed state. Concurrency and asynchronous issues could arise during task execution, and the compilation unit executor is in charge of managing these issues.\vspace{5pt}

\vspace{0.5em}\paragraph*{Fenwick Tree.}
The \textit{Fenwick Tree}~\cite{Fenwick94, Ryabko92}---also known as binary indexed tree---is a data structure that provides search and insert operations in \(O(\log n)\) time complexity. In our approach, we decided to use the Fenwick tree to represent the lexical scopes of the source code.
Basically, the Fenwick tree is a \textit{global shared context} \(\Gamma_{\mathcal{S}}\) and it is paramount to the language server to provide responses to the LSP client efficiently.
The language server leverages the Fenwick tree to provide the \textit{folding range} capability to the client and to efficiently perform scope-based search activities.
During the editing, the client can remove or add new scopes, and the Fenwick tree must be updated accordingly. Note that the removal of a scope causes the removal of one subtree of the Fenwick tree.
Just as all other data structures, the Fenwick tree is populated automatically by the language workbench during the tasks execution.\vspace{5pt}

\vspace{0.5em}\paragraph*{LSP Graph.}
The \textit{LSP Graph} is the second \textit{global shared context} \(\Gamma_{\mathcal{S}}'\) used by the language server.
It is a symbol dependency graph~\cite{Knuth97}. Its nodes represent the symbols within the source code and its edges represent dependencies between symbols. A symbol can be a type, a function, a variable, etc. The dependencies between the symbols are based on the usage and the kind of the symbol in the source code. However, as soon as the name resolution phase is completed, the language workbench starts to populate the LSP Graph with the dependencies between source code symbols. The LSP Graph is used by the language server to provide the \textit{FindReferences}, \textit{GoToDefinition}, and \textit{CodeCompletion} features to the client. It is important to note that the LSP Graph is populated automatically by the language workbench during the tasks execution. Without it, may LSP features would not be available. Note that the edges are not necessarily symmetric. This means that if a symbol A depends on a symbol B, it does not mean that B depends on A\@. The LSP Graph is used by the language server to provide the information to the client efficiently.\vspace{5pt}

\vspace{0.5em}\paragraph{Compilation Helper.}
Data structures face challenges due to the exogenous approach to modularity and a lack of awareness of other components.
Data related to the creation of these data structures is provided directly via \toolname{Typelang} or indirectly via the language workbench\@. The \textit{Compilation Helper}, introduced in this approach, aims at providing a centralized place to manage the data structures and the tasks execution, ensuring that the data structures are populated correctly and that the tasks are executed in the correct order.
Thus, the compilation helper represents the external component declaring the bindings between all the aforementioned (independent) components---\emph{i.e.}:
\begin{compactitem}
    \item the \textit{root} of the compilation units;
    \item the \textit{compilation unit executor};
    \item the \textit{Fenwick tree};
    \item the \textit{LSP Graph}.
\end{compactitem}

The language server uses the compilation helper to interact with the LSP graph and to queue any tasks.
A compilation helper reference is passed to the language server variants to allow variants to interact with the data structures and with the tasks execution.
Additionally, the compilation helper collects error events reported by the execution listener. These may include \toolname{Typelang} exceptions or injected errors detected in the generated code by the language workbench.
The compilation helper is responsible for making error management decisions. For example, if the error is related to the type inference, the compilation helper can decide to re-execute the tasks in the compilation unit, achieving a fully typed state.
As a centralized place in which all the errors are managed, the compilation helper can either choose \begin{inparaenum}
    \item to handle the errors as compilation errors---e.g., stop the compilation and emit the errors in standard error/output---or
    \item to update the data structures, handling the errors as LSP errors---e.g., keeping alive the language server and emit the errors to the client.
\end{inparaenum}
Note that, the \textit{Compilation Helper} is used to manage both the type system errors and reusing the same error handling mechanism for the language server errors.\vspace{5pt}

\noindent\toolname{Typelang} variants are essential to the language server, both for the errors it provides and for the data structures it populates. Each artifact derives its language server variant from the \toolname{Typelang} variant used to implement the language. The \textit{collecting} and \textit{assembling} phases, as well as type definitions, are inherited from those of \toolname{Typelang}. A language server variant is composed by the features provided by the \toolname{Typelang} feature boxes, used by the respective artifacts.
This means that each artifact is unaware of all the features offered by the LSP, knowing only those that are actually needed by the artifact itself.
An artifact with its feature boxes can be shipped to another language variant to correctly generate the language server variant. The reusability is guaranteed by the exogenous feature boxes definition approach: LSP capability implementations and type system components can therefore be reused across different languages.

\subsection{Language Server Integration in Modular language workbenches}\label{lsp:variant_oriented}

By recalling the properties presented in Sect.~\ref{sec:theoretical-overview}, as already done for \toolname{Typelang} (Sect.~\ref{sec:typelang}), we outline how the language server variants can benefit of \textit{variant-oriented programming} paradigm and \textit{cross-artifact coordination} layer principles when integrated within a language workbench\@.
As introduced in Sect.~\ref{sec:theoretical-overview}, let \(\mathcal{S}\) be a modular language workbench (our \textit{variant-oriented software}), and \(P\) be the language server SPL\@.
\(P \Rrightarrow \mathcal{S}\) holds if and only if the four principles of \textit{variant-oriented programming} are observed, meaning that the language server variants are in \(\rightleftharpoons_{\Gamma}^{\mathcal{S}}\)-relationship.

\vspace{0.5em}\paragraph*{Variant-Oriented Programming.}
The LSP features used to generate the respective language server variant \(v_i\) are directly provided by each feature box used by the associated \toolname{Typelang} variant: for any \(v_i, v_j \in P\), \(v_i\) and \(v_j\) are orthogonal to each other, as they are generated by different feature boxes.
There is no restriction imposed by the \textit{variant-oriented programming} paradigm on the generation of variants which depend on another variant, as long as the \(\rightleftharpoons_{\Gamma}^{\mathcal{S}}\)-relationship is preserved.
Therefore, the \textit{independence} principle of variant-oriented programming is upheld.
Unlike \toolname{Typelang}, the granularity of the feature activation is at the feature box level, and the language server variant is composed by the features provided by the \toolname{Typelang} feature boxes useb by an artifact.
This denotes that the simultaneous \textit{coexistence} of \(v_i\) and \(v_j\) in \(\mathcal{S}\) is guaranteed by construction.
As explained in Sect.~\ref{lsp:building_blocks}, \(\mathcal{S}\) provides the necessary data structures---often automatically populated---to the language server variants, providing the information to the LSP client.
\(v_i\) and \(v_j\) must be able to dialogue with each other to provide correct information to the LSP client.
The \textit{semantic interoperability} property is respected; it is a \textit{must-have} property for the language server variants to be able to co-exist in \(\mathcal{S}\).
The \textit{variant-oriented programming} requires the existence of a shared context \(\Gamma \in C_{\mathcal{S}}\) where \(v_i\) and \(v_j\) can co-operate. Lots of shared contexts \(\Gamma\) are provided by the language workbench, such as the \textit{Fenwick tree} and the \textit{LSP graph}. The \textit{co-operation} is also achieved at runtime, where the language server variants use the compilation helper to dynamically interact with these shared contexts \(\Gamma\).

\vspace{0.5em}\paragraph*{Cross-Artifact Coordination.}
Each artifact \(a \in A_{\mathcal{S}}\) can be defined in terms of the language server variant \(v_i\).
Or better, if \(a\) wants to provide the LSP support for its feature boxes, it must allow itself to be defined in terms of the language server variant \(v_i\).
The shared context provided by each artifact are two-fold: the \textit{typing environment} \(\Gamma\)---inherited from the \toolname{Typelang} variant---and the \textit{compilation helper} \(\Gamma'\).
Each \(a\) provide the \textit{typing environment} \(\Gamma\) which is populated by the \toolname{Typelang} variant and it is used by the language server variants to provide the information to the LSP client.
The compilation helper is the shared context \(\Gamma'\) that each artifact \(a\) provides to the language server variants to retrieve information already computed related to the LSP capabilities.
Two global shared contexts  are provided by the language workbench to the language server variants: the \textit{Fenwick tree} \(\Gamma_{\mathcal{S}}\) and the \textit{LSP graph} \(\Gamma_{\mathcal{S}}'\) explained in Sect.~\ref{lsp:building_blocks}.
\(\Gamma_{\mathcal{S}}\) and \(\Gamma_{\mathcal{S}}'\) are populated and interrogated by the language server variants to provide the information to the LSP client.
By clarifying these aspects, the \textit{cross-artifact modularization} layer for the language server is achieved.

\begin{Listing}[t]
    \centering
    \showgroovy*[\linewidth]{build.gradle}
    \caption{\toolname{Gradle} build file to generate LSP plugins and the syntax highlighting.}%
    \label{lst:lsp:build_gradle}
\end{Listing}

\subsection{LSP Plugin and Syntax Highlighting Generation}\label{lsp:client_generation}
Improvements in the editing support are also possible by leveraging the language workbench capabilities. In general, since the language workbenches know the syntax and semantics of the language, generating the LSP plugin and syntax highlighting leads to a reduction in the effort required to provide this support.
To provide a complete editing support, the editors usually require the implementation of LSP plugins for the language server---which is responsible for providing all the LSP capabilities to the client---and the \textit{syntax highlighting} support.
As shown by recent work, Neverlang provides the \textit{categories} introduced by~\citet{Cazzola19} that have been used to generate the syntax highlighting for LPL as a third-party plugin for the Eclipse IDE~\cite{Murphy06}.
In the same work, the authors proposed the generation of the semantic support for the IDEs, such as the \textit{Semantic highlighting} and \textit{Debugging} features by leveraging the language workbench capabilities.
Furthermore, Monticore~\cite{Krahn07}, Rascal~\cite{VanDerStorm11} and Spoofax~\cite{Kats10} provide the editing support for the Eclipse IDE\@.
For more details refer to Table~\ref{ttab:problem-statement:lw-comparison}.

\vspace{0.5em}\paragraph*{Overcoming the IDE Dependency.}
To overcome the direct dependency to the Eclipse IDE, thanks also to the advent of the LSP and the widely used TextMate~\cite{Gray07b, Skiadas07}\footnote{\toolname{VSCode} is the most popular editor according to the Stack Overflow Developer Survey 2021 and it embraces the TextMate grammar implementing the native support for the syntax highlighting (see https://github.com/microsoft/vscode-textmate)} grammar, we extend the proposed approach to provide a methodology to generate the syntax highlighting and the LSP plugin for all the editors that support the LSP\@.
Developer side, a \toolname{Gradle} plugin generates the necessary files for the syntax highlighting and the LSP plugin.
This aims to further reduce the effort towards this support and increase the reusability of the language workbenches.
Listing~\ref{lst:lsp:build_gradle} shows the \toolname{Gradle} build file that generates the LSP plugin and the syntax highlighting.
Developers should only specify the fields shown in Table~\ref{tab:problem-statement:lw-comparison} inside of \texttt{LSPClient} block.
\begin{table}[h!]
    \centering
    \resizebox{\linewidth}{!}{%
        \rowcolors{1}{white}{gray!20}
        \begin{tabular}{ l c c }
            \toprule
            \multicolumn{1}{c}{Name}           & \multicolumn{1}{c}{Line} & \multicolumn{1}{c}{Description}                            \\
            \midrule
            \texttt{generatorVersion}          & 2                        & The version of the plugin                                  \\
            \texttt{clientImplementations}     & 3                        & The list of the desired editors \\
            \texttt{templateGeneratorsClasses} & 10                       & The list of the associated template generators             \\
            \texttt{languageName}              & 17                       & The name of the language                                   \\
            \texttt{launcher}                  & 18                       & The language server launcher                                           \\
            \texttt{fileExt}                   & 19                       & The extension of the language                              \\
            \texttt{binPath}                   & 20                       & An executable file containing the language server\\
            \bottomrule
        \end{tabular}
    }
    \caption{Fields to specify in the \texttt{LSPClient} block of the \toolname{Gradle} build file to generate the LSP plugin and the syntax highlighting.}%
    \label{tab:problem-statement:lw-comparison}
\end{table}

\begin{Listing}[t]
    \centering
    \showneverlang*[\linewidth]{IfStatement.nl}
    \caption{Example of a Neverlang module that provides the implementation of the \textit{IfStatement} artifact with \texttt{if} and \texttt{else} categories made available}%
    \label{lst:lsp:if_statement}
\end{Listing}

\noindent It is important to note the editor list and the associated template generators can be extended by the developers to support more editors.
The support is implemented for the widely used \toolname{VSCode}~\cite{DelSole23}, \toolname{NeoVim}, and \toolname{Vim} editors, including both the syntax highlighting and the LSP plugin.
Sect.~\ref{sec:case-study} will show the implementation effort in terms of LoC and the NoC.
With this approach, we embrace the opportunity offered by language workbenches to provide the syntax highlighting elements.
For instance, the Neverlang's \textit{categories} construct as shown in Listing~\ref{lst:lsp:if_statement} and the Spoofax's ability to provide the syntax highlighting elements for NWL as shown by~\citet{Visser10}.

\vspace{0.5em}\paragraph*{Syntax Highlighting Generation.}
Listing~\ref{lst:lsp:if_statement} shows how the \textit{IfStatement} artifact defines the \textit{if} statement using the terminal symbols \texttt{if} and \texttt{else}. In the \textit{categories} block (line~7), it exports the \texttt{keyword} category (line~8), which includes the terminal symbols \texttt{if} and \texttt{else}.
Two cases can be distinguished:
	1) the syntax element are provided by the already modularized boxes and the language workbench can collect them and perform their union without duplicates (e.g., the Neverlang language workbench\@), and
	2) the syntax elements are provided by the language constructs (e.g., the Spoofax language workbench\@).
The \toolname{Gradle} plugin merely needs the syntax elements to be provided, so that the template generators can use them to generate the syntax highlighting editor support.
The \toolname{Gradle} plugin requires the programmer to specify, among other fields, the \texttt{binPath}---i.e., the path to the jar file containing the implementation of the language server. The server is then launched by the editors according to the \texttt{launcher} field upon opening a file with the extension specified in the \texttt{fileExt} field.
Additionally, template generators can be written in any JVM-compatible language, as the \toolname{Gradle} plugin uses reflection to instantiate each template generator and invoke the methods that conform to the contract defined by the plugin interface.

\begin{figure*}[t]
	\centering
	\begin{subfigure}{0.45\textwidth}
		\centering
		\includegraphics[width=\linewidth]{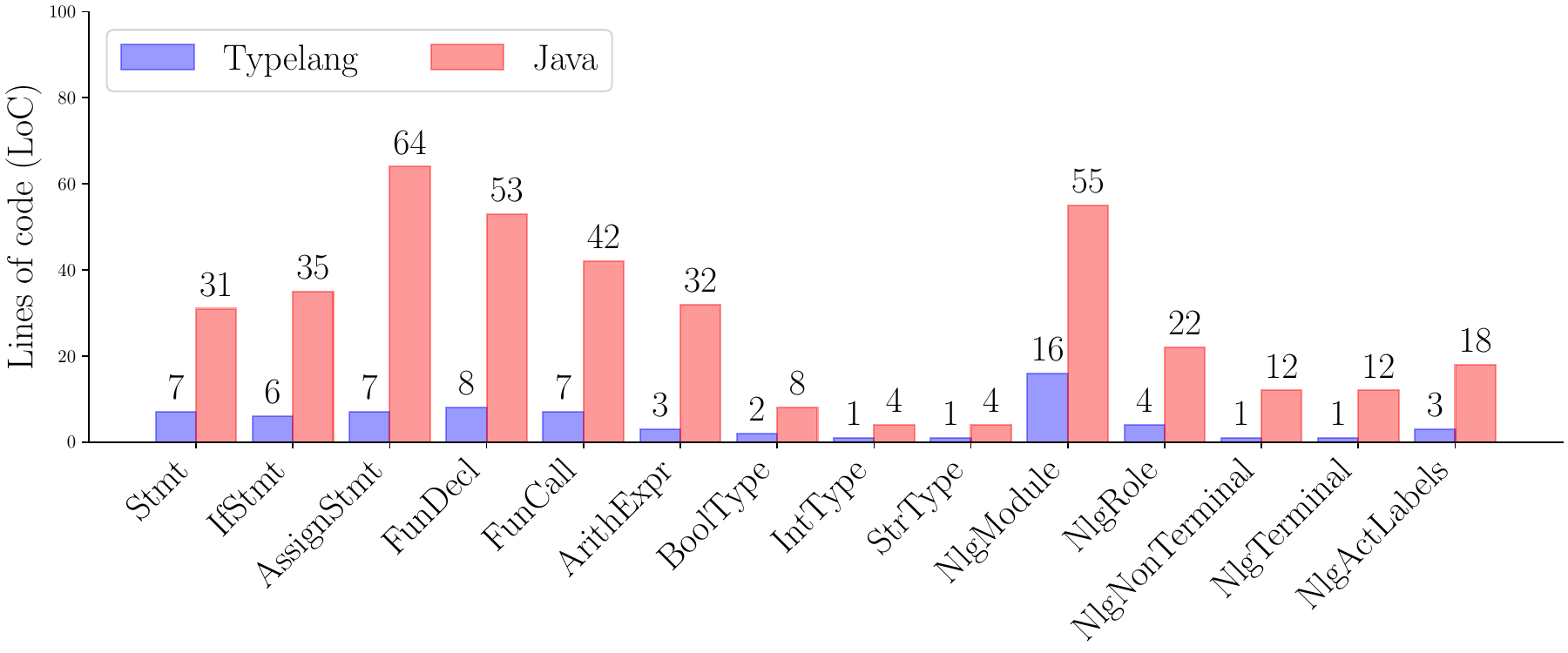}
		\caption{A visualization of the LoC reduction achieved by implementing a type system using \toolname{Typelang} for some Neverlang artifact compared to a traditional Java-based approach. This figure illustrates how the use of \toolname{Typelang} leads to significant simplification, with a reduction of approximately \(89.06\%\).}%
		\label{fig:typelang:typelang_loc}
	\end{subfigure}
	\hspace{.05\textwidth}%
	\begin{subfigure}{0.45\textwidth}
		\centering
		\includegraphics[width=\linewidth]{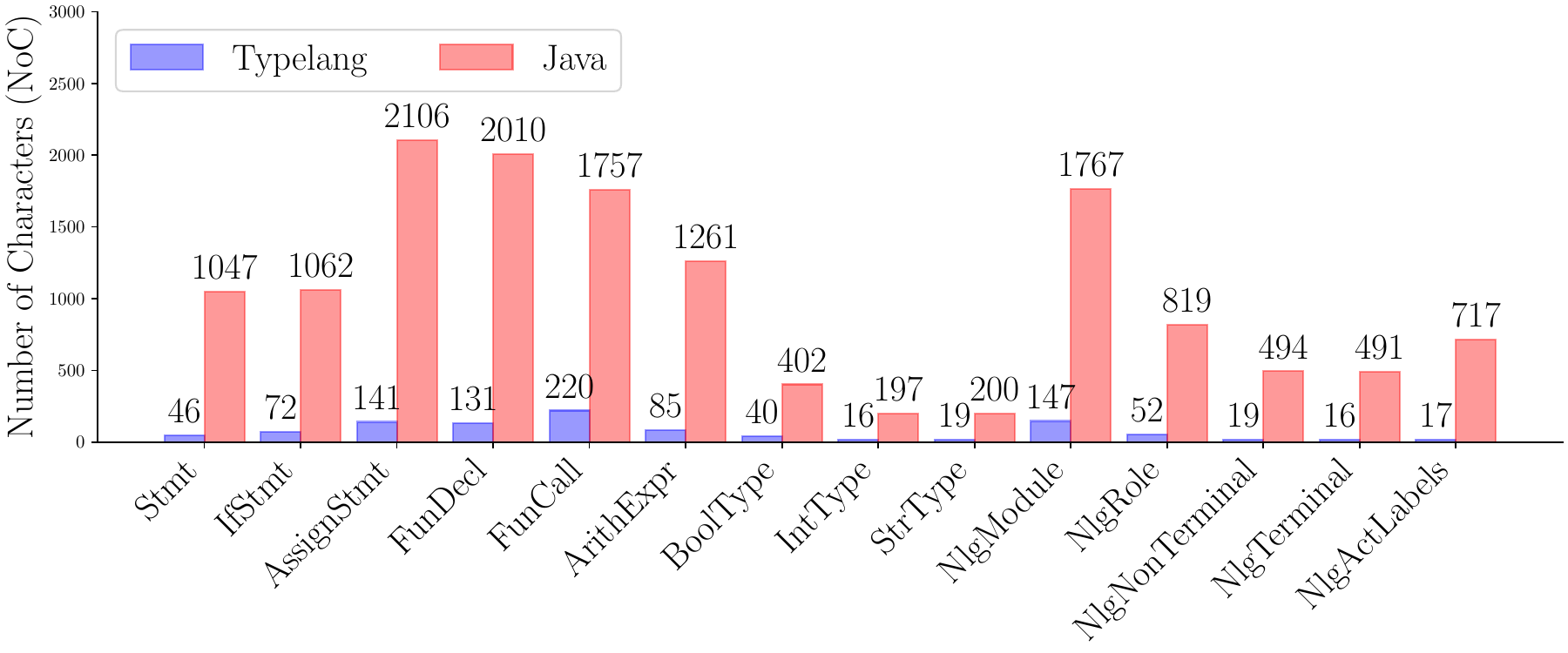}
		\caption{A visualization of the NoC reduction achieved by implementing a type system using \toolname{Typelang} for some Neverlang artifact compared to a traditional Java-based approach. This figure illustrates how the use of \toolname{Typelang} leads to significant simplification, with a reduction of approximately \(93.48\%\).}%
		\label{fig:typelang:typelang_chars}
	\end{subfigure}
	\caption{A comparison of the LoC and NoC needed to implement a type system using \toolname{Typelang} compared to a traditional Java-based approach.}%
	\label{fig:typelang:general}
\end{figure*}

\section{Demonstration Case Study}\label{sec:case-study}
We demonstrate the applicability of our approach through a case study implemented using the Neverlang language workbench. We chose Neverlang due to its compatibility with the flexibility, extensibility and modularity requirements of variant-oriented programming.
The implementation consists of approximately \(13,000\) LoC (approximately \(370,000\) NoC) of Java code and approximately \(2,000\) LoC (approximately \(60,000\) NoC) of Neverlang code including \toolname{Typelang} code.

To assess our approach, we address the following research questions:
\begin{compactdesc}
	\item[\(\fontrq{RQ}\)\textit{\textsubscript{1}}] \textit{To what degree is it possible to streamline by associating variants to language artifacts?}
	\item[\(\fontrq{RQ}\)\textit{\textsubscript{2}}] \textit{To what degree is it possible to automate the generation of LSP clients, lowering \(\fontc{E}\) to \(\mathbf{1}\)?}
	\item[\(\fontrq{RQ}\)\textit{\textsubscript{3}}] \textit{Can the language server be automatically generated starting from the \toolname{Typelang} variants lowering \(\fontc{L}\) to \(\fontc{N}\)?}\smallskip
\end{compactdesc}

These questions assess our system’s modularity and reuse at the artifact level (\(\fontrq{RQ}\)\textit{\textsubscript{1}}), the degree of automation in LSP plugin generation (\(\fontrq{RQ}\)\textit{\textsubscript{2}}), and the feasibility of deriving language servers directly from \toolname{Typelang} variants (\(\fontrq{RQ}\)\textit{\textsubscript{3}}).
Our case study proceeds as follows: we first implement a type system for several Neverlang artifacts using \toolname{Typelang}; then, we assemble these artifacts into two complete language variants---\toolname{SimpleLanguage},\footnote{\url{https://github.com/graalvm/simplelanguage}} a general-purpose language from the GraalVM project~\cite{Wurthinger13},\footnote{\url{http://openjdk.java.net/projects/graal}} and Neverlang itself---and generate their language servers and LSP plugins for Visual Studio Code, NeoVim, and Vim.
We quantify the effort in terms of lines of code (LoC) and number of characters (NoC), offering objective metrics for comparison against prior work~\cite{Bunder19}.
The full source code and case study implementation are publicly available on Zenodo.\footnote{\url{https://doi.org/10.5281/zenodo.15276991}}
The following subsections describe the experimental setup, present results for each metric, and answer \(\fontrq{RQ}\)\textit{\textsubscript{1}}, \(\fontrq{RQ}\)\textit{\textsubscript{2}}, and \(\fontrq{RQ}\)\textit{\textsubscript{3}} based on our empirical findings.

\begin{table}[t]
	\centering
	\resizebox{\linewidth}{!}{%
		\rowcolors{1}{white}{gray!20}
		\begin{tabular}{ll}
			\toprule
			\multicolumn{1}{c}{Abstract Term}            & \multicolumn{1}{c}{Neverlang Term}      \\\midrule
			Language Artifact                            & Module                                  \\
			Feature Box                                  & Annotated Java class                    \\
			Collector                                    & Annotation preprocessor                 \\
			\makecell{Type checking/inference semantics} & \makecell{Type checking/inference role} \\\bottomrule
		\end{tabular}
	}
	\caption{Mapping the defined abstract terms to their counterparts in Neverlang.}%
	\label{tab:typelang:abstract_neverlang}
\end{table}

\subsection{Tool Support}
To have a clear correspondence between the generic terms used previously and the specific Neverlang terms for those concepts, we provide a mapping in Table~\ref{tab:typelang:abstract_neverlang}.

Our \toolname{Typelang} implementation leverages Neverlang and integrates with its parser~\cite{Cazzola12c} to use its polyglot capabilities~\cite{Cazzola15c}.
Prior to the introduction of \toolname{Typelang} in Neverlang, developers had to write significant amounts of Java code to implement custom type checking and type inference rules, with limited reuse for existing type system specifications.
For instance, Listing~\ref{lst:typelang:AssignStatement} demonstrates how the \toolname{Typelang} DSL can define type system rules for the \texttt{stmt.AssignmentStatement} module in Neverlang.
In this example, the assignment is treated as a new declaration and will be statically and strongly typed.

\vspace{0.5em}\paragraph*{Neverlang Integration.}
We extended the Neverlang compiler with an annotation preprocessor capable of handling the \toolname{Typelang} DSL\@. Neverlang modules can include semantic actions annotated with the \inlineneverlang{<typelang>} label; any code within these semantic actions will be written according to the syntax of a \toolname{Typelang} variant.
When the Neverlang parser encounters the \inlineneverlang{<typelang>} annotation in a module, the Neverlang annotation preprocessor gathers all annotated Java classes to generate the \toolname{Typelang} variant needed to compile that specific Neverlang module.
Neverlang delegates the parsing of the code of these semantic actions to the \toolname{Typelang} parser. If any features remain unrecognized after the collection phase, Neverlang will fail during parsing.
Furthermore, Neverlang uses \textit{dependency injection} (via the Google Guice Library~\cite{Vanbrabant08}) to distribute the annotated Java class to internal structures.
Through this injection, Neverlang ensures that all languages containing the \texttt{stmt.AssignStatement} module will also inherit its type checking and type inferencing capabilities.

\begin{Listing}[t]
	\centering
	\showneverlang*[\linewidth]{AssignStatement.nl}
	\caption{An example of \toolname{Typelang} DSL to perform type checking and type inferencing for an assignment statement in the Neverlang language workbench\@.}%
	\label{lst:typelang:AssignStatement}
\end{Listing}

\vspace{0.5em}\paragraph*{Assignment Statement Example.}
Listing~\ref{lst:typelang:AssignStatement} shows the role \texttt{check-infer} (line 10) which uses its own \toolname{Typelang} variant. This variant is deduced from the feature boxes being used---specifically, \texttt{identifier} and \texttt{InferenceException} in this case.
The \texttt{check-infer} role assumes the existence of a variant \(v_i\) from the \toolname{Typelang} family \(P\) capable of compiling the \texttt{stmt.AssignStatement} module; if such a variant exists, it will be used for type checking and type inferencing; otherwise the module cannot be compiled.
The \texttt{check-infer} role uses the keyword \inlinetl{try} to infer the type of the right-hand side expression (line 13) and \inlinetl{check}s that the types of the left-hand side and right-hand side are compatible (line 14). Line 15 declares the effective usage of the left-hand side identifier with the inferred type. If an \textit{InferenceException} is raised during these phases, it indicates that the type associated with the identifier cannot be inferred. Consequently, line 17 is executed to define a new identifier whose type corresponds to the type of the right-hand side expression.
This approach assumes that each time an \textit{InferenceException} occurs, a \textit{new variable declaration} rather than the \textit{re-assignment} of an existing one is performed. While this may not always be true depending of the language (the exception could arise for several reasons, for instance, because the type of the right-hand side expression is not visible in the current scope, or the type of the right-hand side expression is not compatible with the type of the left-hand side identifier previously declared), but it can be readily covered.
It is reasonable to assume that the \texttt{stmt.AssignStatement} module will be used in conjunction with another Neverlang module defining a declared \textit{scope} and a \textit{priority level}, as shown in Fig.~\ref{fig:typelang:modular_type_system}.

\vspace{0.5em}\paragraph*{Typing Environment.}
Each \textit{typing environment} is associated to a scope defined by a Neverlang module.
The \textit{typing environment} admits the \texttt{stmt.AssignStatement} module will contain unordered pairs of the form \((ID, Table Entry)\) for all identifiers declared within its scope, as shown by the blue boxes in Fig.~\ref{fig:typelang:type_system}.
Given the variety of type inference strategies available---including, \textit{Hindley-Milner}~\cite{Hindley69, Milner78}, \textit{constraint-based}~\cite{Pierce02}, and \textit{unification-based}~\cite{Robinson65} algorithms---Neverlang employs a strategy pattern~\cite{Gamma95} to select the algorithm for the type inference phase.
Furthermore, additional and custom type inference strategies can be integrated later without modifying the existing ones, ensuring the type system's extensibility.

\vspace{0.5em}\paragraph*{Reduction in LoC and NoC.}
Implementing the \texttt{check-infer} role in Java required \(64\) LoC and \(2,106\) NoC (excluding whitespace, newline, and tab characters). The same role in \toolname{Typelang} achieves a LoC reduction of approximately \(89.06\%\) (see Fig.~\ref{fig:typelang:typelang_loc}) and a NoC reduction of approximately \(93.48\%\) (see Fig.~\ref{fig:typelang:typelang_chars}), resulting in a total of \(7\) LoC and \(141\) NoC\@.
The summary in Fig.~\ref{fig:typelang:general} demonstrates that the \toolname{Typelang} DSL effectively reduces the amount of code needed to implement a type system for several artifacts in Neverlang.
We calculated the overall LoC and NoC reduction percentages using the following formulas:
\begin{gather*}
	\textrm{LoC} = \frac{\textrm{LoC}_{\textrm{java}} - \textrm{LoC}_{\textrm{typelang}}}{\textrm{LoC}_{\textrm{java}}} \times 100\\
	\textrm{NoC} = \frac{\textrm{NoC}_{\textrm{java}} - \textrm{NoC}_{\textrm{typelang}}}{\textrm{NoC}_{\textrm{java}}} \times 100
\end{gather*}
and the average per-artifact LoC and NoC percent reduction using the following formulas:
\begin{align*}
	\overline{\mathrm{LoC}} = \frac{\sum_{i=1}^{n} \mathrm{LoC}_i}{n}
	\quad\quad
	\overline{\mathrm{NoC}} = \frac{\sum_{i=1}^{n} \mathrm{NoC}_i}{n}
\end{align*}
The results of our analysis are:
\begin{compactitem}
	\item overall and average LoC reduction percentage are approximately \(82.90\%\) and \(82.32\%\), respectively;
	\item overall and average NoC reduction percentage are approximately \(93.87\%\) and \(93.18\%\), respectively.
\end{compactitem}

\subsection{Degree of Type System Reuse}

\vspace{0.5em}\paragraph*{Overview.}
In recent decades, reuse has gained recognition as a key factor in software development, encompassing both software engineering~\cite{Makitalo20, Griss00, Bassett96, Bezerra15} and programming language design~\cite{Mendez-Acuna16b, Batory98, Hudak98, Cazzola23b}.
This work aims to reduce the code needed to implement a programming language type system. This section discusses the extent of type system reuse within Neverlang and how \toolname{Typelang} facilitates it.
The \toolname{Typelang} DSL empowers language developers to define type systems modularly, thereby enabling the reuse of these type systems across various languages.

\begin{Listing}[t]
	\centering
	\showneverlang*[\linewidth]{Expressions.nl}
	\caption{A Neverlang \texttt{\textbf{bundle}} to define the \texttt{Expressions} module.}%
	\label{lst:case-study:expressions}
\end{Listing}

To illustrate the degree of reuse of type systems in Neverlang, an example of a Neverlang \texttt{bundle} is needed.
A \inlineneverlang{bundle}---shown in Fig.~\ref{lst:case-study:expressions}---is a Neverlang construct that contains a set of \inlineneverlang{slice}s intended for inclusion within a \inlineneverlang{language} construct.
In this example, the \texttt{Expressions} bundle defines the type system for expressions in Neverlang, specifically for \textit{additive} and \textit{multiplicative} expressions over \texttt{integer} and \texttt{double} types.
A Neverlang \inlineneverlang{language} can import multiple \inlineneverlang{bundle}s, enabling the reuse of type systems across different languages.

\begin{figure*}[p!]
	\centering
	\begin{subfigure}{0.45\textwidth}
		\centering
		\includegraphics[width=\linewidth]{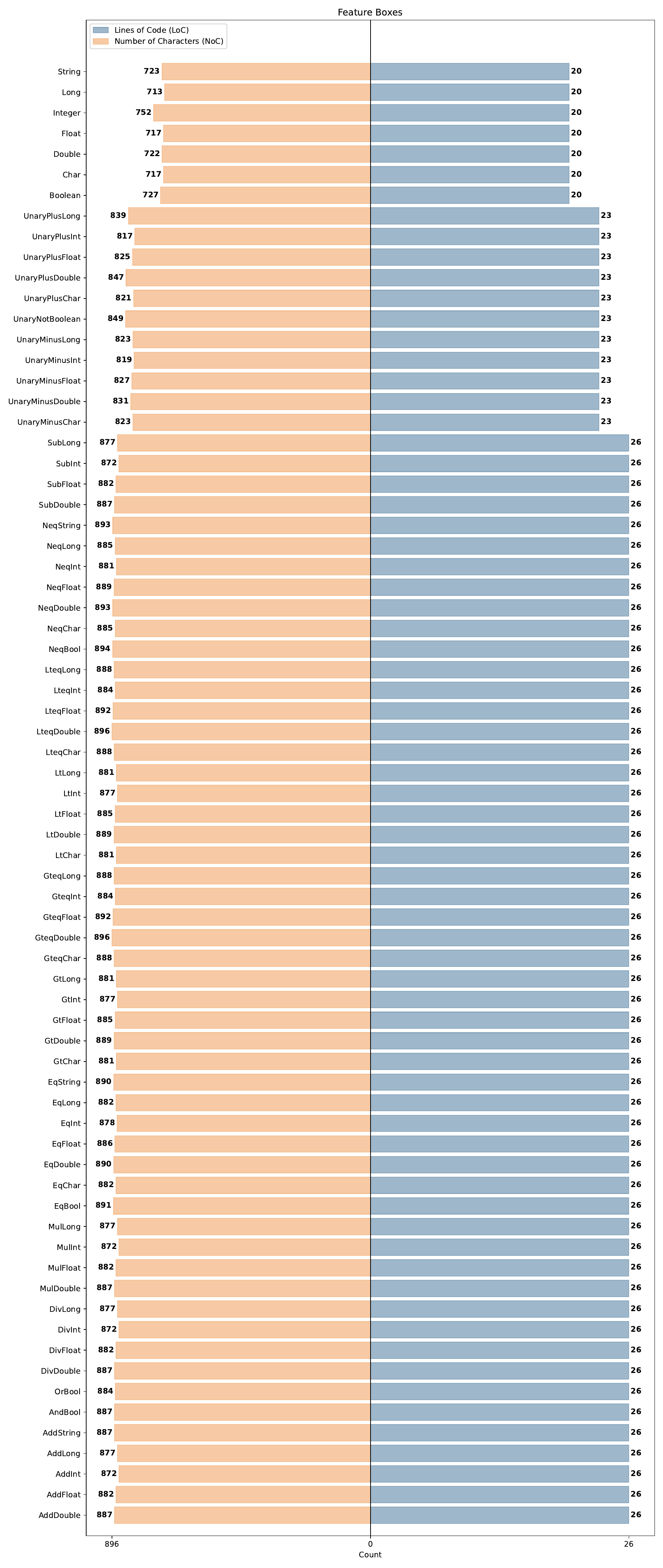}
		\caption{A visualization of the LoC and NoC needed to implement the \textit{feature boxes} for the Java primitive types and the operators defined on them.}%
		\label{lst:case-study:java_feature_boxes}
	\end{subfigure}
	\hspace{.05\textwidth}%
	\begin{subfigure}{0.45\textwidth}
		\centering
		\includegraphics[width=\linewidth]{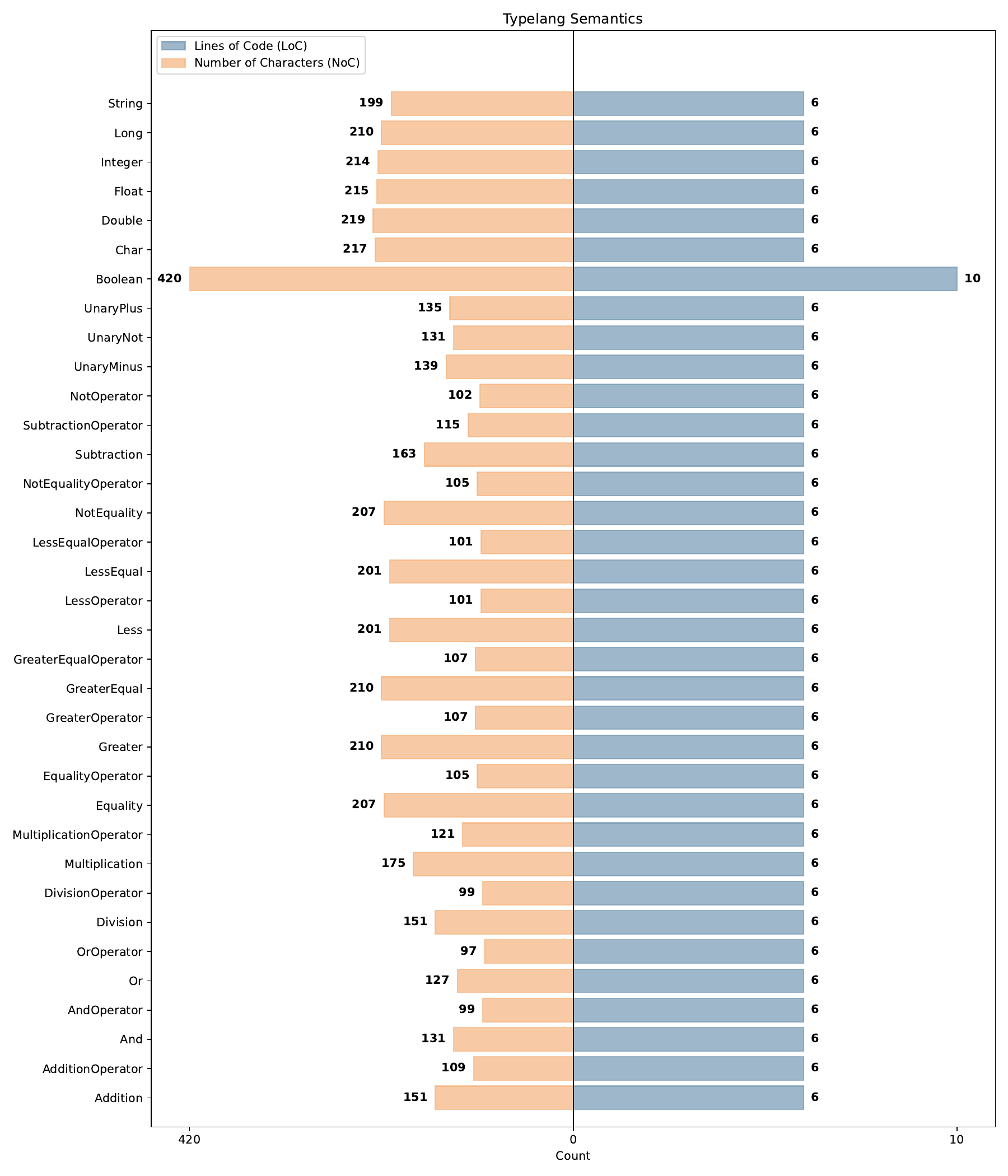}
		\caption{A visualization of the LoC and NoC needed to implement the \textit{semantic actions} for the \toolname{Typelang} roles in Neverlang. The \toolname{Typelang} roles are used to define the type system for the expressions in Neverlang.}%
		\label{lst:case-study:nl_lsp_roles}
	\end{subfigure}
	\caption{A comparison of the LoC and NoC needed to implement a type system using \toolname{Typelang} compared to a traditional Java-based approach.}%
	\label{lst:case-study:all}
\end{figure*}

To demonstrate the extent of type system reuse in Neverlang, we implemented a set of \textit{feature boxes} (see Fig.~\ref{lst:case-study:java_feature_boxes}) for:\smallskip
\begin{compactitem}
	\item the most common primitive types, and
	\item the most common operators defined on these types.\smallskip
\end{compactitem}
Furthermore, we implemented a set of \textit{semantic actions} for the \toolname{Typelang} roles in Neverlang for the \texttt{Expressions} (see Fig.~\ref{lst:case-study:nl_lsp_roles}).

\vspace{0.5em}\paragraph*{Metrics for the Degree of Reuse.}
This section aims to quantify the number of expression languages in which a type system can be reused.
We define the following sets:\smallskip
\begin{compactitem}
	\item \(U\), a set of \(N\) primitive types,
	\item \(O\), a set of \(M\) operators defined on these primitive types.\smallskip
\end{compactitem}
An expression language is defined by two subsets: \(u\subseteq U\) and \(o\subseteq O_{U}\), where \(O_{U}=\{o\in O\mid\text{types}(o)\subseteq u\}\).\footnote{The types of an operator are the types on which the operator is defined.}
Operators are dependent on types; an operator can only be included if it is defined on the types present in the expression language.
Our strategy to count the number of expression languages involves counting the combinations of primitive types and the operators that can be defined on them.
There are \(2^N\) possible combinations of primitive types. For each combination of primitive types, we can include only the operators defined on those types. Thus, for each \(u \subseteq U\) we can construct \(2^{|O_{u}|}\) combinations of operators, where \(|O_{u}|\) is the number of operators defined on the types in \(u\).
The total number of expression languages \(\fontc{L}\) is given by:
\begin{align*}
	\fontc{L} = \sum_{u\,\subseteq\,U\,|\,u\,\neq\,\emptyset}\left(2^{\left|O_{u}\right|} - 1\right)
\end{align*}
where \(-1\) is used to exclude the empty set of operators. That is, languages with no operators are not considered valid expression languages.

Consider a concrete example where \(U = \{int, double\}\) and \(O = \{+, *\}\) in which:\vspace*{-.25cm}
\begin{multicols}{2}
	\small
	\noindent\hspace*{.25cm}\(+: int \times int \to int\),\\
	\hspace*{.25cm}\(+: double \times double \to double\),\\
	\hspace*{.25cm}\(*: int \times int \to int\),\\
	\hspace*{.25cm}\(*: double \times double \to double\).
\end{multicols}
\noindent The possible combinations of primitive types and operators are:\vspace*{-.25cm}
\begin{multicols}{2}
	\small
	\noindent
	\hspace*{.25cm}\(u=\{int, double\},\:o=\{+, *\}\)\\
	\hspace*{.25cm}\(u=\{int\},\:o=\{+, *\}\)\\
	\hspace*{.25cm}\(u=\{double\},\:o=\{+, *\}\)\\
	\hspace*{.25cm}\(u=\{int, double\},\:o=\{+\}\)\\
	\hspace*{.25cm}\(u=\{int\},\:o=\{+\}\)\\
	\hspace*{.25cm}\(u=\{double\},\:o=\{+\}\)\\
	\hspace*{.25cm}\(u=\{int, double\},\:o=\{*\}\)\\
	\hspace*{.25cm}\(u=\{int\},\:o=\{*\}\)\\
	\hspace*{.25cm}\(u=\{double\},\:o=\{*\}\)\\
	\hspace*{.25cm}\(u=\{int, double\},\:o=\emptyset\)\\
	\hspace*{.25cm}\(u=\{int\},\:o=\emptyset\)\\
	\hspace*{.25cm}\(u=\{double\},\:o=\emptyset\)
\end{multicols}
\vspace*{-.25cm}\noindent The total number of expression languages (\(\fontc{L}\)) is calculated by:
\begin{align*}
	\fontc{L} & = \left(2^{\left|O_{\{int,\,double\}}\right|}-1\right) + \left(2^{\left|O_{\{int\}}\right|}-1\right) + \left(2^{\left|O_{\{double\}}\right|}-1\right) \\
	          & = \left(2^{\left|\{+,\,*\}\right|}-1\right) + \left(2^{\left|\{+,\,*\}\right|}-1\right) + \left(2^{\left|\{+,\,*\}\right|}-1\right)                   \\
	          & = \left(2^2 - 1\right) + \left(2^2 - 1\right) + \left(2^2 - 1\right)                                                                                  \\
	          & = 9
\end{align*}
We introduce two metrics to evaluate the degree of type system reuse in Neverlang, drawing inspiration from the software reuse literature~\cite{Krueger92, Frakes96, Bay04}: the normalized absolute reuse degree and the operator conditional reuse degree.

\textbf{Normalized absolute reuse degree (NAR).} It measures how many times a type system component is reused across the expression language implementation. It is defined as:\[NAR(c) = \frac{|\{l \in L \mid c \in l\}|}{\fontc{L}}\] where \(c\) is a component (a type or an operator) of the type system, and \(L\) is the set of all expression languages.

\textbf{Operator conditional reuse degree (OCR).} It estimates the reuse of an operator within the expression languages that include a specific primitive type. It is defined as:\[OCR(o\,|\,t) = \frac{|\{l \in L \mid t \in l \land o \in l\}|}{|\{l \in L \mid t \in L\}|}\] where \(t\) is a primitive type, \(o\) is an operator defined on it, and \(L\) is the set of all expression languages.

\vspace{0.5em}\noindent\textbf{Concrete Example.}
Consider the set of expression language \(L = \{l_1, l_2, l_3\}\) where:\smallskip
\begin{compactitem}
	\item \(L_1 = \{int, +\}\),
	\item \(L_2 = \{bool, int, +, *, ==\}\),
	\item \(L_3 = \{double, +, *\}\).\smallskip
\end{compactitem}
The NAR for each operator is calculated as:\smallskip
\begin{compactitem}
	\item \(NAR(+) = \displaystyle\frac{|\{l_1, l_2, l_3\}|}{3} = \frac{3}{3} = 1\),
	\item \(NAR(*) = \displaystyle\frac{|\{l_2, l_3\}|}{3} = \frac{2}{3} \approx 0.67\), and
	\item \(NAR(==) = \displaystyle\frac{|\{l_2\}|}{3} = \frac{1}{3} \approx 0.33\).\smallskip
\end{compactitem}
On the other hand, the OCR for each operator is calculated as:
\begin{compactitem}
	\item  \(OCR(+, int) = \displaystyle\frac{|\{l_1, l_2\}|}{|\{l_1, l_2\}|} = \frac{2}{2} = 1\),
	\item  \(OCR(*, int) = \displaystyle\frac{|\{l_2\}|}{|\{l_1, l_2\}|} = \frac{1}{2} = 0.5\), and
	\item  \(OCR(==, int) = \displaystyle\frac{|\{l_2\}|}{|\{l_1, l_2\}|} = \frac{1}{2} = 0.5\).\smallskip
\end{compactitem}

The LoC and NoC saved for the \textit{feature boxes} and the \textit{semantic actions} implementing the \toolname{Typelang} roles for each \(l \in L\) can be trivially calculated by summing respective values reported in Fig.~\ref{lst:case-study:java_feature_boxes} and Fig.~\ref{lst:case-study:nl_lsp_roles}.

The NAR and OCR metrics provide a quantitative assessment of type system reuse. While NAR reflects the absolute reuse of a component across expression languages, OCR captures conditional reuse based on the presence of a given primitive type.
Importantly, these metrics are not limited to primitive operators of the expression languages---they can be generalized to any programming language component defined in a modular way. That is, the NAR and OCR metrics offer a general method for evaluating type system reuse across modular language implementations.
By using this approach, we also demonstrate that the number of combinations is reduced from \(\fontc{T} \times \mathbf{1}\) to \(\fontc{N} \times \mathbf{1}\), where \(\fontc{T} = \fontc{L}\) represents the number of type systems and \(\fontc{N} << \fontc{T}\).

\subsection{Language Server Generation}
We developed a set of default feature boxes for Neverlang, implemented as annotated Java classes. These classes are designed to be used within the \toolname{Typelang}-related semantic actions to perform type checking and type inference.
In particular, we provide:
\begin{compactitem}
	\item approximately \(20\) annotated Java classes for types;
	\item approximately \(20\) annotated Java classes for signatures;
	\item approximately \(10\) annotated Java classes for scopes;
	\item approximately \(5\) annotated Java classes for exceptions.\smallskip
\end{compactitem}
For each component described in Sect.~\ref{lsp:building_blocks} we provide a corresponding set of default implementations that conform to the respective interfaces.
The orchestration and communication among these components are handled by the Neverlang runtime,  relieving the language developer from the burden of implementing additional coordination logic.
For the sake of completeness, the components for which we provide default implementations include:\vspace{-.25cm}
\begin{multicols}{2}
	\begin{compactitem}
		\item Compilation Unit,
		\item Compilation Unit Task,
		\item Compilation Unit Executor,
		\item Table Entry,
		\item Fenwick Tree,
		\item LSP Graph,
		\item Compilation Helper.\vspace*{-.45cm}
	\end{compactitem}
\end{multicols}
Language developers can choose to either adopt the provided default implementations or define custom ones from scratch to suit specific needs.
This design allows language developers to obtain a fully functional language server out of the box, while still maintaining the flexibility to tailor it to their language’s requirements.
The only exception is the \textit{workspace handler}, for which Neverlang does not provide a default implementation. This is because the \textit{workspace handler} handles pieces of information that are strictly related to the implemented language and cannot be predetermined.
Instead, Neverlang provides an abstract class that developers can extend to implement the required behavior.
Listing~\ref{lst:case-study:simplelang-workspace-handler} shows an example \textit{workspace handler} implementation for \toolname{SimpleLanguage}.
\begin{Listing}[t]
	\centering
	\showjava*[\linewidth]{SimpleLangWorkspaceHandler.java}
	\caption{The \textit{workspace handler} implementation for \textit{SimpleLanguage}.}%
	\label{lst:case-study:simplelang-workspace-handler}
\end{Listing}
An implementation for the \textit{workspace handler} abstraction involves defining the methods:\smallskip
\begin{compactitem}
	\item \inlinejava{getSourceSet} (line 10) returns the source set of the language based on the given file extension; that is, the collection of files that must be processed by the LSP\@;
	\item \inlinejava{language} (line 15) returns an instance of the language;
	\item \inlinejava{lspRoles} (line 19) returns a stream of roles associated to the LSP\@; \emph{i.e.}, the names of all parse tree visitors needed to populate the LSP-related data structures;
	\item \inlinejava{compilationHelper} (line 24) returns the class of the concrete implementation of the compilation helper;
	\item \inlinejava{priorities} (line 29) returns a list of priorities for the compilation units, sorted according to the desired visit sequence.\smallskip
\end{compactitem}
Notice how the \textit{workspace handler} connects all components and embodies the exogenous approach to dependencies management. For instance, by specifying priorities within the \textit{workspace handler}, each priority definition remains unaware of the others, as well as of their relative order.
Aside from the Neverlang code that defines the language---optionally leveraging the reusable default implementations---the \textit{workspace handler} is the only component that must be implemented by the developer to obtain a fully functional language server for a given language, such as \toolname{SimpleLanguage} in this case.

\subsection{LSP Client Generation}
\begin{figure}[t]
	\centering
	\begin{subfigure}{0.45\textwidth}
		\includegraphics[width=1\linewidth]{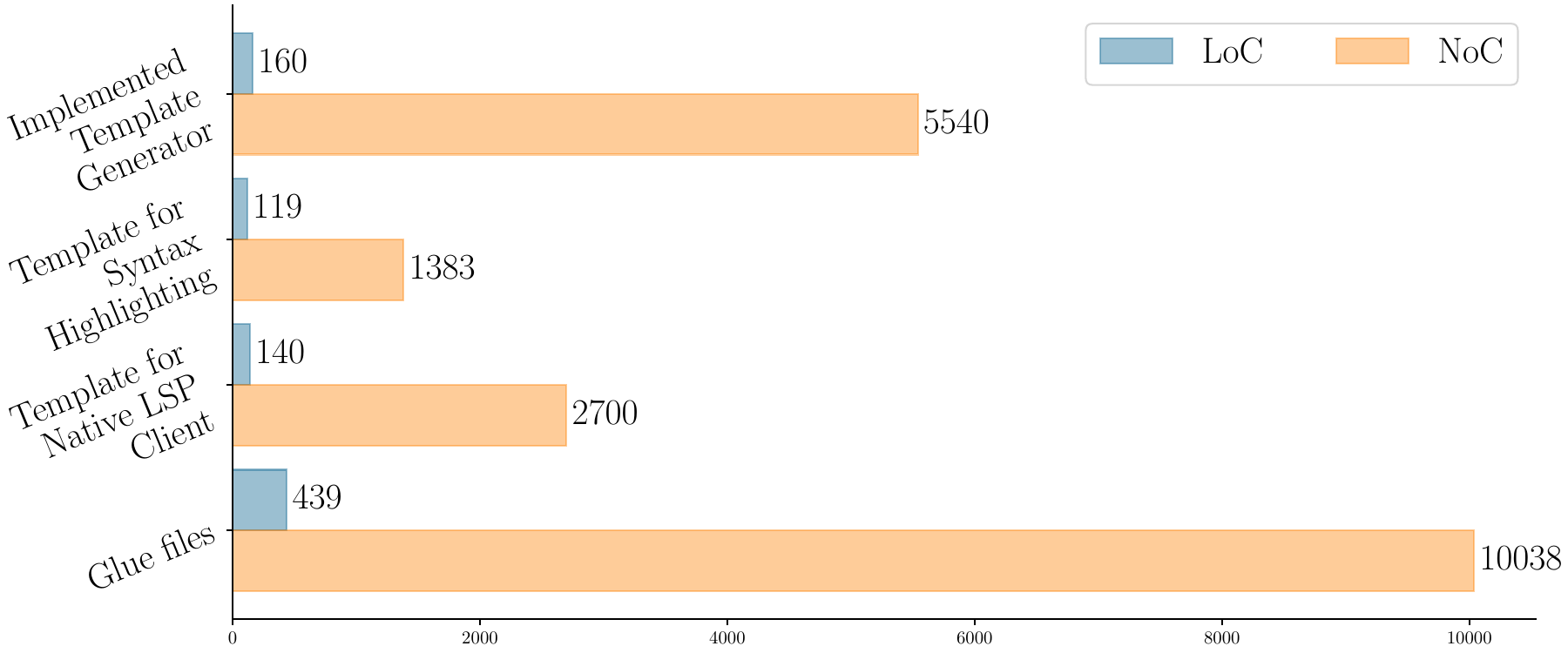}
		\caption{LoCs and NoCs required to implement the template generators for \toolname{VSCode}. The \toolname{VSCode} generator leverages the existence of annotated templates for syntax highlighting and LSP plugin. Additional glue files specific to \toolname{VSCode} are needed to complete the support.}%
		\label{fig:vscode}
	\end{subfigure}

	\hspace{.05\textwidth}%

	\begin{subfigure}{0.45\textwidth}
		\includegraphics[width=1\linewidth]{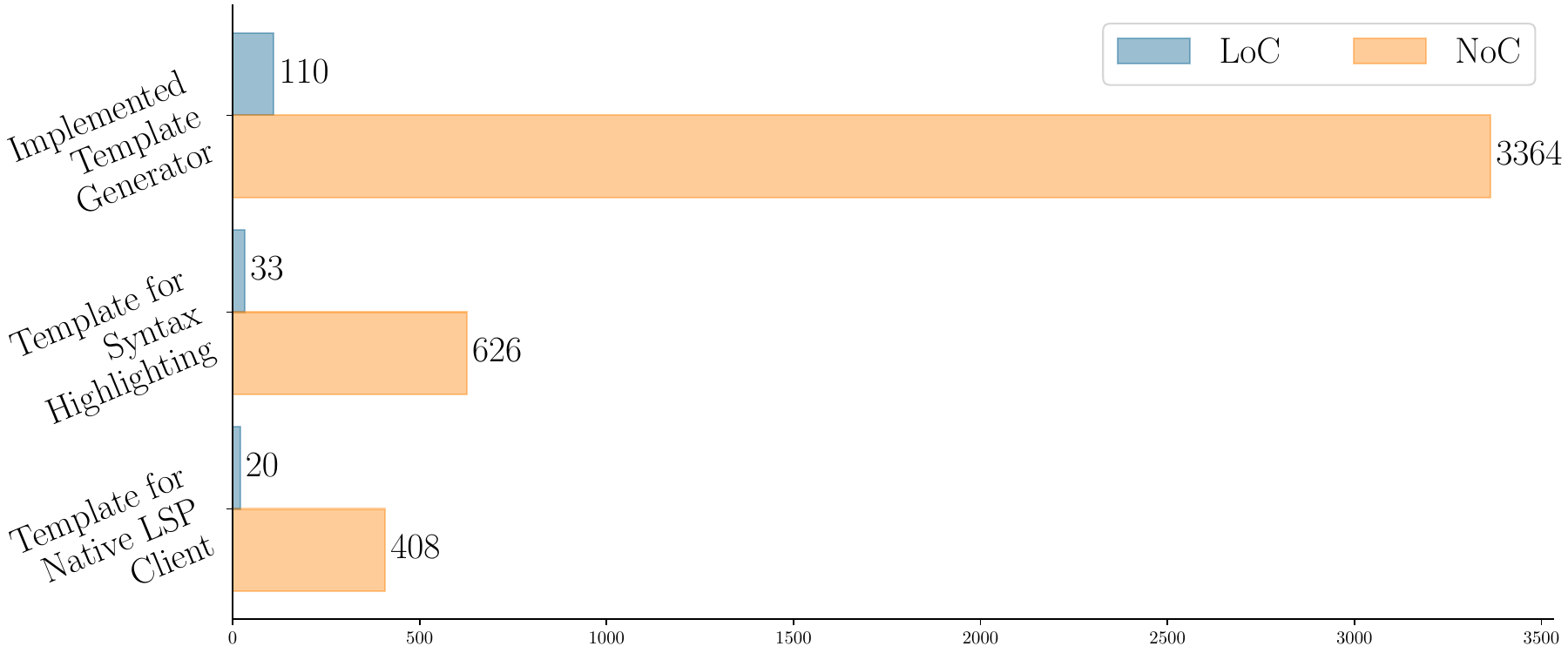}
		\caption{LoCs and NoCs required to implement the template generators for \toolname{NeoVim}. The \toolname{NeoVim} generator leverages the existence of annotated templates for syntax highlighting and LSP plugin.}%
		\label{fig:nvim}
	\end{subfigure}

	\hspace{.05\textwidth}%

	\begin{subfigure}{0.45\textwidth}
		\includegraphics[width=1\linewidth]{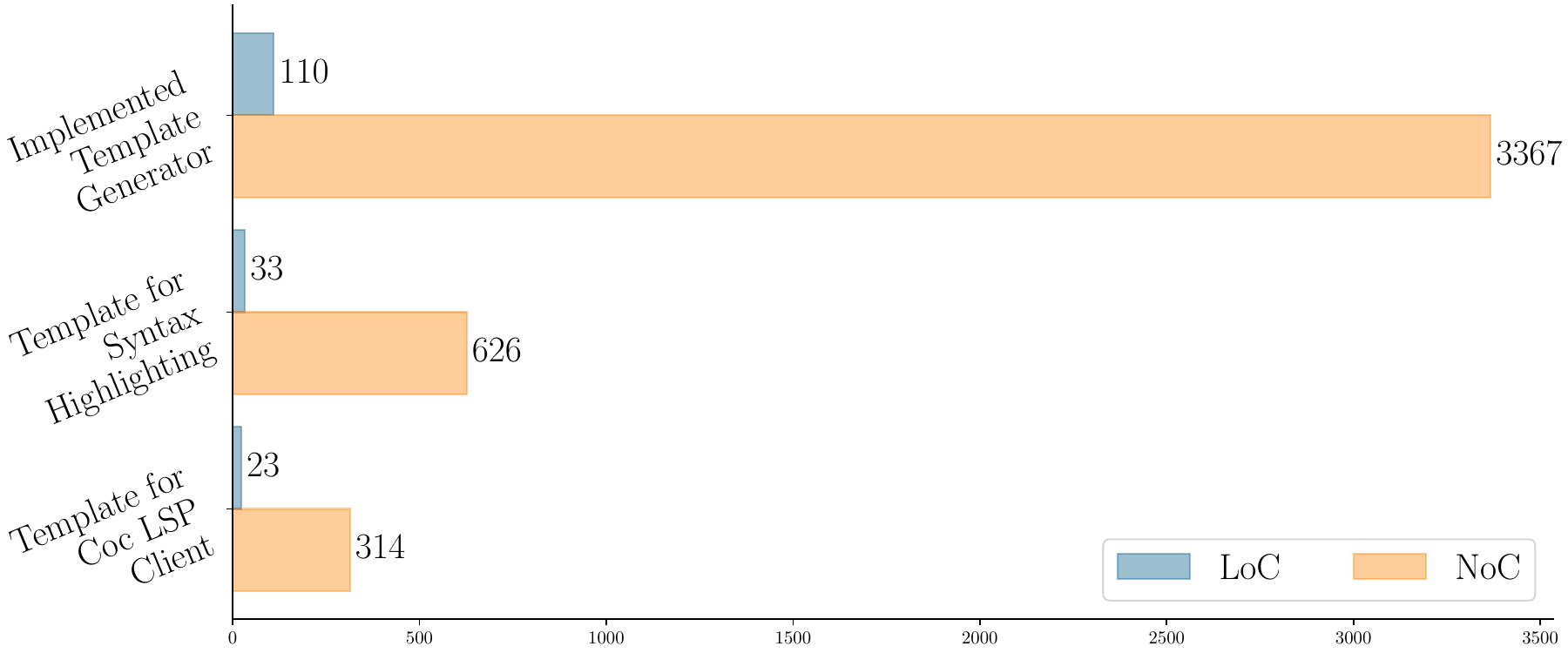}
		\caption{LoCs and NoCs required to implement the template generators for \toolname{Vim}. The \toolname{Vim} generator leverages the existence of annotated templates for syntax highlighting and LSP plugin.}%
		\label{fig:vim}
	\end{subfigure}
	\caption{Comparative analysis of LoCs and NoCs required to implement the template generators for \toolname{VSCode}, \toolname{NeoVim} and \toolname{Vim}. The figure highlights the differences in implementation complexity, demonstrating the flexibility and adaptability of the \toolname{Gradle} plugin in supporting multi editor ecosystems.}%
	\label{ffig:editors}
\end{figure}

The LSP is completed by the \textit{LSP client generator}. This generator is based on a system of \textit{template generators} responsible for generating syntax highlighting rules and LSP configuration files for supported editors.
To demonstrate the flexibility and simplicity of our approach, we implemented the aforementioned \toolname{Gradle} plugin, which currently supports \toolname{Vim}, \toolname{NeoVim} and \toolname{VSCode}.
Support for additional editors can easily be added in the future.
Each plugin has been used to generate the corresponding LSP plugin for both \toolname{SimpleLanguage} and Neverlang. Fig.~\ref{ffig:editors} shows the LoC and the NoC needed to implement the template generators for the three editors.
The implementation is relatively straightforward: template strings contain placeholders that are filled with values specific to the implemented language. Each placeholder follows the format \texttt{\$\{placeholder\}}.
Figs.~\ref{fig:vscode},~\ref{fig:nvim} and~\ref{fig:vim} report the LoC and NoC for the \toolname{VSCode}, \toolname{NeoVim}, and \toolname{Vim} plugins respectively, grouped by the following categories:\smallskip
\begin{compactitem}
	\item \textit{implemented template generator}, the Java code, integrated with the \toolname{Gradle} plugin, that implements the template generator;
	\item \textit{template for syntax highlighting}, the annotated template defining syntax highlighting;
	\item \textit{template for LSP plugin}, the annotated template for the LSP plugin;
	\item \textit{glue files}, additional files required only for full support in \toolname{VSCode}.\smallskip
\end{compactitem}
Each template needs to be annotated only once per editor. After that, the \toolname{Gradle} plugin uses language-specific input to populate the placeholders and generate a concrete plugin for any supported language.

Note that in our analysis, the NoC excludes whitespace, newline and tab characters.
According to Fig.~\ref{fig:vscode}, the \textit{implemented template generator} for \toolname{VSCode} has been implemented in \(160\) LoC and \(5,540\) NoC. For \toolname{Vim} (Fig.~\ref{fig:vim}) and \toolname{NeoVim} (Fig.~\ref{fig:nvim}), the template generators have been implemented in \(110\) LoC and \(3,364\) NoC.
\toolname{Vim} and \toolname{NeoVim} share the same file format for syntax highlighting, we could reuse the same code for both (see Fig.~\ref{fig:nvim} and Fig.~\ref{fig:vim}).
Moreover, \toolname{NeoVim} has built-in support for the LSP protocol, whereas \toolname{Vim} needs an external plugin. We chose to adopt CoC (\textit{Conquer of Completion}), one of the most widely used LSP clients for \toolname{Vim}. However, similar template generators could be created for other \toolname{Vim} plugins with minimal effort.
In our implementation, using CoC required slightly less code (20 LoC) than supporting the native \toolname{NeoVim} client (23 LoC).
According to~\citet{Bunder19}, the development effort for \toolname{VSCode} support is significant (around \(725\) minutes), as is the effort for implementing the language server (around \(350\) minutes). These figures are consistent with our findings: the amount of code required for the \toolname{VSCode} client is noticeably higher compared to the other editors. Specifically, the \textit{template for syntax highlighting} consists of 119 LoC and 1,383 NoC, while the \textit{template for LSP plugin} comprises 140 LoC and 2,700 NoC.

\begin{figure*}[t]
	\centering
	\begin{subfigure}{0.45\textwidth}
		\centering
		\includegraphics[width=\linewidth]{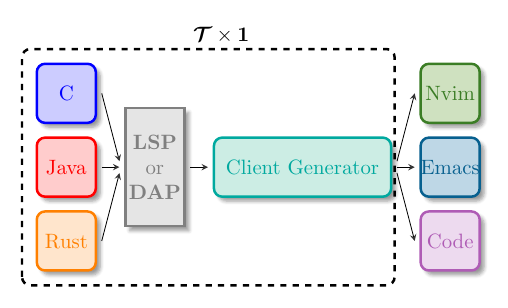}
		\caption{Reduction to \(\fontc{T} \times \mathbf{1}\) where \(\fontc{T}\) is the number of type systems implemented and holds \(\fontc{T} = \fontc{L}\). It is achieved by the generation of the language server leveraging the \toolname{Typelang} definitions and the editing support.}%
		\label{fig:lsp:reduction_lx1}
	\end{subfigure}
	\hspace{.05\textwidth}%
	\centering
	\begin{subfigure}{0.45\textwidth}
		\centering
		\includegraphics[width=\linewidth]{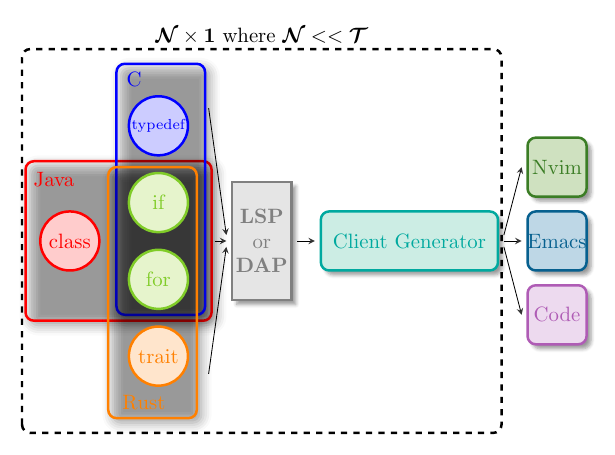}
		\caption{Reduction to \(\fontc{N} \times \mathbf{1}\) where \(\fontc{N} << \fontc{T}\) and \(\fontc{N}\) is the number of type systems without overlaps. It is achieved by reusing the artifacts across multiple languages.}%
		\label{fig:lsp:reduction_nx1}
	\end{subfigure}
	\caption{A graphical representation of the \(\fontc{T} \times \mathbf{1}\) and \(\fontc{N} \times \mathbf{1}\) reduction using the client generator.}%
	\label{fig:lsp:reduction}
\end{figure*}

\subsection{Discussion}
This section presents the results of the case study and addresses the research questions introduced in Sect.~\ref{sec:introduction}. We then discuss the limitations of our approach and its applicability to other language workbenches. Finally, we explore the implications of our approach for the future of the LSP\@.

\subsubsection{Research Questions}
In this study, we presented a demonstration case study to showcase the effectiveness of our approach. We addressed the three research questions introduced in Sect.~\ref{sec:introduction}, providing a detailed answer to each.

\begin{description}
	\item[\(\fontrq{RQ}\)\textit{\textsubscript{1}}] \textit{To what degree is it possible to streamline by associating variants to language artifacts?}
\end{description}

We demonstrated that by associating \toolname{Typelang} variants with language artifacts, it is possible to define type checking and type inference semantics in a modular way.
The \textit{variant-oriented programming} paradigm, along with the \textit{cross-artifact coordination} layer, played a key role in enabling different variants to coexist and span across multiple language artifacts. This results in a flexible and modular type system.
The benefits of this modularization are reflected in our results: the amount of code required to implement a type system was reduced by approximately \(82.90\%\) in terms of LoC and \(93.87\%\) in terms of NoC. On a per-artifact basis, the average reduction was approximately \(82.32\%\) for LoC and \(93.18\%\) for NoC. Moreover, the developed artifacts are reusable across different languages, which further reduces the overall development effort.

\begin{itemize}
	\item[\(\fontrq{RQ}\)\textit{\textsubscript{2}}] \textit{To what degree is it possible to automate the generation of LSP clients, lowering \(\fontc{E}\) to \(\mathbf{1}\)?}
\end{itemize}

LSP clients for different editors can be generated using the \toolname{Gradle} plugin introduced in Sect.~\ref{lsp:client_generation}. This plugin is responsible for generating both syntax highlighting and LSP configuration files for the supported editors.
The amount of code needed to implement the template generators varies across editors: the \toolname{VSCode} generator needs \(160\) LoC and \(5,540\) NoC, while the \toolname{NeoVim} and \toolname{Vim} generators need \(110\) LoC and \(3,364\) NoC. The higher code requirements for the \toolname{VSCode} generator stem from the need for additional glue files to achieve full language support.
In contrast, \toolname{NeoVim} provides native support for the LSP protocol, and \toolname{Vim} generators need less code because \toolname{NeoVim} offers a native support for the LSP protocol, meanwhile \toolname{Vim} relies on an external plugin. Since \toolname{NeoVim} uses the same syntax highlighting format as \toolname{Vim}, the same code can be reused for both.
Each template needs to be annotated only once per editor, after which the plugin can be reused to generate LSP clients for all supported editors and for any language. As a result, the effort required to support multiple editors is effectively reduced from \(\fontc{E}\) to \(\mathbf{1}\), thanks to the automation provided by the plugin.

\begin{itemize}
	\item[\(\fontrq{RQ}\)\textit{\textsubscript{3}}] \textit{Can the language server be automatically generated starting from the \toolname{Typelang} variants lowering \(\fontc{L}\) to \(\fontc{N}\)?}
\end{itemize}

In Sect.~\ref{lsp:building_blocks} and Sect.~\ref{lsp:variant_oriented}, we highlighted the possibility of generating the language server implementation by leveraging the data structures provided by the language workbench. We also demonstrated that the properties of \textit{variant-oriented programming} and \textit{cross-artifact modularization} layer have been crucial in ensuring that these data structures can be accessed across different variants.
By implementing the language servers according to the proposed methodology, we effectively reduced the number of language server implementations from \(\fontc{L}\) to \(\fontc{N}\) where \(\fontc{N} << \fontc{L}\).
The first reduction---from \(\fontc{L} \times \mathbf{1}\) to \(\fontc{T} \times \mathbf{1}\), as  shown in Fig.~\ref{fig:lsp:reduction_lx1}---is achieved by generating the language server from \toolname{Typelang} definitions. Even though \(\fontc{L} = \fontc{T}\), this results in a significant reduction in development time and effort, as defining the type system in \toolname{Typelang} is substantially simpler than implementing a full language server. This is supported by our case study and the answer to \(\fontrq{RQ}\)\textit{\textsubscript{1}}.
To the best of our knowledge, the best reduction in the literature so far is \(\fontc{T} \times \fontc{E}\), as achieved by Xtext. However, Xtext enforces a monolithic implementation of languages and does not support automated generation of LSP client plugins, making it impossible to reduce \(\fontc{E}\) and, consequently, offering no gains on the client side.
Client implementation is often overlooked in LSP development, yet it can represent a significant portion of the effort, as shown by~\citet{Bunder19}.
Our proposal not only improves upon the current state of the art by reducing the cost to \(\fontc{T}\times\mathbf{1}\), but it also achieves a further reduction to \(\fontc{N}\times\mathbf{1}\), where \(\fontc{N} << \fontc{T}\) represents the number of distinct, non-overlapping type systems. This is made possible through artifact reuse across multiple languages, as illustrated in Fig.~\ref{fig:lsp:reduction_nx1}.

\subsubsection{Limitations}
In this section, we discuss the limitations of our approach to generating LSP clients and language servers. Specifically, we address the limitations concerning: \begin{inparaenum}
	\item the proposed data structures (i.e., the \textit{Fenwick Tree} and the \textit{LSP Graph}), and
	\item the \toolname{Gradle} plugin.
\end{inparaenum}

The \textit{Fenwick tree} and the \textit{LSP graph} are data structures introduced in the context of LSP\@. While the former is a well-established data structure, it is not commonly used in LSP scenarios and could be replaced with alternatives such as the \textit{segment tree}~\cite{DeBerg08}. The segment tree is less efficient in terms of memory usage—\(\mathcal{O}(2n)\) instead of \(\mathcal{O}(n)\)—and has slightly higher time complexity—\(\log n + k\) instead of \(\log n\), where \(n\) is the number of elements and \(k\) the number of retrieved segments. However, unlike the Fenwick tree, which supports only prefix sums or point updates, the segment tree can support any associative function (e.g., \texttt{sum}, \texttt{min}, \texttt{max}, \texttt{gcd}) and enables efficient range updates via lazy propagation.

The \textit{LSP graph} is a novel data structure introduced specifically for the LSP domain. To the best of our knowledge, no alternative data structure currently exists that serves the same purpose. However, our implementation does not yet support \textit{code completion}—a feature that can be trivially derived from the LSP Graph. For example, the neighbors of a node in the graph can serve as valid completion suggestions. We chose not to implement this feature in our case study, as it was not strictly necessary for demonstration purposes; nonetheless, it is a clear direction for future work.

\toolname{Typelang} is a domain-specific language designed to define type systems within the LSP context. It is not a general-purpose language and is not intended for use outside the domain of type system definition. However, as a family of DSLs, a full implementation of \toolname{Typelang} could be used to define type systems for a wide range of programming languages. By leveraging the \textit{variant-oriented programming} paradigm and the \textit{cross-artifact coordination} layer, developers can reuse type system components across multiple languages.

The \toolname{Gradle} plugin is a specialized tool designed for LSP-related tasks. Although it is not intended as a general-purpose Gradle plugin, it can be used to generate LSP clients for various programming languages, as demonstrated in our case study. At present, it supports a limited number of editors, but it is easily extensible to support additional ones.

\subsubsection{Applicability}
The applicability of our approach extends beyond the Neverlang language workbench. It is suitable for any language workbench that supports the modularization of linguistic components. As shown in Table~\ref{ttab:problem-statement:lw-comparison}, other popular language workbenches—such as Melange, MPS, and Spoofax—also support this feature. Therefore, our approach can be applied to these language workbenches as well.

Both the \textit{variant-oriented programming} paradigm and the \textit{cross-artifact coordination} layer are generic mechanisms, applicable to any programming language. From one perspective, \toolname{Typelang} is a DSL designed to demonstrate the effectiveness of these paradigms; from another, it is a DSL for implementing type systems. Any DSL that supports the modularization of linguistic components could replace \toolname{Typelang} without compromising the benefits provided by our approach.

The same holds true for the \toolname{Gradle} plugin. While originally developed for Neverlang, its applicability is not restricted to language workbenches with modular linguistic components. Instead, it can be used with any language workbench that supports generating a compiler or interpreter targeting the JVM\@. Extending the plugin to support non-JVM-based languages is part of our planned future work. In conclusion, our approach is not confined to Neverlang: it generalizes to any language workbench with support for linguistic modularization and JVM-based compilation or interpretation.

\subsubsection{Implications}
The progressive reduction from \(\fontc{L} \times \fontc{E}\) to \(\fontc{T} \times \mathbf{1}\), and ultimately to \(\fontc{N} \times \mathbf{1}\), carries significant implications for the future of the LSP\@. By minimizing the number of required language servers and clients, our approach simplifies the development process and reduces the inherent complexity of LSP implementations. This simplification translates into faster development cycles, lower maintenance overhead, and improved runtime performance for both servers and clients.

Moreover, this reduction facilitates greater modularity and reusability of components across tools and languages. Developers can focus on implementing language-specific logic once, rather than repeatedly adapting it to fit a matrix of editors and environments. This not only lowers the barrier to entry for supporting new languages but also enhances the consistency and reliability of the user experience across tools.

From an ecosystem perspective, our model supports a more sustainable and scalable architecture. As the number of programming languages and development tools continues to grow, a leaner interaction model helps prevent the combinatorial explosion in LSP implementations. It also paves the way for richer interoperability and tool composition, potentially enabling new workflows and integrations that were previously too costly or complex to implement.

We believe that this streamlined architecture has the potential to make LSP more approachable, practical, and powerful for developers. In the long term, it may pave the way for a more unified and efficient tooling landscape across programming languages and environments.

\section{Related Work}\label{sec:related-work}
\paragraph*{Programming Paradigms.}
Programming paradigms have long been an established topic of research in the software engineering community~\cite{Rentsch82, Cardozo11b, Hughes89, Kumar16, Hu16}. Many paradigms~\cite{Prehofer97, Kiczales97} are specifically designed to address core software development concerns such as modularity, reusability, and maintainability. In \textit{feature-oriented programming}~\cite{Apel13, Czarnecki04, Prehofer01}, a feature module is a unit of composition that encapsulates specific functionality. It is treated as a first-class entity and can be composed with other feature modules to build complete software systems. Feature-oriented programming is typically employed in the development of software product lines and for the incremental development of programs. Similarly, \textit{delta-oriented programming}~\cite{Schaefer10, Damiani14, Koscielny14} focuses on the dynamic and incremental application of changes (called deltas) to a core module. Both paradigms share conceptual similarities with our approach in that they aim to modularize software systems into units that encapsulate distinct functionality. However, our approach required the ability to explicitly define and compose variants within a complex software system. This need is not fully addressed by either feature-oriented or delta-oriented paradigms, as they are primarily concerned with the incremental evolution of programs or the modification of a central core module. In contrast, our variants are not necessarily incremental nor must they relate to a single core; moreover, the overall system may or may not form a product line.

A related paradigm is \textit{aspect-oriented programming}~\cite{Kiczales01, Laddad03}, which introduces aspect modules to encapsulate crosscutting concerns—those that affect multiple parts of a system. Aspects define pointcuts and advices, where pointcuts identify join points in the execution of a program and advices specify the behavior to execute at those points. While the similarities to our approach are subtle, our notion of shared contexts can be considered analogous to join points: they define the interaction boundaries where variants can communicate and coordinate.

\vspace{0.5em}\paragraph*{Variability in Language Variants.}
\citet{deLara22} employ graph transformation techniques to model and compose language variants. While their approach excels at visualizing and managing interdependencies among language components, it does not directly address the automation of type system integration or the generation of LSPs. In contrast, \toolname{Typelang} leverages a variant-oriented programming paradigm and a cross-artifact coordination layer to modularize type system definitions. These definitions are then automatically integrated into the language server generation process, thereby significantly reducing manual effort across different editors.
\citet{deLara21} adopt multi-level modeling to capture variability within language families, facilitating the systematic reuse of language artifacts. While their framework provides robust support for language evolution and family engineering, it is primarily focused on high-level models and lacks dedicated mechanisms for modularizing type system components or automating LSP plugin generation.\ \toolname{Typelang} addresses this gap by providing a family of DSLs that not only encapsulate type definitions and checking rules in a modular fashion but also directly support the generation of editor integrations across multiple language artifacts.

\textit{FunKons}~\cite{Mosses14}, developed within the K framework, proposes a component-based approach to achieve modularity in semantics. Its main strength lies in composing semantic rules from independent components to ensure reusability at the semantic level. While \toolname{Typelang} shares the goal of modularity, it extends the focus to type systems, offering integrated support for type inference, type checking, and error reporting within a modular architecture that also facilitates LSP generation.
\citet{Mosses04} promote a modular formulation of operational semantics using structured rules. Though this work significantly advances the treatment of semantic variability, its scope remains confined to operational semantics. In contrast, \toolname{Typelang} captures semantic variability within its DSLs dedicated to type system definitions, addressing not only the correctness of language behavior but also the practical integration of type systems into development environments.

\vspace{0.5em}\paragraph*{Multi Product Lines.}
In the context of software product lines~\cite{Clements01, Pohl05, Krueger06}, the \textit{multi-product line} methodology~\cite{Rosenmuller10, Rosenmuller08, Rosenmuller11} refers to a special case in which multiple product lines are integrated into a unified product line. Our approach aims to operationalize this concept by offering a programming paradigm that enables the specification and composition of multiple variants within a complex software system, which may itself be a product line. Several methodological approaches have been proposed to support the development of multi-product lines. Notably, in 2011, \citet{ElSharkawy11} introduced a methodology to support the heterogeneous composition of multi-product lines. Similarly, \citet{Hartmann08} proposed the \textit{Context Variability Model}, a technique designed to constrain feature models, thereby enabling multi-product line support and facilitating staged configuration within software supply chains.

\vspace{0.5em}\paragraph*{Language Workbenches and IDEs.}
The development of DSLs has become a prominent area of research, attracting considerable attention from the software engineering community~\cite{Mendez-Acuna16, Kosar16}. Several language workbenches—such as Spoofax~\cite{Wachsmuth14}, MPS~\cite{Voelter12}, MontiCore~\cite{Krahn10}, and Melange~\cite{Degueule15}—have been proposed to facilitate the creation of DSLs. These platforms typically support the generation of IDE features; however, such support is often generic and fails to leverage the specific characteristics of individual DSLs. Most rely on static templates that overlook the modularity of DSL features, making it challenging to specify and reuse IDE services in a composable fashion, as enabled by our approach.
While MontiCore~\cite{Butting18, Butting18b} and Melange~\cite{Mendez-Acuna17} offer support for LPLs, their IDE support is generally limited to basic features like syntax highlighting and code completion. EMFText~\cite{Heidenreich11} also deserves mention as an EMF-based tool~\cite{Steinberg08}—similar to Xtext—that supports modular language development and facilitates IDE generation. Although EMFText does not explicitly address IDE variability, it shares several foundational concepts with our LPL-driven IDE approach, such as the use of attribute grammars to propagate information between languages and their IDEs, and a dedicated DSL for IDE description.
Among available workbenches, Xtext~\cite{Bunder19a} remains, to the best of our knowledge, the only tool that supports automatic generation of language servers conforming to the LSP\@. However, Xtext is monolithic and does not support modularity in the development of language syntax, semantics, type systems, or language server implementations, as summarized in Table~\ref{ttab:problem-statement:lw-comparison}.
More recently, \citet{Mosses19b} has proposed a formal, component-based methodology for building language workbenches that emphasizes correctness and reuse. While their framework systematically supports language development, it does not address the modularization of type systems or the generation of editor support. In contrast, Typelang is designed to operate within a modular language workbench, offering reusable type system fragments and enabling the automated generation of LSP clients for multiple editors.
Langium~\cite{Jordan24}, a modern monolithic language engineering tool, also supports LSP generation through a grammar-centric design. Although it provides robust LSP integration, it does not natively support modular type system definitions or the reuse of language components. Typelang, in contrast, introduces a family of DSLs designed specifically to modularize type systems and streamline the generation of LSP plugins for various editors, thereby enhancing both flexibility and reusability in language engineering.

\vspace{0.5em}\paragraph*{Languages for Type Systems.}
Bettini et al.~\cite{Bettini16, Bettini16b, Bettini19} introduced XSemantics, a DSL for specifying type systems and operational semantics in XText. Their work demonstrates how a language's type system can enhance language server generation. While their focus is on improving language server generation through the type system, our approach extends this idea by modularizing both the type system and the language server implementation, providing greater flexibility and reusability.

\citet{Beguet23} employ satisfiability modulo theories (SMT) to express and verify type system properties. Their approach uses logical DSLs to ensure type rules adhere to soundness criteria, with the SMT solver validating these constraints. In contrast, \toolname{Typelang} is designed to promote modularity, composability, and reusability within a language workbench. Instead of focusing on automated reasoning about type correctness, Typelang integrates type system definitions into the development pipeline, automatically generating language server support and streamlining editor integration. This approach is particularly useful for environments where rapid reassembly and deployment of language artifacts across multiple editors are required.

In ``Type Systems as Macros,'' \citet{Chang17b} treat type system definitions as syntactic extensions, allowing for automatic expansion of type constructs during compilation. While macros offer abstraction and reuse, their expansion is often closely tied to compiler infrastructure, limiting portability across different language workbenches or editors. Typelang, by contrast, is built around modular type system implementation through DSLs. This approach not only encapsulates type inference and checking but also integrates them into variant-oriented programming, enabling seamless composition of language artifacts and automatic generation of language server plugins.

Pacak et al.~\cite{Pacak20} focus on performance optimizations for type checking, particularly through incremental computation. Their method reduces recomputation by updating only the affected portions of the type checker. While this is essential for scaling type checking in large codebases and enhancing interactive development, it does not align directly with the goals of \toolname{Typelang}. Typelang’s primary focus is on the reuse of type system components and the automation of language server generation. The variant-oriented design in Typelang emphasizes modularity and cross-artifact coordination, which could potentially incorporate incremental techniques as an optimization, though the core strength of Typelang lies in reusing type system fragments across diverse language variants.

Language workbenches like Spoofax and MPS provide end-to-end tooling for language development, including syntax definition, semantic analysis, and editor integration. However, they often suffer from monolithic type system implementations that are tightly coupled to the host environment. For example, MPS supports projectional editing and reusability at the syntax level but lacks dedicated mechanisms for modularizing type system definitions. Similarly, Spoofax, while powerful in transformations, does not offer explicit constructs for composing and reusing type checking rules. In contrast, Typelang addresses these limitations by introducing a family of DSLs that modularize type definitions, type inference, and type checking. Its variant-oriented programming model separates concerns and promotes the reuse of common type system components. Moreover, its integration with automated LSP plugin generation significantly reduces the development overhead of supporting multiple editors.

\section{Conclusion}\label{sec:conclusion}
In this paper, we introduced \toolname{Typelang}, a family of DSLs for modular, composable, and reusable type system development. Leveraging \textit{variant-oriented programming} and a \textit{cross-artifact modularization} layer, \toolname{Typelang} enables the reuse of type system variants across artifacts and supports automated generation of language servers in modular workbenches. To reduce the LSP plugin implementation effort from \(\fontc{E}\) to \(\mathbf{1}\), we developed a \toolname{Gradle} plugin that automates plugin generation, reducing the effort further to \(\fontc{T} \times \mathbf{1}\) and then to \(\fontc{N} \times \mathbf{1}\), where \(\fontc{N} << \fontc{T}\)  through artifact reuse.
Our approach streamlines the creation of type systems and LSP plugins for language families, improving efficiency in modular language workbenches. Empirical results show that our approach can reduce the effort required to implement type systems by 93.48\% in terms of NoC and LSP plugins by 100\%.

\section*{Acknowledgments}
This work was partly supported by the MUR project ``T-LADIES'' (PRIN 2020TL3X8X). 

\bibliographystyle{elsarticle-harv}
\bibliography{local,strings,metrics,programming,software_engineering,data_structures,software_architecture,dsl,pl,splc,oolanguages,my_work,grammars,security,roles,learning,cop,testing,dsu,distributed_systems,foundations,reflection,aosd,pattern,tools,logic}

\newcommand\sortnoop[1]{}\newcommand{\splps}{\textsf{SP$\!$\reflectbox{L}L$\!$\reflectbox{SP}}}\newcommand{\starpiler}{\ensuremath{\displaystyle\bigstar}piler}
\begin{thebibliography}{143}
\expandafter\ifx\csname natexlab\endcsname\relax\def\natexlab#1{#1}\fi
\providecommand{\url}[1]{\texttt{#1}}
\providecommand{\href}[2]{#2}
\providecommand{\path}[1]{#1}
\providecommand{\DOIprefix}{doi:}
\providecommand{\ArXivprefix}{arXiv:}
\providecommand{\URLprefix}{URL: }
\providecommand{\Pubmedprefix}{pmid:}
\providecommand{\doi}[1]{\href{http://dx.doi.org/#1}{\path{#1}}}
\providecommand{\Pubmed}[1]{\href{pmid:#1}{\path{#1}}}
\providecommand{\bibinfo}[2]{#2}
\ifx\xfnm\relax \def\xfnm[#1]{\unskip,\space#1}\fi
\bibitem[{Aho et~al.(1986)Aho, Sethi and Ullman}]{Aho86}
\bibinfo{author}{Aho, A.V.}, \bibinfo{author}{Sethi, R.},
  \bibinfo{author}{Ullman, J.D.}, \bibinfo{year}{1986}.
\newblock \bibinfo{title}{{Compilers: Principles, Techniques, and Tools}}.
\newblock \bibinfo{publisher}{Addison Wesley}, \bibinfo{address}{Reading,
  Massachusetts}.
\bibitem[{Ammann(1978)}]{Ammann78}
\bibinfo{author}{Ammann, U.}, \bibinfo{year}{1978}.
\newblock \bibinfo{title}{{Error Recovery in Recursive Descent Parsers and
  Run-Time Storage Organization}}.
\newblock \bibinfo{type}{Technical Report} \bibinfo{number}{D-INFK}. ETH
  Z\"urich, Department of Computer Science. \bibinfo{address}{Z\"urich,
  Switzerland}.
\bibitem[{Apel et~al.(2008)Apel, Leich and Saake}]{Apel08}
\bibinfo{author}{Apel, S.}, \bibinfo{author}{Leich, T.},
  \bibinfo{author}{Saake, G.}, \bibinfo{year}{2008}.
\newblock \bibinfo{title}{{Aspectual Feature Modules}}.
\newblock \bibinfo{journal}{IEEE Transactions on Software Engineering}
  \bibinfo{volume}{34}, \bibinfo{pages}{162--180}.
\bibitem[{Apel et~al.(2013)Apel, von Thein, Wendler, Gr\"o{\ss}linger and
  Beyer}]{Apel13}
\bibinfo{author}{Apel, S.}, \bibinfo{author}{von Thein, A.},
  \bibinfo{author}{Wendler, P.}, \bibinfo{author}{Gr\"o{\ss}linger, A.},
  \bibinfo{author}{Beyer, F.}, \bibinfo{year}{2013}.
\newblock \bibinfo{title}{{Strategies for Product-Line Verification: Case
  Studies and Experiments}}, in: \bibinfo{editor}{Chang, B.H.},
  \bibinfo{editor}{Pohl, K.} (Eds.), \bibinfo{booktitle}{{Proceedings of the
  35th International Conference on Software Engineering (ICSE'13)}},
  \bibinfo{publisher}{IEEE}, \bibinfo{address}{San Francisco, CA, USA}. pp.
  \bibinfo{pages}{482--491}.
\bibitem[{Barros et~al.(2022)Barros, Peldszus, Assun\c{c}\~ao and
  Berger}]{Barros22}
\bibinfo{author}{Barros, D.}, \bibinfo{author}{Peldszus, S.},
  \bibinfo{author}{Assun\c{c}\~ao, W.K.G.}, \bibinfo{author}{Berger, T.},
  \bibinfo{year}{2022}.
\newblock \bibinfo{title}{{Editing Support for Software Languages:
  Implementation Practices in Language Server Protocols}}, in:
  \bibinfo{editor}{Wimmer, M.} (Ed.), \bibinfo{booktitle}{{Proceedings of the
  25th International Conference on Model Driven Engineering Langauges and
  Systems (MoDELS'22)}}, \bibinfo{publisher}{ACM},
  \bibinfo{address}{Montr\'eal, Canada}. pp. \bibinfo{pages}{232--243}.
\bibitem[{Bassett(1996)}]{Bassett96}
\bibinfo{author}{Bassett, P.}, \bibinfo{year}{1996}.
\newblock \bibinfo{title}{{Framing Software Reuse: Lessons from the Real
  World}}.
\newblock \bibinfo{publisher}{Prentice Hall}.
\bibitem[{Basten et~al.(2015)Basten, van~den Bos, Hills, Klint, Lankamp,
  Lisser, van~der Ploeg, van~der Storm and Vinju}]{Basten15}
\bibinfo{author}{Basten, B.}, \bibinfo{author}{van~den Bos, J.},
  \bibinfo{author}{Hills, M.}, \bibinfo{author}{Klint, P.},
  \bibinfo{author}{Lankamp, A.}, \bibinfo{author}{Lisser, B.},
  \bibinfo{author}{van~der Ploeg, A.}, \bibinfo{author}{van~der Storm, T.},
  \bibinfo{author}{Vinju, J.}, \bibinfo{year}{2015}.
\newblock \bibinfo{title}{{Modular Language Implementation in
  Rascal---Experience Report}}.
\newblock \bibinfo{journal}{{Science of Computer Programming}}
  \bibinfo{volume}{114}, \bibinfo{pages}{7--19}.
\bibitem[{Batory et~al.(1998)Batory, Lofaso and Smaragdakis}]{Batory98}
\bibinfo{author}{Batory, D.}, \bibinfo{author}{Lofaso, B.},
  \bibinfo{author}{Smaragdakis, Y.}, \bibinfo{year}{1998}.
\newblock \bibinfo{title}{{JTS: Tools for Implementing Domain-Specific
  Languages}}, in: \bibinfo{booktitle}{{Proceedings of the 5th International
  Conference on Software Reuse (ICSR'98)}}, \bibinfo{publisher}{IEEE Computer
  Society}, \bibinfo{address}{Victoria, BC, Canada}. pp.
  \bibinfo{pages}{143--153}.
\bibitem[{Batory et~al.(2004)Batory, Sarvela and Rauschmayer}]{Batory04}
\bibinfo{author}{Batory, D.}, \bibinfo{author}{Sarvela, J.N.},
  \bibinfo{author}{Rauschmayer, A.}, \bibinfo{year}{2004}.
\newblock \bibinfo{title}{{Scaling Step-Wise Refinement}}.
\newblock \bibinfo{journal}{IEEE Transactions on Software Engineering}
  \bibinfo{volume}{30}, \bibinfo{pages}{355--371}.
\bibitem[{Bay and Pauls(2004)}]{Bay04}
\bibinfo{author}{Bay, T.G.}, \bibinfo{author}{Pauls, K.}, \bibinfo{year}{2004}.
\newblock \bibinfo{title}{{Reuse Frequency as Metric for Component
  Assessment}}.
\newblock \bibinfo{type}{Technical Report} \bibinfo{number}{464}. ETH,
  Department of Computer Science. \bibinfo{address}{Z\"urich, Switzerland}.
\bibitem[{B{\'e}guet and Amiard(2023)}]{Beguet23}
\bibinfo{author}{B{\'e}guet, R.}, \bibinfo{author}{Amiard, R.},
  \bibinfo{year}{2023}.
\newblock \bibinfo{title}{{Application of SMT in a Meta-Compiler: A Logic DSL
  for Specifying Type Systems}}, in: \bibinfo{editor}{Graham-Lengrand, S.},
  \bibinfo{editor}{Preiner, M.} (Eds.), \bibinfo{booktitle}{{Proceedings of the
  21st International Workshop on Satisfiability Modulo Theories (SMT'23)}},
  \bibinfo{publisher}{CEUR}, \bibinfo{address}{Haifa, Israel}. pp.
  \bibinfo{pages}{46--61}.
\bibitem[{{\sortnoop{Berg}}de~Berg et~al.(2008){\sortnoop{Berg}}de~Berg,
  Cheong, van Kreveld and Overmars}]{DeBerg08}
\bibinfo{author}{{\sortnoop{Berg}}de~Berg, M.}, \bibinfo{author}{Cheong, O.},
  \bibinfo{author}{van Kreveld, M.}, \bibinfo{author}{Overmars, M.},
  \bibinfo{year}{2008}.
\newblock \bibinfo{title}{{Computational Geometry: Algorithms and
  Applications}}.
\newblock \bibinfo{edition}{3rd} ed., \bibinfo{publisher}{Springer}.
\bibitem[{Bertolotti et~al.(2023a)Bertolotti, Cazzola and Favalli}]{Cazzola23b}
\bibinfo{author}{Bertolotti, F.}, \bibinfo{author}{Cazzola, W.},
  \bibinfo{author}{Favalli, L.}, \bibinfo{year}{2023}a.
\newblock \bibinfo{title}{{On the Granularity of Linguistic Reuse}}.
\newblock \bibinfo{journal}{{Journal of Systems and Software}}
  \bibinfo{volume}{202}.
\newblock \DOIprefix\doi{10.1016/j.jss.2023.111704}.
\bibitem[{Bertolotti et~al.(2023b)Bertolotti, Cazzola and Favalli}]{Cazzola23d}
\bibinfo{author}{Bertolotti, F.}, \bibinfo{author}{Cazzola, W.},
  \bibinfo{author}{Favalli, L.}, \bibinfo{year}{2023}b.
\newblock \bibinfo{title}{{\splps: Software Product Lines Extraction Driven by
  Language Server Protocol}}.
\newblock \bibinfo{journal}{{Journal of Systems and Software}}
  \bibinfo{volume}{205}.
\newblock \DOIprefix\doi{10.1016/j.jss.2023.111809}.
\bibitem[{Bettini(2011)}]{Bettini11}
\bibinfo{author}{Bettini, L.}, \bibinfo{year}{2011}.
\newblock \bibinfo{title}{{A DSL for Writing Type Systems for Xtext
  Languages}}, in: \bibinfo{editor}{Probst, C.W.}, \bibinfo{editor}{Wimmer, C.}
  (Eds.), \bibinfo{booktitle}{{Proceedings of the 9th International Conference
  on Principles and Practice of Programming in Java (PPPJ'11)}},
  \bibinfo{publisher}{ACM}, \bibinfo{address}{Kongens Lyngby, Denmark}. pp.
  \bibinfo{pages}{31--40}.
\bibitem[{Bettini(2013a)}]{Bettini13b}
\bibinfo{author}{Bettini, L.}, \bibinfo{year}{2013}a.
\newblock \bibinfo{title}{{Implementing Domain-Specific Languages with Xtext
  and Xtend}}.
\newblock \bibinfo{publisher}{{PACKT Publishing Ltd}}.
\bibitem[{Bettini(2013b)}]{Bettini13}
\bibinfo{author}{Bettini, L.}, \bibinfo{year}{2013}b.
\newblock \bibinfo{title}{{Implementing Java-like Languages in Xtext with
  Xsemantics}}, in: \bibinfo{booktitle}{{Proceedings of the 28th Annual ACM
  Symposium on Applied Computing (SAC'13)}}, \bibinfo{organization}{ACM},
  \bibinfo{address}{Coimbra, Portugal}. pp. \bibinfo{pages}{1559--1564}.
\bibitem[{Bettini(2016)}]{Bettini16}
\bibinfo{author}{Bettini, L.}, \bibinfo{year}{2016}.
\newblock \bibinfo{title}{{Implementing Type Systems for the IDE with
  Xsemantics}}.
\newblock \bibinfo{journal}{{Journal of Logical and Algebraic Methods in
  Programming}} \bibinfo{volume}{85}, \bibinfo{pages}{655--680}.
\bibitem[{Bettini(2019)}]{Bettini19}
\bibinfo{author}{Bettini, L.}, \bibinfo{year}{2019}.
\newblock \bibinfo{title}{{Type Errors for the IDE with Xtext and Xsemantics}}.
\newblock \bibinfo{journal}{{Journal Open Computer Science}}
  \bibinfo{volume}{9}, \bibinfo{pages}{52--79}.
\bibitem[{Bettini et~al.(2016)Bettini, von Pilgrim and Reiser}]{Bettini16b}
\bibinfo{author}{Bettini, L.}, \bibinfo{author}{von Pilgrim, J.},
  \bibinfo{author}{Reiser, M.O.}, \bibinfo{year}{2016}.
\newblock \bibinfo{title}{{Implementing a Typed Javascript and Its IDE: A
  CASE-Study with Xsemantics}}.
\newblock \bibinfo{journal}{{Journal on Advances in Software}}
  \bibinfo{volume}{9}, \bibinfo{pages}{283--303}.
\bibitem[{Bezerra et~al.(2015)Bezerra, Andrade and Monteiro}]{Bezerra15}
\bibinfo{author}{Bezerra, C.I.M.}, \bibinfo{author}{Andrade, R.M.C.},
  \bibinfo{author}{Monteiro, J.M.}, \bibinfo{year}{2015}.
\newblock \bibinfo{title}{{Measures for Quality Evaluation of Feature Models}},
  in: \bibinfo{booktitle}{{Proceedings of the 9th International Conference on
  Software and Software Reuse (ICSR'15)}}, \bibinfo{publisher}{Springer},
  \bibinfo{address}{Miami, FL, USA}. pp. \bibinfo{pages}{282--297}.
\bibitem[{Booch et~al.(2007)Booch, Maksimchuk, Engle, Young, Conallen and
  Houston}]{Booch07}
\bibinfo{author}{Booch, G.}, \bibinfo{author}{Maksimchuk, R.},
  \bibinfo{author}{Engle, M.}, \bibinfo{author}{Young, B.},
  \bibinfo{author}{Conallen, J.}, \bibinfo{author}{Houston, K.},
  \bibinfo{year}{2007}.
\newblock \bibinfo{title}{{Object-Oriented Analysis and Design with
  Applications}}.
\newblock \bibinfo{edition}{Third} ed., \bibinfo{publisher}{Addison-Wesley}.
\bibitem[{Briand et~al.(1999)Briand, Daly and W{\"u}st}]{Briand99}
\bibinfo{author}{Briand, L.C.}, \bibinfo{author}{Daly, J.W.},
  \bibinfo{author}{W{\"u}st, J.}, \bibinfo{year}{1999}.
\newblock \bibinfo{title}{{A Unified Framework for Coupling Measurement in
  Object-Oriented Systems}}.
\newblock \bibinfo{journal}{IEEE Transactions on Software Engineering}
  \bibinfo{volume}{25}, \bibinfo{pages}{91--121}.
\bibitem[{B\"under(2019)}]{Bunder19a}
\bibinfo{author}{B\"under, H.}, \bibinfo{year}{2019}.
\newblock \bibinfo{title}{{Decoupling Language and Editor: The Impact of the
  Language Server Protocol on Textual Domain-Specific Languages}}, in:
  \bibinfo{editor}{Hammoudi, S.}, \bibinfo{editor}{Ferreira~Pires, L.},
  \bibinfo{editor}{Seli\c{c}, B.} (Eds.), \bibinfo{booktitle}{{Proceedings of
  the 7th International Conference on Model-Driven Engineering and Software
  Development (MODELWARD'19)}}, \bibinfo{publisher}{SciTe Press},
  \bibinfo{address}{Prague, Czech Republic}. pp. \bibinfo{pages}{129--140}.
\bibitem[{B\"under and Kuchen(2019)}]{Bunder19}
\bibinfo{author}{B\"under, H.}, \bibinfo{author}{Kuchen, H.},
  \bibinfo{year}{2019}.
\newblock \bibinfo{title}{{Towards Multi-editor Support for Domain-Specific
  Languages Utilizing the Language Server Protocol}}, in:
  \bibinfo{editor}{Hammoudi, S.}, \bibinfo{editor}{Pires, L.F.},
  \bibinfo{editor}{Seli\'c, B.} (Eds.), \bibinfo{booktitle}{{Proceedings of the
  7th International Conference on Model-Driven Engineering and Software
  Development (MODELSWARD'19)}}, \bibinfo{publisher}{Springer},
  \bibinfo{address}{Prague, Czech Republic}. pp. \bibinfo{pages}{225--245}.
\bibitem[{Butting et~al.(2018a)Butting, Eikermann, Kautz, Rumpe and
  Wortmann}]{Butting18}
\bibinfo{author}{Butting, A.}, \bibinfo{author}{Eikermann, R.},
  \bibinfo{author}{Kautz, O.}, \bibinfo{author}{Rumpe, B.},
  \bibinfo{author}{Wortmann, A.}, \bibinfo{year}{2018}a.
\newblock \bibinfo{title}{{Controlled and Extensible Variability of Concrete
  and Abstract Syntax with Independent Language Features}}, in:
  \bibinfo{booktitle}{{Proceedings of the 12th International Workshop on
  Variability Modelling of Software Intensive Systems (VAMOS'18)}},
  \bibinfo{publisher}{ACM}, \bibinfo{address}{Madrid, Spain}. pp.
  \bibinfo{pages}{75--82}.
\bibitem[{Butting et~al.(2018b)Butting, Eikermann, Kautz, Rumpe and
  Wortmann}]{Butting18b}
\bibinfo{author}{Butting, A.}, \bibinfo{author}{Eikermann, R.},
  \bibinfo{author}{Kautz, O.}, \bibinfo{author}{Rumpe, B.},
  \bibinfo{author}{Wortmann, A.}, \bibinfo{year}{2018}b.
\newblock \bibinfo{title}{{Modeling Language Variability with Reusable Language
  Components}}, in: \bibinfo{editor}{Berger, T.}, \bibinfo{editor}{Borba, P.}
  (Eds.), \bibinfo{booktitle}{{Proceedings of the 22nd International Systems
  and Software Product Line Conference (SPLC'18)}}, \bibinfo{publisher}{ACM},
  \bibinfo{address}{Gothenburg, Sweden}. pp. \bibinfo{pages}{65--75}.
\bibitem[{Cardelli(1988)}]{Cardelli88}
\bibinfo{author}{Cardelli, L.}, \bibinfo{year}{1988}.
\newblock \bibinfo{title}{{Structural Subtyping and the Notion of Power Type}},
  in: \bibinfo{editor}{Ferrante, J.}, \bibinfo{editor}{Peter, M.} (Eds.),
  \bibinfo{booktitle}{{Proceedings of the 15th ACM Symposium on Principles of
  Programming Languages (POPL'88)}}, \bibinfo{publisher}{ACM},
  \bibinfo{address}{San Diego, CA, USA}. pp. \bibinfo{pages}{70--79}.
\bibitem[{Cardozo et~al.(2011)Cardozo, G\"unther, D'Hondt and
  Mens}]{Cardozo11b}
\bibinfo{author}{Cardozo, N.}, \bibinfo{author}{G\"unther, S.},
  \bibinfo{author}{D'Hondt, T.}, \bibinfo{author}{Mens, K.},
  \bibinfo{year}{2011}.
\newblock \bibinfo{title}{{Feature-Oriented Programming and Context-Oriented
  Programming: Comparing Paradigm Characteristics by Example Implementations}},
  in: \bibinfo{editor}{Hartmann, H.}, \bibinfo{editor}{Breivold, H.P.} (Eds.),
  \bibinfo{booktitle}{{Proceedings of the 6th International Conference on
  Software Engineering Advances (ICSEA'11)}}, \bibinfo{publisher}{IARIA},
  \bibinfo{address}{Barcelona, Spain}. pp. \bibinfo{pages}{130--135}.
\bibitem[{Cazzola(2012)}]{Cazzola12c}
\bibinfo{author}{Cazzola, W.}, \bibinfo{year}{2012}.
\newblock \bibinfo{title}{{Domain-Specific Languages in Few Steps: The
  Neverlang Approach}}, in: \bibinfo{editor}{Gschwind, T.},
  \bibinfo{editor}{De~Paoli, F.}, \bibinfo{editor}{Gruhn, V.},
  \bibinfo{editor}{Book, M.} (Eds.), \bibinfo{booktitle}{{Proceedings of the
  11\textsuperscript{th} International Conference on Software Composition
  (SC'12)}}, \bibinfo{publisher}{Springer}, \bibinfo{address}{Prague, Czech
  Republic}. pp. \bibinfo{pages}{162--177}.
\bibitem[{Cazzola and Favalli(2022)}]{Cazzola21b}
\bibinfo{author}{Cazzola, W.}, \bibinfo{author}{Favalli, L.},
  \bibinfo{year}{2022}.
\newblock \bibinfo{title}{{Towards a Recipe for Language Decomposition: Quality
  Assessment of Language Product Lines}}.
\newblock \bibinfo{journal}{{Empirical Software Engineering}}
  \bibinfo{volume}{27}.
\newblock \DOIprefix\doi{10.1007/s10664-021-10074-6}.
\bibitem[{Cazzola and Favalli(2024)}]{Cazzola24d}
\bibinfo{author}{Cazzola, W.}, \bibinfo{author}{Favalli, L.},
  \bibinfo{year}{2024}.
\newblock \bibinfo{title}{{Software Modernization Powered by Dynamic Language
  Product Lines}}.
\newblock \bibinfo{journal}{{Journal of Systems and Software}}
  \bibinfo{volume}{218}.
\newblock \DOIprefix\doi{10.1016/j.jss.2024.112188}.
\bibitem[{Cazzola and Olivares(2016)}]{Cazzola16}
\bibinfo{author}{Cazzola, W.}, \bibinfo{author}{Olivares, D.M.},
  \bibinfo{year}{2016}.
\newblock \bibinfo{title}{{Gradually Learning Programming Supported by a
  Growable Programming Language}}.
\newblock \bibinfo{journal}{{IEEE Transactions on Emerging Topics in
  Computing}} \bibinfo{volume}{4}, \bibinfo{pages}{404--415}.
\newblock \DOIprefix\doi{10.1109/TETC.2015.2446192}. \bibinfo{note}{special
  Issue on Emerging Trends in Education}.
\bibitem[{Cazzola and Poletti(2010)}]{Cazzola10b}
\bibinfo{author}{Cazzola, W.}, \bibinfo{author}{Poletti, D.},
  \bibinfo{year}{2010}.
\newblock \bibinfo{title}{{DSL Evolution through Composition}}, in:
  \bibinfo{booktitle}{{Proceedings of the 7\textsuperscript{th} ECOOP Workshop
  on Reflection, AOP and Meta-Data for Software Evolution (RAM-SE'10)}},
  \bibinfo{publisher}{ACM}, \bibinfo{address}{Maribor, Slovenia}.
\bibitem[{Cazzola and Vacchi(2013)}]{Cazzola13e}
\bibinfo{author}{Cazzola, W.}, \bibinfo{author}{Vacchi, E.},
  \bibinfo{year}{2013}.
\newblock \bibinfo{title}{{Neverlang 2: Componentised Language Development for
  the JVM}}, in: \bibinfo{editor}{Binder, W.}, \bibinfo{editor}{Bodden, E.},
  \bibinfo{editor}{L\"owe, W.} (Eds.), \bibinfo{booktitle}{{Proceedings of the
  12\textsuperscript{th} International Conference on Software Composition
  (SC'13)}}, \bibinfo{publisher}{Springer}, \bibinfo{address}{Budapest,
  Hungary}. pp. \bibinfo{pages}{17--32}.
\bibitem[{Chang et~al.(2017)Chang, Knauth and Greenman}]{Chang17b}
\bibinfo{author}{Chang, S.}, \bibinfo{author}{Knauth, A.},
  \bibinfo{author}{Greenman, B.}, \bibinfo{year}{2017}.
\newblock \bibinfo{title}{{Type Systems as Macros}}, in:
  \bibinfo{editor}{Gordon, A.D.} (Ed.), \bibinfo{booktitle}{{Proceedings of the
  44th Symposium on Principles of Programming Languages (PoPL'17)}},
  \bibinfo{publisher}{ACM}, \bibinfo{address}{Paris, France}. pp.
  \bibinfo{pages}{694--705}.
\bibitem[{Clements and Northrop(2001)}]{Clements01}
\bibinfo{author}{Clements, P.}, \bibinfo{author}{Northrop, L.},
  \bibinfo{year}{2001}.
\newblock \bibinfo{title}{{Software Product Lines: Practices and Patterns}}.
\newblock \bibinfo{publisher}{{Addison-Wesley}}.
\bibitem[{Cook et~al.(1990)Cook, Hill and Canning}]{Cook90}
\bibinfo{author}{Cook, W.R.}, \bibinfo{author}{Hill, W.},
  \bibinfo{author}{Canning, P.S.}, \bibinfo{year}{1990}.
\newblock \bibinfo{title}{{Inheritance Is Not Subtyping}}, in:
  \bibinfo{editor}{Allen, F.E.} (Ed.), \bibinfo{booktitle}{{Proceedings of the
  17th ACM SIGPLAN-SIGACT Symposium on Principles of Programming Languages
  (POPL'90)}}, \bibinfo{publisher}{ACM}, \bibinfo{address}{San Francisco, CA,
  USA}. pp. \bibinfo{pages}{125--135}.
\bibitem[{Cooper and Torczon(2022)}]{Cooper22}
\bibinfo{author}{Cooper, K.D.}, \bibinfo{author}{Torczon, L.},
  \bibinfo{year}{2022}.
\newblock \bibinfo{title}{{Engineering a Compiler}}.
\newblock \bibinfo{publisher}{Morgan Kaufmann}.
\bibitem[{Crane and Dingel(2005)}]{Crane05}
\bibinfo{author}{Crane, M.L.}, \bibinfo{author}{Dingel, J.},
  \bibinfo{year}{2005}.
\newblock \bibinfo{title}{{UML vs. Classical vs. Rhapsody Statecharts: Not All
  Models Are Created Equal}}, in: \bibinfo{editor}{Briand, L.},
  \bibinfo{editor}{Williams, C.} (Eds.), \bibinfo{booktitle}{{Proceedings of
  the 8th International Conference on Model Driven Engineering Languages and
  Systems (MoDELS'05)}}, \bibinfo{publisher}{Springer},
  \bibinfo{address}{Montego Bay, Jamaica}. pp. \bibinfo{pages}{97--112}.
\bibitem[{Czarnecki et~al.(2004)Czarnecki, Helsen and Eisenecker}]{Czarnecki04}
\bibinfo{author}{Czarnecki, K.}, \bibinfo{author}{Helsen, S.},
  \bibinfo{author}{Eisenecker, U.}, \bibinfo{year}{2004}.
\newblock \bibinfo{title}{{Staged Configuration Using Feature Models}}, in:
  \bibinfo{editor}{Weiss, D.}, \bibinfo{editor}{van Ommering, R.} (Eds.),
  \bibinfo{booktitle}{{Proceedings of the 3rd International Conference on
  Software Product-Line (SPLC'04)}}, \bibinfo{publisher}{Springer},
  \bibinfo{address}{Boston, MA, USA}. pp. \bibinfo{pages}{266--283}.
\bibitem[{Damiani et~al.(2014)Damiani, Schaefer and Winkelmann}]{Damiani14}
\bibinfo{author}{Damiani, F.}, \bibinfo{author}{Schaefer, I.},
  \bibinfo{author}{Winkelmann, T.}, \bibinfo{year}{2014}.
\newblock \bibinfo{title}{{Delta-Oriented Multi Software Product Lines}}, in:
  \bibinfo{editor}{Heymans, P.}, \bibinfo{editor}{Rubin, J.} (Eds.),
  \bibinfo{booktitle}{{Proceedings of 18th International Software Product Line
  Conference (SPLC'14)}}, \bibinfo{publisher}{ACM}, \bibinfo{address}{Florence,
  Italy}. pp. \bibinfo{pages}{232--236}.
\bibitem[{Damini et~al.(2019)Damini, Lienhardt and Paolini}]{Damiani19}
\bibinfo{author}{Damini, F.}, \bibinfo{author}{Lienhardt, M.},
  \bibinfo{author}{Paolini, L.}, \bibinfo{year}{2019}.
\newblock \bibinfo{title}{{A Formal Model for Multi Software Product Lines}}.
\newblock \bibinfo{journal}{{Science of Computer Programming}}
  \bibinfo{volume}{172}, \bibinfo{pages}{203--231}.
\bibitem[{Deelstra et~al.(2005)Deelstra, Sinnema and Bosch}]{Deelstra05}
\bibinfo{author}{Deelstra, S.}, \bibinfo{author}{Sinnema, M.},
  \bibinfo{author}{Bosch, J.}, \bibinfo{year}{2005}.
\newblock \bibinfo{title}{{Product Derivation in Software Product Families: A
  Case Study}}.
\newblock \bibinfo{journal}{{Journal of Systems and Software}}
  \bibinfo{volume}{74}, \bibinfo{pages}{173--194}.
\bibitem[{Degueule et~al.(2015)Degueule, Combemale, Blouin, Barais and
  J\'ez\'equel}]{Degueule15}
\bibinfo{author}{Degueule, T.}, \bibinfo{author}{Combemale, B.},
  \bibinfo{author}{Blouin, A.}, \bibinfo{author}{Barais, O.},
  \bibinfo{author}{J\'ez\'equel, J.M.}, \bibinfo{year}{2015}.
\newblock \bibinfo{title}{{Melange: a Meta-Language for Modular and Reusable
  Development of DSLs}}, in: \bibinfo{editor}{Di~Ruscio, D.},
  \bibinfo{editor}{V\"olter, M.} (Eds.), \bibinfo{booktitle}{{Proceedings of
  the 8th International Conference on Software Language Engineering (SLE'15)}},
  \bibinfo{publisher}{ACM}, \bibinfo{address}{Pittsburgh, PA, USA}. pp.
  \bibinfo{pages}{25--36}.
\bibitem[{Del~Sole(2023)}]{DelSole23}
\bibinfo{author}{Del~Sole, A.}, \bibinfo{year}{2023}.
\newblock \bibinfo{title}{{Visual Studio Distilled}}.
\newblock \bibinfo{edition}{Third} ed., \bibinfo{publisher}{Apress}.
\bibitem[{Ekman and Hedin(2007)}]{Ekman07b}
\bibinfo{author}{Ekman, T.}, \bibinfo{author}{Hedin, G.}, \bibinfo{year}{2007}.
\newblock \bibinfo{title}{{The JastAdd System --- Modular Extensible Compiler
  Construction}}.
\newblock \bibinfo{journal}{{Science of Computer Programming}}
  \bibinfo{volume}{69}, \bibinfo{pages}{14--26}.
\bibitem[{El-Sharkawy et~al.(2011)El-Sharkawy, Kr\"oher and
  Schmid}]{ElSharkawy11}
\bibinfo{author}{El-Sharkawy, S.}, \bibinfo{author}{Kr\"oher, C.},
  \bibinfo{author}{Schmid, K.}, \bibinfo{year}{2011}.
\newblock \bibinfo{title}{{Supporting Heterogeneous Compositional Multi
  Software Product Lines}}, in: \bibinfo{editor}{Santana~de Almeida, E.},
  \bibinfo{editor}{Kishi, T.} (Eds.), \bibinfo{booktitle}{{Proceedings of the
  15th International Software Product Line Conference (SPLC'11)}},
  \bibinfo{publisher}{IEEE}, \bibinfo{address}{M\"unich, Germany}. pp.
  \bibinfo{pages}{1--4}.
\bibitem[{Erdweg et~al.(2013)Erdweg, van~der Storm, V\"olter, Boersma, Bosman,
  Cook, Gerrtsen, Hulshout, Kelly, Loh, Konat, Molina, Palatnik, Pohjonen,
  Schindler, Schindler, Solmi, Vergu and Visser}]{Erdweg13b}
\bibinfo{author}{Erdweg, S.}, \bibinfo{author}{van~der Storm, T.},
  \bibinfo{author}{V\"olter, M.}, \bibinfo{author}{Boersma, M.},
  \bibinfo{author}{Bosman, R.}, \bibinfo{author}{Cook, W.R.},
  \bibinfo{author}{Gerrtsen, A.}, \bibinfo{author}{Hulshout, A.},
  \bibinfo{author}{Kelly, S.}, \bibinfo{author}{Loh, A.},
  \bibinfo{author}{Konat, G.D.P.}, \bibinfo{author}{Molina, P.J.},
  \bibinfo{author}{Palatnik, M.}, \bibinfo{author}{Pohjonen, R.},
  \bibinfo{author}{Schindler, E.}, \bibinfo{author}{Schindler, K.},
  \bibinfo{author}{Solmi, R.}, \bibinfo{author}{Vergu, V.A.},
  \bibinfo{author}{Visser, E.}, \bibinfo{year}{2013}.
\newblock \bibinfo{title}{{The State of the Art in Language Workbenches}}, in:
  \bibinfo{editor}{Erwig, M.}, \bibinfo{editor}{Paige, R.F.},
  \bibinfo{editor}{Van~Wyk, E.} (Eds.), \bibinfo{booktitle}{{Proceedings of the
  6th International Conference on Software Language Engineering (SLE'13)}},
  \bibinfo{publisher}{Springer}, \bibinfo{address}{Indianapolis, USA}. pp.
  \bibinfo{pages}{197--217}.
\bibitem[{Favalli et~al.(2020)Favalli, K\"uhn and Cazzola}]{Cazzola20}
\bibinfo{author}{Favalli, L.}, \bibinfo{author}{K\"uhn, T.},
  \bibinfo{author}{Cazzola, W.}, \bibinfo{year}{2020}.
\newblock \bibinfo{title}{{Neverlang and FeatureIDE Just Married: Integrated
  Language Product Line Development Environment}}, in: \bibinfo{editor}{Collet,
  P.}, \bibinfo{editor}{Nadi, S.} (Eds.), \bibinfo{booktitle}{{Proceedings of
  the 24th International Software Product Line Conference (SPLC'20)}},
  \bibinfo{publisher}{ACM}, \bibinfo{address}{Montr\'eal, Canada}. pp.
  \bibinfo{pages}{285--295}.
\bibitem[{Fenwick(1994)}]{Fenwick94}
\bibinfo{author}{Fenwick, P.M.}, \bibinfo{year}{1994}.
\newblock \bibinfo{title}{{A New Data Structure for Cumulative Frequency
  Tables}}.
\newblock \bibinfo{journal}{{Software: Practice and Experience}}
  \bibinfo{volume}{24}, \bibinfo{pages}{327--336}.
\bibitem[{Fowler(2005)}]{Fowler05c}
\bibinfo{author}{Fowler, M.}, \bibinfo{year}{2005}.
\newblock \bibinfo{title}{{Inversion of Control}}.
\newblock \bibinfo{howpublished}{Martin Fowler's Blog}.
\newblock \URLprefix
  \url{https://martinfowler.com/bliki/InversionOfControl.html}.
\bibitem[{Frakes and Terry(1996)}]{Frakes96}
\bibinfo{author}{Frakes, W.}, \bibinfo{author}{Terry, C.},
  \bibinfo{year}{1996}.
\newblock \bibinfo{title}{{Software Reuse: Metrics and Models}}.
\newblock \bibinfo{journal}{{ACM Computing Surveys}} \bibinfo{volume}{28},
  \bibinfo{pages}{415--435}.
\bibitem[{Gamma et~al.(1995)Gamma, Helm, Johnson and Vlissides}]{Gamma95}
\bibinfo{author}{Gamma, E.}, \bibinfo{author}{Helm, R.},
  \bibinfo{author}{Johnson, R.}, \bibinfo{author}{Vlissides, J.},
  \bibinfo{year}{1995}.
\newblock \bibinfo{title}{{Design Patterns: Elements of Reusable
  Object-Oriented Software}}.
\newblock Professional Computing Series, \bibinfo{publisher}{Addison-Wesley},
  \bibinfo{address}{Reading, Ma, USA}.
\bibitem[{Graham et~al.(1979)Graham, Haley and Joy}]{Graham79}
\bibinfo{author}{Graham, S.L.}, \bibinfo{author}{Haley, C.B.},
  \bibinfo{author}{Joy, W.N.}, \bibinfo{year}{1979}.
\newblock \bibinfo{title}{{Practical LR Error Recovery}}, in:
  \bibinfo{editor}{Johnson, S.C.} (Ed.), \bibinfo{booktitle}{{Proceedings of
  the 1979 Sigplan Symposium on Compiler Construction (CC'79)}},
  \bibinfo{publisher}{ACM}, \bibinfo{address}{Denver, CO, USA}. pp.
  \bibinfo{pages}{168--175}.
\bibitem[{Gray(2007)}]{Gray07b}
\bibinfo{author}{Gray, II, E.J.}, \bibinfo{year}{2007}.
\newblock \bibinfo{title}{{TextMate: Power Editing for the Mac}}.
\newblock \bibinfo{publisher}{Pragmatic Bookshelf}.
\bibitem[{Griss(2000)}]{Griss00}
\bibinfo{author}{Griss, M.L.}, \bibinfo{year}{2000}.
\newblock \bibinfo{title}{{Implementing Product-Line Features with Component
  Reuse}}, in: \bibinfo{editor}{Frakes, W.B.} (Ed.),
  \bibinfo{booktitle}{{Proceedings of the 6th International Conference on
  Software Reuse (ICSR'00)}}, \bibinfo{publisher}{Springer},
  \bibinfo{address}{Vienna, Austria}. pp. \bibinfo{pages}{137--151}.
\bibitem[{Gr\"onninger et~al.(2008)Gr\"onninger, Krahn, Rumpe, Schindler and
  V\"olkel}]{Gronninger08}
\bibinfo{author}{Gr\"onninger, H.}, \bibinfo{author}{Krahn, H.},
  \bibinfo{author}{Rumpe, B.}, \bibinfo{author}{Schindler, M.},
  \bibinfo{author}{V\"olkel, S.}, \bibinfo{year}{2008}.
\newblock \bibinfo{title}{{MontiCore: A Framework for the Development of
  Textual Domain Specific Languages}}, in: \bibinfo{editor}{Sch\"afer, W.},
  \bibinfo{editor}{Dwyer, M.}, \bibinfo{editor}{Gruhn, V.} (Eds.),
  \bibinfo{booktitle}{{Companion Proceedings of the 30th International
  Conference on Software Enginering (Companion ICSE'08)}},
  \bibinfo{publisher}{IEEE}, \bibinfo{address}{Leipzig, Germany}. pp.
  \bibinfo{pages}{925--926}.
\bibitem[{Gunasinghe and Marcus(2022)}]{Gunasinghe21}
\bibinfo{author}{Gunasinghe, N.}, \bibinfo{author}{Marcus, N.},
  \bibinfo{year}{2022}.
\newblock \bibinfo{title}{{Language Server Protocol and Implementation:
  Supporting Language-Smart Editing and Programming Tools}}.
\newblock \bibinfo{publisher}{Apress}.
\bibitem[{Hartmann and Trew(2008)}]{Hartmann08}
\bibinfo{author}{Hartmann, H.}, \bibinfo{author}{Trew, T.},
  \bibinfo{year}{2008}.
\newblock \bibinfo{title}{{Using Feature Diagrams with Context Variability to
  Model Multiple Product Lines for Software Supply Chains}}, in:
  \bibinfo{editor}{Geppert, B.} (Ed.), \bibinfo{booktitle}{{Proceedings of the
  12th International Software Product Line Conference (SPLC'08)}},
  \bibinfo{publisher}{IEEE}, \bibinfo{address}{Limerick, Ireland}. pp.
  \bibinfo{pages}{12--21}.
\bibitem[{Haugen et~al.(2008)Haugen, M{\o}ller-Pedersen, Oldevik, Olsen and
  Svendsen}]{Haugen08}
\bibinfo{author}{Haugen, {\O}.}, \bibinfo{author}{M{\o}ller-Pedersen, B.},
  \bibinfo{author}{Oldevik, J.}, \bibinfo{author}{Olsen, G.K.},
  \bibinfo{author}{Svendsen, A.}, \bibinfo{year}{2008}.
\newblock \bibinfo{title}{{Adding Standardized Variability to Domain Specific
  Languages}}, in: \bibinfo{editor}{Pohl, K.}, \bibinfo{editor}{Geppert, B.}
  (Eds.), \bibinfo{booktitle}{{Proceedings of the 12th International Software
  Product Line Conference (SPLC'08)}}, \bibinfo{publisher}{IEEE},
  \bibinfo{address}{Limerick, Ireland}. pp. \bibinfo{pages}{139--148}.
\bibitem[{Heidenreich et~al.(2011)Heidenreich, Johannes, Karol, Seifert and
  Wende}]{Heidenreich11}
\bibinfo{author}{Heidenreich, F.}, \bibinfo{author}{Johannes, J.},
  \bibinfo{author}{Karol, S.}, \bibinfo{author}{Seifert, M.},
  \bibinfo{author}{Wende, C.}, \bibinfo{year}{2011}.
\newblock \bibinfo{title}{{Model-Based Language Engineering with EMFText}}, in:
  \bibinfo{editor}{L\"{a}mmel, R.}, \bibinfo{editor}{Visser, J.},
  \bibinfo{editor}{Saraiva, J.} (Eds.), \bibinfo{booktitle}{{Proceedings of the
  International Summer School on Generative and Transformational Techniques in
  Software Engineering (GTTSE'11)}}, \bibinfo{publisher}{Springer},
  \bibinfo{address}{Braga, Portugal}. pp. \bibinfo{pages}{322--345}.
\bibitem[{Hindley(1969)}]{Hindley69}
\bibinfo{author}{Hindley, R.}, \bibinfo{year}{1969}.
\newblock \bibinfo{title}{{The Principal Type-Scheme of an Object in
  Combinatory Logic}}.
\newblock \bibinfo{journal}{{Transactions of the America Mathematical Society}}
  \bibinfo{volume}{146}, \bibinfo{pages}{29--60}.
\bibitem[{Holl et~al.(2012)Holl, Gr\"unbacher and Rabiser}]{Holl12}
\bibinfo{author}{Holl, G.}, \bibinfo{author}{Gr\"unbacher, P.},
  \bibinfo{author}{Rabiser, R.}, \bibinfo{year}{2012}.
\newblock \bibinfo{title}{{A Systematic Review and an Expert Survey on
  Capabilities Supporting Multi-Product Lines}}.
\newblock \bibinfo{journal}{{Information and Software Technology}}
  \bibinfo{volume}{54}, \bibinfo{pages}{828--852}.
\bibitem[{Hotz et~al.(2006)Hotz, Wolter, Krebs, Deelstra, Sinnema, Nijhuis and
  MacGregor}]{Hotz06}
\bibinfo{author}{Hotz, L.}, \bibinfo{author}{Wolter, K.},
  \bibinfo{author}{Krebs, T.}, \bibinfo{author}{Deelstra, S.},
  \bibinfo{author}{Sinnema, M.}, \bibinfo{author}{Nijhuis, G.J.},
  \bibinfo{author}{MacGregor, J.}, \bibinfo{year}{2006}.
\newblock \bibinfo{title}{{Configuration in Industrial Product Families: The
  ConIPF Methodology}}.
\newblock \bibinfo{publisher}{IOS Press}.
\bibitem[{Hu et~al.(2016)Hu, Shinde, Adrian, Chua, Saxena and Liang}]{Hu16}
\bibinfo{author}{Hu, H.}, \bibinfo{author}{Shinde, S.},
  \bibinfo{author}{Adrian, S.}, \bibinfo{author}{Chua, Z.L.},
  \bibinfo{author}{Saxena, P.}, \bibinfo{author}{Liang, Z.},
  \bibinfo{year}{2016}.
\newblock \bibinfo{title}{{Data-Oriented Programming: On the Expressiveness of
  Non-control Data Attacks}}, in: \bibinfo{editor}{Shmatikov, V.},
  \bibinfo{editor}{Erlingsson, U.} (Eds.), \bibinfo{booktitle}{{Proceedings of
  the 2016 IEEE Symposium on Security and Privacy (SP'16)}},
  \bibinfo{publisher}{IEEE}, \bibinfo{address}{San Jos\`e, CA, USA}. pp.
  \bibinfo{pages}{969--986}.
\bibitem[{Hudak(1998)}]{Hudak98}
\bibinfo{author}{Hudak, P.}, \bibinfo{year}{1998}.
\newblock \bibinfo{title}{{Modular Domain Specific Languages and Tools}}, in:
  \bibinfo{editor}{Devanbu, P.}, \bibinfo{editor}{Poulin, J.} (Eds.),
  \bibinfo{booktitle}{{Proceedings of the 5th International Conference on
  Software Reuse (ICSR'98)}}, \bibinfo{publisher}{IEEE},
  \bibinfo{address}{Victoria, BC, Canada}. pp. \bibinfo{pages}{134--142}.
\bibitem[{Hughes(1989)}]{Hughes89}
\bibinfo{author}{Hughes, J.}, \bibinfo{year}{1989}.
\newblock \bibinfo{title}{{Why Functional Programming Matters}}.
\newblock \bibinfo{journal}{{The Computer Journal}} \bibinfo{volume}{32},
  \bibinfo{pages}{98--107}.
\bibitem[{H\"ursch and Videira~Lopes(1995)}]{Videira-Lopes95}
\bibinfo{author}{H\"ursch, W.}, \bibinfo{author}{Videira~Lopes, C.},
  \bibinfo{year}{1995}.
\newblock \bibinfo{title}{{Separation of Concerns}}.
\newblock \bibinfo{type}{Technical Report} \bibinfo{number}{NU-CCS-95-03}.
  Northeastern University, Boston.
\bibitem[{Jordan and Zib(2024)}]{Jordan24}
\bibinfo{author}{Jordan, T.}, \bibinfo{author}{Zib, S.}, \bibinfo{year}{2024}.
\newblock \bibinfo{title}{{A Langium-Based Approach to BigER}}.
\newblock \bibinfo{type}{Bachelor's thesis}. Technische Universit\"at Wien.
  \bibinfo{address}{Wien, Austria}.
\bibitem[{Kang et~al.(1990)Kang, Cohen, Hess, Novak and Peterson}]{Kang90}
\bibinfo{author}{Kang, K.C.}, \bibinfo{author}{Cohen, S.G.},
  \bibinfo{author}{Hess, J.A.}, \bibinfo{author}{Novak, W.E.},
  \bibinfo{author}{Peterson, A.S.}, \bibinfo{year}{1990}.
\newblock \bibinfo{title}{{Feature-Oriented Domain Analysis (FODA) Feasibility
  Study}}.
\newblock \bibinfo{type}{Technical Report} \bibinfo{number}{CMU/SEI-90-TR-21}.
  {Carnegie Mellon University}. \bibinfo{address}{Pittsburgh, Pennsylvania,
  USA}.
\bibitem[{K\"astner et~al.(2006)K\"astner, Apel and Ostermann}]{Kastner11}
\bibinfo{author}{K\"astner, C.}, \bibinfo{author}{Apel, S.},
  \bibinfo{author}{Ostermann, K.}, \bibinfo{year}{2006}.
\newblock \bibinfo{title}{{The Road to Feature Modularity?}}, in:
  \bibinfo{editor}{Schäfer, I.}, \bibinfo{editor}{John, I.},
  \bibinfo{editor}{Schmid, K.} (Eds.), \bibinfo{booktitle}{{Proceedings of the
  3rd Workshop on Feature-Oriented Software Development (FOSD'11)}},
  \bibinfo{publisher}{ACM}, \bibinfo{address}{M\"unich, Germany}.
\bibitem[{Kats and Visser(2010)}]{Visser10}
\bibinfo{author}{Kats, L.C.L.}, \bibinfo{author}{Visser, E.},
  \bibinfo{year}{2010}.
\newblock \bibinfo{title}{{The Spoofax Language Workbench: Rules for
  Declarative Specification of Languages and IDEs}}, in:
  \bibinfo{editor}{Rinard, M.}, \bibinfo{editor}{Sullivan, K.J.},
  \bibinfo{editor}{Steinberg, D.H.} (Eds.), \bibinfo{booktitle}{{Proceedings of
  the ACM International Conference on Object Oriented Programming Systems
  Languages and Applications (OOPSLA'10)}}, \bibinfo{publisher}{ACM},
  \bibinfo{address}{Reno, Nevada, USA}. pp. \bibinfo{pages}{444--463}.
\bibitem[{Kats et~al.(2010)Kats, Visser and Wachsmuth}]{Kats10}
\bibinfo{author}{Kats, L.C.L.}, \bibinfo{author}{Visser, E.},
  \bibinfo{author}{Wachsmuth, G.}, \bibinfo{year}{2010}.
\newblock \bibinfo{title}{{Pure and Declarative Syntax Definition: Paradise
  Lost and Regained}}, in: \bibinfo{booktitle}{{Proceedings of ACM Conference
  on New Ideas in Programming and Reflections on Software (Onward! 2010)}},
  \bibinfo{organization}{ACM}, \bibinfo{address}{Reno-Tahoe, Nevada, USA}.
\bibitem[{Kiczales et~al.(2001)Kiczales, Hilsdale, Hugunin, Kersten, Palm and
  Griswold}]{Kiczales01}
\bibinfo{author}{Kiczales, G.}, \bibinfo{author}{Hilsdale, E.},
  \bibinfo{author}{Hugunin, J.}, \bibinfo{author}{Kersten, M.},
  \bibinfo{author}{Palm, J.}, \bibinfo{author}{Griswold, B.},
  \bibinfo{year}{2001}.
\newblock \bibinfo{title}{{An Overview of AspectJ}}, in:
  \bibinfo{editor}{Knudsen, J.L.} (Ed.), \bibinfo{booktitle}{Proceedings of the
  15th European Conference on Object-Oriented Programming (ECOOP'01)},
  \bibinfo{publisher}{Springer-Verlag}, \bibinfo{address}{Budapest, Hungary}.
  pp. \bibinfo{pages}{327--353}.
\bibitem[{Kiczales et~al.(1997)Kiczales, Lamping, Mendhekar, Maeda,
  Videira~Lopes, Loingtier and Irwin}]{Kiczales97}
\bibinfo{author}{Kiczales, G.}, \bibinfo{author}{Lamping, J.},
  \bibinfo{author}{Mendhekar, A.}, \bibinfo{author}{Maeda, C.},
  \bibinfo{author}{Videira~Lopes, C.}, \bibinfo{author}{Loingtier, J.M.},
  \bibinfo{author}{Irwin, J.}, \bibinfo{year}{1997}.
\newblock \bibinfo{title}{{Aspect-Oriented Programming}}, in:
  \bibinfo{editor}{Ak\c{s}it, M.}, \bibinfo{editor}{Matsuoka, S.} (Eds.),
  \bibinfo{booktitle}{11th European Conference on Object Oriented Programming
  (ECOOP'97)}, \bibinfo{publisher}{Springer-Verlag},
  \bibinfo{address}{Helsinki, Finland}. pp. \bibinfo{pages}{220--242}.
\bibitem[{Klint et~al.(2009)Klint, van~der Storm and Vinju}]{Klint09}
\bibinfo{author}{Klint, P.}, \bibinfo{author}{van~der Storm, T.},
  \bibinfo{author}{Vinju, J.}, \bibinfo{year}{2009}.
\newblock \bibinfo{title}{{EASY Meta-Programming with Rascal}}, in:
  \bibinfo{editor}{Fernandes, J.M.}, \bibinfo{editor}{L\"ammel, R.},
  \bibinfo{editor}{Visser, J.}, \bibinfo{editor}{Saraiva, J.} (Eds.),
  \bibinfo{booktitle}{{Proceedings of the International Summer School on
  Generative and Transformational Techniques in Software Engineering III
  (GTTSE'09)}}, \bibinfo{publisher}{Springer}, \bibinfo{address}{Braga,
  Portugal}. pp. \bibinfo{pages}{222--289}.
\bibitem[{Knuth(1997)}]{Knuth97}
\bibinfo{author}{Knuth, D.E.}, \bibinfo{year}{1997}.
\newblock \bibinfo{title}{{The Art of Computer Programming: Fundamental
  Algorithms}}.
\newblock \bibinfo{edition}{third} ed., \bibinfo{publisher}{Addison Wesley}.
\bibitem[{Kosar et~al.(2016)Kosar, Bohra and Mernik}]{Kosar16}
\bibinfo{author}{Kosar, T.}, \bibinfo{author}{Bohra, S.},
  \bibinfo{author}{Mernik, M.}, \bibinfo{year}{2016}.
\newblock \bibinfo{title}{{Domain Specific Languages: A Systematic Mapping
  Study}}.
\newblock \bibinfo{journal}{{Information and Software Technology}}
  \bibinfo{volume}{71}, \bibinfo{pages}{77--91}.
\bibitem[{Koscielny et~al.(2014)Koscielny, Holthusen, Schaefer, Schulze,
  Bettini and Ferruccio}]{Koscielny14}
\bibinfo{author}{Koscielny, J.}, \bibinfo{author}{Holthusen, S.},
  \bibinfo{author}{Schaefer, I.}, \bibinfo{author}{Schulze, S.},
  \bibinfo{author}{Bettini, L.}, \bibinfo{author}{Ferruccio, D.},
  \bibinfo{year}{2014}.
\newblock \bibinfo{title}{{DeltaJ 1.5: Delta-Oriented Programming for Java
  1.5}}, in: \bibinfo{editor}{Childers, B.} (Ed.),
  \bibinfo{booktitle}{{Proceedings of the 2014 International Conference on
  Principles and Practices of Programming on the Java platform: Virtual
  machines, Languages, and Tools (PPPJ'14)}}, \bibinfo{publisher}{ACM},
  \bibinfo{address}{Cracow, Poland}. pp. \bibinfo{pages}{63--74}.
\bibitem[{Krahn et~al.(2007)Krahn, Rumpe and V\"olkel}]{Krahn07}
\bibinfo{author}{Krahn, H.}, \bibinfo{author}{Rumpe, B.},
  \bibinfo{author}{V\"olkel, S.}, \bibinfo{year}{2007}.
\newblock \bibinfo{title}{{Efficient Editor Generation for Compositional DSLs
  in Eclipse}}, in: \bibinfo{editor}{Tolvanen, J.P.}, \bibinfo{editor}{Gray,
  J.}, \bibinfo{editor}{Rossi, M.}, \bibinfo{editor}{Sprinkle, J.} (Eds.),
  \bibinfo{booktitle}{{Proceedings of the 7th OOPSLA Workshop on
  Domain-Specific Modeling (DSM'07)}}, \bibinfo{address}{Montr\'eal, Canada}.
\bibitem[{Krahn et~al.(2010)Krahn, Rumpe and V{\"o}lkel}]{Krahn10}
\bibinfo{author}{Krahn, H.}, \bibinfo{author}{Rumpe, B.},
  \bibinfo{author}{V{\"o}lkel, S.}, \bibinfo{year}{2010}.
\newblock \bibinfo{title}{{MontiCore: A Framework for Compositional Development
  of Domain Specific Languages}}.
\newblock \bibinfo{journal}{{International Journal on Software Tools for
  Technology Transfer}} \bibinfo{volume}{12}, \bibinfo{pages}{353--372}.
\bibitem[{Krueger(1992)}]{Krueger92}
\bibinfo{author}{Krueger, C.W.}, \bibinfo{year}{1992}.
\newblock \bibinfo{title}{{Software Reuse}}.
\newblock \bibinfo{journal}{{ACM Computing Surveys}} \bibinfo{volume}{24},
  \bibinfo{pages}{131--183}.
\bibitem[{Krueger(2006)}]{Krueger06}
\bibinfo{author}{Krueger, C.W.}, \bibinfo{year}{2006}.
\newblock \bibinfo{title}{{New Methods in Software Product Line Practice}}.
\newblock \bibinfo{journal}{Communications of the ACM} \bibinfo{volume}{49},
  \bibinfo{pages}{37--40}.
\bibitem[{K\"uhn and Cazzola(2016)}]{Cazzola16i}
\bibinfo{author}{K\"uhn, T.}, \bibinfo{author}{Cazzola, W.},
  \bibinfo{year}{2016}.
\newblock \bibinfo{title}{{Apples and Oranges: Comparing Top-Down and Bottom-Up
  Language Product Lines}}, in: \bibinfo{editor}{Rabiser, R.},
  \bibinfo{editor}{Xie, B.} (Eds.), \bibinfo{booktitle}{{Proceedings of the
  20th International Software Product Line Conference (SPLC'16)}},
  \bibinfo{publisher}{ACM}, \bibinfo{address}{Beijing, China}. pp.
  \bibinfo{pages}{50--59}.
\bibitem[{K\"uhn et~al.(2015)K\"uhn, Cazzola and Olivares}]{Cazzola15f}
\bibinfo{author}{K\"uhn, T.}, \bibinfo{author}{Cazzola, W.},
  \bibinfo{author}{Olivares, D.M.}, \bibinfo{year}{2015}.
\newblock \bibinfo{title}{{Choosy and Picky: Configuration of Language Product
  Lines}}, in: \bibinfo{editor}{Botterweck, G.}, \bibinfo{editor}{White, J.}
  (Eds.), \bibinfo{booktitle}{{Proceedings of the 19th International Software
  Product Line Conference (SPLC'15)}}, \bibinfo{publisher}{ACM},
  \bibinfo{address}{Nashville, TN, USA}. pp. \bibinfo{pages}{71--80}.
\bibitem[{K\"uhn et~al.(2019)K\"uhn, Cazzola, Pirritano~Giampietro and
  Poggi}]{Cazzola19}
\bibinfo{author}{K\"uhn, T.}, \bibinfo{author}{Cazzola, W.},
  \bibinfo{author}{Pirritano~Giampietro, N.}, \bibinfo{author}{Poggi, M.},
  \bibinfo{year}{2019}.
\newblock \bibinfo{title}{{Piggyback IDE Support for Language Product Lines}},
  in: \bibinfo{editor}{Th\"um, T.}, \bibinfo{editor}{Duchien, L.} (Eds.),
  \bibinfo{booktitle}{{Proceedings of the 23rd International Software Product
  Line Conference (SPLC'19)}}, \bibinfo{publisher}{ACM},
  \bibinfo{address}{Paris, France}. pp. \bibinfo{pages}{131--142}.
\bibitem[{K\"uhn et~al.(2014)K\"uhn, Leuth\"auser, G\"otz, Seidl and
  A{\ss}mann}]{Kuehn14}
\bibinfo{author}{K\"uhn, T.}, \bibinfo{author}{Leuth\"auser, M.},
  \bibinfo{author}{G\"otz, S.}, \bibinfo{author}{Seidl, C.},
  \bibinfo{author}{A{\ss}mann, U.}, \bibinfo{year}{2014}.
\newblock \bibinfo{title}{{A Metamodel Family for Role-Based Modeling and
  Programming Languages}}, in: \bibinfo{editor}{Combemale, B.},
  \bibinfo{editor}{Pearce, D.J.}, \bibinfo{editor}{Barais, O.},
  \bibinfo{editor}{Vinju, J.} (Eds.), \bibinfo{booktitle}{{Proceedings of the
  7th International Conference Software Language Engineering (SLE'14)}},
  \bibinfo{publisher}{Springer}, \bibinfo{address}{V\"aster\r{a}s, Sweden}. pp.
  \bibinfo{pages}{141--160}.
\bibitem[{Kumar et~al.(2016)Kumar, Kumar and Iyyappan}]{Kumar16}
\bibinfo{author}{Kumar, A.}, \bibinfo{author}{Kumar, A.},
  \bibinfo{author}{Iyyappan, M.}, \bibinfo{year}{2016}.
\newblock \bibinfo{title}{{Applying Separation of Concern for Developing
  Softwares Using Aspect Oriented Programming Concepts}}.
\newblock \bibinfo{journal}{{Procedia Computer Science}} \bibinfo{volume}{85},
  \bibinfo{pages}{906--914}.
\bibitem[{Laddad(2003)}]{Laddad03}
\bibinfo{author}{Laddad, R.}, \bibinfo{year}{2003}.
\newblock \bibinfo{title}{{AspectJ in Action: Pratical Aspect-Oriented
  Programming}}.
\newblock \bibinfo{publisher}{Manning Pubblications Company}.
\bibitem[{de~Lara and Guerra(2021)}]{deLara21}
\bibinfo{author}{de~Lara, J.}, \bibinfo{author}{Guerra, E.},
  \bibinfo{year}{2021}.
\newblock \bibinfo{title}{{Language Family Engineering With Product Lines of
  Multi-Level Models}}.
\newblock \bibinfo{journal}{{Formal Aspects of Computing}}
  \bibinfo{volume}{33}, \bibinfo{pages}{1173--1208}.
\bibitem[{de~Lara et~al.(2022)de~Lara, Guerra and Bottoni}]{deLara22}
\bibinfo{author}{de~Lara, J.}, \bibinfo{author}{Guerra, E.},
  \bibinfo{author}{Bottoni, P.}, \bibinfo{year}{2022}.
\newblock \bibinfo{title}{{Modular Language Product Lines: A Graph
  Transformation Approach}}, in: \bibinfo{editor}{Bencomo, N.},
  \bibinfo{editor}{Wimmer, M.} (Eds.), \bibinfo{booktitle}{{Proceedings of the
  25th International Conference on Model Driven Engineering Languages and
  Systems (MoDELS'22)}}, \bibinfo{publisher}{ACM},
  \bibinfo{address}{Montr\'eal, Canada}. pp. \bibinfo{pages}{334--344}.
\bibitem[{Liebig et~al.(2013)Liebig, Daniel and Apel}]{Liebig13}
\bibinfo{author}{Liebig, J.}, \bibinfo{author}{Daniel, R.},
  \bibinfo{author}{Apel, S.}, \bibinfo{year}{2013}.
\newblock \bibinfo{title}{{Feature-Oriented Language Families: A Case Study}},
  in: \bibinfo{editor}{Collet, P.}, \bibinfo{editor}{Schmid, K.} (Eds.),
  \bibinfo{booktitle}{{Proceedings of the 7th International Workshop on
  Variability Modelling of Software-intensive Systems (VaMoS'13)}},
  \bibinfo{publisher}{ACM}, \bibinfo{address}{Pisa, Italy}.
\bibitem[{\sortnoop{Linden}van~der Linden et~al.(2007)\sortnoop{Linden}van~der
  Linden, Schmid and Rommes}]{VanDerLinden07}
\bibinfo{author}{\sortnoop{Linden}van~der Linden, F.}, \bibinfo{author}{Schmid,
  K.}, \bibinfo{author}{Rommes, E.}, \bibinfo{year}{2007}.
\newblock \bibinfo{title}{{Software Product Lines in Action: The Best
  Industrial Practice in Product Line Engineering}}.
\newblock \bibinfo{publisher}{Springer}.
\bibitem[{M\"akitalo et~al.(2020)M\"akitalo, Taivalsaari, Kiviluoto, Mikkonen
  and Capilla}]{Makitalo20}
\bibinfo{author}{M\"akitalo, N.}, \bibinfo{author}{Taivalsaari, A.},
  \bibinfo{author}{Kiviluoto, A.}, \bibinfo{author}{Mikkonen, T.},
  \bibinfo{author}{Capilla, R.}, \bibinfo{year}{2020}.
\newblock \bibinfo{title}{{On Opportunistic Software Reuse}}.
\newblock \bibinfo{journal}{{Computing}} \bibinfo{volume}{102},
  \bibinfo{pages}{2385--2408}.
\bibitem[{M\'endez-Acu\~na et~al.(2016)M\'endez-Acu\~na, Galindo, Combemale,
  Blouin and Baudry}]{Mendez-Acuna16b}
\bibinfo{author}{M\'endez-Acu\~na, D.}, \bibinfo{author}{Galindo, J.A.},
  \bibinfo{author}{Combemale, B.}, \bibinfo{author}{Blouin, A.},
  \bibinfo{author}{Baudry, B.}, \bibinfo{year}{2016}.
\newblock \bibinfo{title}{{Puzzle: A Tool for Analyzing and Extracting
  Specification Clones in DSLs}}, in: \bibinfo{editor}{Kapitsaki, G.M.},
  \bibinfo{editor}{Santana~de Almeida, E.} (Eds.),
  \bibinfo{booktitle}{{Proceedings of the International Conference on Software
  Reuse (ICSR'16)}}, \bibinfo{publisher}{Springer}, \bibinfo{address}{Limassol,
  Cyprus}. pp. \bibinfo{pages}{393--396}.
\bibitem[{M\'endez-Acu{\~n}a et~al.(2017)M\'endez-Acu{\~n}a, Galindo,
  Combemale, Blouin and Baudry}]{Mendez-Acuna17}
\bibinfo{author}{M\'endez-Acu{\~n}a, D.}, \bibinfo{author}{Galindo, J.A.},
  \bibinfo{author}{Combemale, B.}, \bibinfo{author}{Blouin, A.},
  \bibinfo{author}{Baudry, B.}, \bibinfo{year}{2017}.
\newblock \bibinfo{title}{{Reverse Engineering Language Product Lines from
  Existing DSL Variants}}.
\newblock \bibinfo{journal}{{Journal of Systems and Software}}
  \bibinfo{volume}{133}, \bibinfo{pages}{145--158}.
\bibitem[{M\'endez-Acu{\~n}a et~al.(2016)M\'endez-Acu{\~n}a, Galindo, Degueule,
  Combemale and Baudry}]{Mendez-Acuna16}
\bibinfo{author}{M\'endez-Acu{\~n}a, D.}, \bibinfo{author}{Galindo, J.A.},
  \bibinfo{author}{Degueule, T.}, \bibinfo{author}{Combemale, B.},
  \bibinfo{author}{Baudry, B.}, \bibinfo{year}{2016}.
\newblock \bibinfo{title}{{Leveraging Software Product Lines Engineering in the
  Development of External DSLs: A Systematic Literature Review}}.
\newblock \bibinfo{journal}{{Computer Languages, Systems \& Structures}}
  \bibinfo{volume}{46}, \bibinfo{pages}{206--235}.
\bibitem[{Mernik et~al.(2005)Mernik, Heering and Sloane}]{Mernik05}
\bibinfo{author}{Mernik, M.}, \bibinfo{author}{Heering, J.},
  \bibinfo{author}{Sloane, A.M.}, \bibinfo{year}{2005}.
\newblock \bibinfo{title}{{When and How to Develop Domain Specific Languages}}.
\newblock \bibinfo{journal}{ACM Computing Surveys} \bibinfo{volume}{37},
  \bibinfo{pages}{316--344}.
\bibitem[{Metzger and Pohl(2014)}]{Metzger14}
\bibinfo{author}{Metzger, A.}, \bibinfo{author}{Pohl, K.},
  \bibinfo{year}{2014}.
\newblock \bibinfo{title}{{Software Product Line Engineering and Variability
  Management: Achievements and Challenges}}, in: \bibinfo{editor}{Dwyer, M.B.},
  \bibinfo{editor}{Herbsleb, J.} (Eds.), \bibinfo{booktitle}{{Proceedings of
  Future of Software Engineering (FoSE'14)}}, \bibinfo{publisher}{ACM},
  \bibinfo{address}{Hyderabad, India}. pp. \bibinfo{pages}{70--84}.
\bibitem[{Milner(1978)}]{Milner78}
\bibinfo{author}{Milner, R.}, \bibinfo{year}{1978}.
\newblock \bibinfo{title}{{A Theory of Type Polymorphism in Programming}}.
\newblock \bibinfo{journal}{{Journal of Computer and System Sciences}}
  \bibinfo{volume}{17}, \bibinfo{pages}{348--375}.
\bibitem[{Mosses(2004)}]{Mosses04}
\bibinfo{author}{Mosses, P.D.}, \bibinfo{year}{2004}.
\newblock \bibinfo{title}{{Modular Structural Operational Semantics}}.
\newblock \bibinfo{journal}{{The Journal of Logic and Algebraic Programming}}
  \bibinfo{volume}{60-61}, \bibinfo{pages}{195--228}.
\bibitem[{Mosses(2019)}]{Mosses19b}
\bibinfo{author}{Mosses, P.D.}, \bibinfo{year}{2019}.
\newblock \bibinfo{title}{{A Component-Based Formal Language Workbench}}, in:
  \bibinfo{editor}{Monahan, R.}, \bibinfo{editor}{Prevosto, V.},
  \bibinfo{editor}{Proen\c{c}a, J.} (Eds.), \bibinfo{booktitle}{{Proceedings of
  the 5th Workshop on Formal Integrated Development Environment (F-IDE'19)}},
  \bibinfo{address}{Porto, Portugal}. pp. \bibinfo{pages}{29--34}.
\bibitem[{Mosses and Vesely(2014)}]{Mosses14}
\bibinfo{author}{Mosses, P.D.}, \bibinfo{author}{Vesely, F.},
  \bibinfo{year}{2014}.
\newblock \bibinfo{title}{{FunKons: Component-Based Semantics in K}}, in:
  \bibinfo{editor}{Escobar, S.} (Ed.), \bibinfo{booktitle}{{Proceedings of the
  10th International Workshop on Rewriting Logic and its Applications
  (WRLA'14)}}, \bibinfo{publisher}{Springer}, \bibinfo{address}{Grenoble,
  France}. pp. \bibinfo{pages}{213--229}.
\bibitem[{Murphy et~al.(2006)Murphy, Kersten and Findlater}]{Murphy06}
\bibinfo{author}{Murphy, G.C.}, \bibinfo{author}{Kersten, M.},
  \bibinfo{author}{Findlater, L.}, \bibinfo{year}{2006}.
\newblock \bibinfo{title}{{How Are Java Software Developers Using the Eclipse
  IDE?}}
\newblock \bibinfo{journal}{{IEEE Software}} \bibinfo{volume}{23},
  \bibinfo{pages}{76--83}.
\bibitem[{Ng et~al.(2011)Ng, Warren, Golde and Hejlberg}]{Ng11}
\bibinfo{author}{Ng, K.}, \bibinfo{author}{Warren, M.}, \bibinfo{author}{Golde,
  P.}, \bibinfo{author}{Hejlberg, A.}, \bibinfo{year}{2011}.
\newblock \bibinfo{title}{{The Roslyn Project: Exposing the C\# and VB
  Compiler's Code Analysis}}.
\newblock \bibinfo{type}{White Paper}. Microsoft.
\bibitem[{Niephaus et~al.(2020)Niephaus, Rein, Edding, Hering, K\"onig, Opahle,
  Scordialo and Hirschfeld}]{Niephaus20}
\bibinfo{author}{Niephaus, F.}, \bibinfo{author}{Rein, P.},
  \bibinfo{author}{Edding, J.}, \bibinfo{author}{Hering, J.},
  \bibinfo{author}{K\"onig, B.}, \bibinfo{author}{Opahle, K.},
  \bibinfo{author}{Scordialo, N.}, \bibinfo{author}{Hirschfeld, R.},
  \bibinfo{year}{2020}.
\newblock \bibinfo{title}{{Example-Based Live Programming for Everyone:
  Building Language-Agnostic Tools for Live Programming with LSP and GraalVM}},
  in: \bibinfo{booktitle}{{Proceedings of the ACM International Symposium on
  New Ideas, New Paradigms and Reflection on Programming and Software
  (Onward!'20)}}, \bibinfo{publisher}{ACM}, \bibinfo{address}{Virtual, USA}.
  pp. \bibinfo{pages}{1--17}.
\bibitem[{\sortnoop{Ommering}van Ommering(2001)}]{VanOmmering01}
\bibinfo{author}{\sortnoop{Ommering}van Ommering, R.}, \bibinfo{year}{2001}.
\newblock \bibinfo{title}{{Configuration Management in Compoent Based Product
  Populations}}, in: \bibinfo{editor}{Westfechtel, B.}, \bibinfo{editor}{Hoek,
  A.} (Eds.), \bibinfo{booktitle}{{Proceedings of the Internationa Workshop on
  Software Configuration Management (SCM'01)}}, \bibinfo{publisher}{Springer},
  \bibinfo{address}{Toronto, Canada}. pp. \bibinfo{pages}{16--23}.
\bibitem[{Pacak et~al.(2020)Pacak, Erdweg and Szab\'o}]{Pacak20}
\bibinfo{author}{Pacak, A.}, \bibinfo{author}{Erdweg, S.},
  \bibinfo{author}{Szab\'o, T.}, \bibinfo{year}{2020}.
\newblock \bibinfo{title}{{A Systematic Approach to Deriving Incremental Type
  Checkers}}, in: \bibinfo{editor}{Grove, D.} (Ed.),
  \bibinfo{booktitle}{{Proceedings of the 35th Conference on Object-Oriented
  Programming, Systems, Languages, and Applications (OOPSLA'20)}},
  \bibinfo{publisher}{ACM}, \bibinfo{address}{Chicago, IL, USA}. pp.
  \bibinfo{pages}{1--28}.
\bibitem[{Parr(2009)}]{Parr09}
\bibinfo{author}{Parr, T.}, \bibinfo{year}{2009}.
\newblock \bibinfo{title}{{Language Implementation Patterns: Create Your Own
  Domain-Specific and General Programming Languages}}.
\newblock \bibinfo{publisher}{Pragmatic Bookshelf}.
\bibitem[{Pierce(2002)}]{Pierce02}
\bibinfo{author}{Pierce, B.C.}, \bibinfo{year}{2002}.
\newblock \bibinfo{title}{{Types and Programming Languages}}.
\newblock \bibinfo{publisher}{MIT Press}.
\bibitem[{Pohl et~al.(2005)Pohl, B\"ockle and van~der Linden}]{Pohl05}
\bibinfo{author}{Pohl, K.}, \bibinfo{author}{B\"ockle, K.},
  \bibinfo{author}{van~der Linden, F.J.}, \bibinfo{year}{2005}.
\newblock \bibinfo{title}{{Software Product Line Engineering: Foundations,
  Principles and Techniques}}.
\newblock \bibinfo{publisher}{Springer}.
\bibitem[{Prehofer(1997)}]{Prehofer97}
\bibinfo{author}{Prehofer, C.}, \bibinfo{year}{1997}.
\newblock \bibinfo{title}{{Feature-Oriented Programming: A Fresh Look at
  Objects}}, in: \bibinfo{editor}{Ak\c{s}it, M.}, \bibinfo{editor}{Matsuoka,
  S.} (Eds.), \bibinfo{booktitle}{{Proceedings of the 11th European Conference
  on Object-Oriented Programming (ECOOP'97)}}, \bibinfo{publisher}{Springer},
  \bibinfo{address}{Helsinki, Finland}. pp. \bibinfo{pages}{419--443}.
\bibitem[{Prehofer(2001)}]{Prehofer01}
\bibinfo{author}{Prehofer, C.}, \bibinfo{year}{2001}.
\newblock \bibinfo{title}{{Feature-Oriented Programming: A New Way of Object
  Composition}}.
\newblock \bibinfo{journal}{{Concurency and Computation: Practice and
  Experience}} \bibinfo{volume}{13}, \bibinfo{pages}{465--501}.
\bibitem[{Rabiser et~al.(2010)Rabiser, Gr\"{u}nbacher and Dhungana}]{Rabiser10}
\bibinfo{author}{Rabiser, R.}, \bibinfo{author}{Gr\"{u}nbacher, P.},
  \bibinfo{author}{Dhungana, D.}, \bibinfo{year}{2010}.
\newblock \bibinfo{title}{{Requirements for Product Derivation Support: Results
  from a Systematic Literature Review and an Expert Survey}}.
\newblock \bibinfo{journal}{{Journal of Information and Software Technology}}
  \bibinfo{volume}{52}.
\bibitem[{Rabiser et~al.(2011)Rabiser, O'Leary and Richardsonl}]{Rabiser11}
\bibinfo{author}{Rabiser, R.}, \bibinfo{author}{O'Leary, P.},
  \bibinfo{author}{Richardsonl, I.}, \bibinfo{year}{2011}.
\newblock \bibinfo{title}{{Key Activities for Product Derivation in Software
  Product Lines}}.
\newblock \bibinfo{journal}{{Journal of Sustems and Software}}
  \bibinfo{volume}{84}, \bibinfo{pages}{285--300}.
\bibitem[{Rask et~al.(2021)Rask, Madsen, Battle, Macedo and Larsen}]{Rask21}
\bibinfo{author}{Rask, J.K.}, \bibinfo{author}{Madsen, F.P.},
  \bibinfo{author}{Battle, N.}, \bibinfo{author}{Macedo, H.D.},
  \bibinfo{author}{Larsen, P.G.}, \bibinfo{year}{2021}.
\newblock \bibinfo{title}{{The Specification Language Server Protocol: A
  Proposal for Standardised LSP Extensions}}, in: \bibinfo{editor}{Paskevich,
  A.}, \bibinfo{editor}{Proen\c{c}a, J.} (Eds.),
  \bibinfo{booktitle}{{Proceedings of the 6th Workshop on Formal Integrated
  Development Environment (F-IDE'21)}}, \bibinfo{publisher}{Elsevier},
  \bibinfo{address}{Held On-Line}. pp. \bibinfo{pages}{3--18}.
\bibitem[{Rentsch(1982)}]{Rentsch82}
\bibinfo{author}{Rentsch, T.}, \bibinfo{year}{1982}.
\newblock \bibinfo{title}{{Object Oriented Programming}}.
\newblock \bibinfo{journal}{{Sigplan Notices}} \bibinfo{volume}{17},
  \bibinfo{pages}{51--57}.
\bibitem[{Richter(1985)}]{Richter85}
\bibinfo{author}{Richter, H.}, \bibinfo{year}{1985}.
\newblock \bibinfo{title}{{Noncorrecting Syntax Error Recovery}}.
\newblock \bibinfo{journal}{ACM Transactions on Programming Languages and
  Systems} \bibinfo{volume}{7}, \bibinfo{pages}{478--489}.
\bibitem[{Robinson(1965)}]{Robinson65}
\bibinfo{author}{Robinson, J.A.}, \bibinfo{year}{1965}.
\newblock \bibinfo{title}{{A Machine-Oriented Logic Based on the Resolution
  Principle}}.
\newblock \bibinfo{journal}{J.~ACM} \bibinfo{volume}{12},
  \bibinfo{pages}{23--41}.
\bibitem[{Rodriguez-Echeverr{\'\i}a et~al.(2018a)Rodriguez-Echeverr{\'\i}a,
  C\'anovas~Izquierdo, Wimmer and Cabot}]{Rodriguez-Echeverria18}
\bibinfo{author}{Rodriguez-Echeverr{\'\i}a, R.},
  \bibinfo{author}{C\'anovas~Izquierdo, J.L.}, \bibinfo{author}{Wimmer, M.},
  \bibinfo{author}{Cabot, J.}, \bibinfo{year}{2018}a.
\newblock \bibinfo{title}{{An LSP Infrastructure to Build EMF Language Servers
  for Web-Deployable Model Editors}}, in: \bibinfo{editor}{Hebig, R.},
  \bibinfo{editor}{Berger, T.} (Eds.), \bibinfo{booktitle}{{Proceedings of the
  2nd International Workshop on Model-Driven Engineering Tools
  (MDE-Tools'18)}}, \bibinfo{publisher}{CEUR}, \bibinfo{address}{Copenhage,
  Denmark}. pp. \bibinfo{pages}{1--10}.
\bibitem[{Rodriguez-Echeverr{\'\i}a et~al.(2018b)Rodriguez-Echeverr{\'\i}a,
  C\'anovas~Izquierdo, Wimmer and Cabot}]{Rodriguez-Echeverria18b}
\bibinfo{author}{Rodriguez-Echeverr{\'\i}a, R.},
  \bibinfo{author}{C\'anovas~Izquierdo, J.L.}, \bibinfo{author}{Wimmer, M.},
  \bibinfo{author}{Cabot, J.}, \bibinfo{year}{2018}b.
\newblock \bibinfo{title}{{Towards a Language Server Protocol Infrastructure
  for Graphical Modeling}}, in: \bibinfo{editor}{Paige, R.},
  \bibinfo{editor}{Haugen, {\O}.} (Eds.), \bibinfo{booktitle}{{Proceedings of
  the 21st International Conference on Model Driven Engineering Languages and
  Systems (MoDELS'18)}}, \bibinfo{publisher}{ACM},
  \bibinfo{address}{Copenhagen, Denmark}. pp. \bibinfo{pages}{370--380}.
\bibitem[{Rosenm\"uller and Siegmund(2010)}]{Rosenmuller10}
\bibinfo{author}{Rosenm\"uller, M.}, \bibinfo{author}{Siegmund, N.},
  \bibinfo{year}{2010}.
\newblock \bibinfo{title}{{Automating the Configuration of Multi Software
  Product Lines}}, in: \bibinfo{editor}{David~Benavides, D.},
  \bibinfo{editor}{Batory, D.S.}, \bibinfo{editor}{Gr\"unbacher, P.} (Eds.),
  \bibinfo{booktitle}{{Proceedings of the 4th International Workshop on
  Variability Modelling of Software-Intensive Systems (VaMoS'10)}},
  \bibinfo{publisher}{Universit\"at Duisburg-Essen}, \bibinfo{address}{Linz,
  Austria}. pp. \bibinfo{pages}{123--130}.
\bibitem[{Rosenm\"uller et~al.(2008)Rosenm\"uller, Siegmund, K\"astner and
  ur~Rahman}]{Rosenmuller08}
\bibinfo{author}{Rosenm\"uller, M.}, \bibinfo{author}{Siegmund, N.},
  \bibinfo{author}{K\"astner, C.}, \bibinfo{author}{ur~Rahman, S.S.},
  \bibinfo{year}{2008}.
\newblock \bibinfo{title}{{Modeling Dependent Software Product Lines}}, in:
  \bibinfo{editor}{Loughran, N.}, \bibinfo{editor}{Groher, I.},
  \bibinfo{editor}{Lopez-Herrejon, R.}, \bibinfo{editor}{Apel, S.},
  \bibinfo{editor}{Schwanninger, C.} (Eds.), \bibinfo{booktitle}{{Proceedings
  of the GPCE Workshop on Modularization, Composition and Generative Techniques
  for Product Line Engineering (McGPLE'08)}}, \bibinfo{publisher}{University of
  Passau}, \bibinfo{address}{Nashville, TN, USA}. pp. \bibinfo{pages}{13--18}.
\bibitem[{Rosenm\"uller et~al.(2011)Rosenm\"uller, Siegmund, Th\"um and
  Saake}]{Rosenmuller11}
\bibinfo{author}{Rosenm\"uller, M.}, \bibinfo{author}{Siegmund, N.},
  \bibinfo{author}{Th\"um}, \bibinfo{author}{Saake, G.}, \bibinfo{year}{2011}.
\newblock \bibinfo{title}{{Multi-Dimensional Variability Modeling}}, in:
  \bibinfo{editor}{Czarnecki, K.}, \bibinfo{editor}{Eisenecker, U.W.} (Eds.),
  \bibinfo{booktitle}{{Proceedings of the 5th Workshop on Variability Modeling
  of Software-Intensive Systems (VaMoS'11)}}, \bibinfo{publisher}{ACM},
  \bibinfo{address}{Namur, Belgium}. pp. \bibinfo{pages}{11--20}.
\bibitem[{Ryabko(1992)}]{Ryabko92}
\bibinfo{author}{Ryabko, B.}, \bibinfo{year}{1992}.
\newblock \bibinfo{title}{{A Fast On-Line Adaptive Code}}.
\newblock \bibinfo{journal}{{IEEE Transactions on Information Theory}}
  \bibinfo{volume}{38}, \bibinfo{pages}{1400--1404}.
\bibitem[{Schaefer et~al.(2010)Schaefer, Bettini, Bono, Damiani and
  Tanzarella}]{Schaefer10}
\bibinfo{author}{Schaefer, I.}, \bibinfo{author}{Bettini, L.},
  \bibinfo{author}{Bono, V.}, \bibinfo{author}{Damiani, F.},
  \bibinfo{author}{Tanzarella, N.}, \bibinfo{year}{2010}.
\newblock \bibinfo{title}{{Delta-Oriented Programming of Software Product
  Lines}}, in: \bibinfo{editor}{Bosch, J.}, \bibinfo{editor}{Lee, J.} (Eds.),
  \bibinfo{booktitle}{{Proceedings of the 14th International Software Product
  Line Conference (SPLC'10)}}, \bibinfo{publisher}{Springer},
  \bibinfo{address}{Jeju Island, South Korea}. pp. \bibinfo{pages}{77--91}.
\bibitem[{Skiadas and Kjosmoen(2007)}]{Skiadas07}
\bibinfo{author}{Skiadas, C.}, \bibinfo{author}{Kjosmoen, T.},
  \bibinfo{year}{2007}.
\newblock \bibinfo{title}{{\LaTeX{ing} with TextMate}}.
\newblock \bibinfo{journal}{{The Prac\TeX\ Journal}} \bibinfo{volume}{3}.
\bibitem[{Steinberg et~al.(2008)Steinberg, Budinsky, Paternostro and
  Merks}]{Steinberg08}
\bibinfo{author}{Steinberg, D.}, \bibinfo{author}{Budinsky, D.},
  \bibinfo{author}{Paternostro, M.}, \bibinfo{author}{Merks, E.},
  \bibinfo{year}{2008}.
\newblock \bibinfo{title}{{EMF: Eclipse Modeling Framework}}.
\newblock \bibinfo{publisher}{{Addison-Wesley}}.
\bibitem[{\sortnoop{Storm}van~der Storm(2011)}]{VanDerStorm11}
\bibinfo{author}{\sortnoop{Storm}van~der Storm, T.}, \bibinfo{year}{2011}.
\newblock \bibinfo{title}{{The Rascal Language Workbench}}.
\newblock \bibinfo{type}{Technical Report} \bibinfo{number}{SEN-1111}. CWI.
\bibitem[{Sweet(1985)}]{Sweet85}
\bibinfo{author}{Sweet, R.E.}, \bibinfo{year}{1985}.
\newblock \bibinfo{title}{{The Mesa Programming Environment}}.
\newblock \bibinfo{journal}{{ACM Sigplan Notice}} \bibinfo{volume}{20},
  \bibinfo{pages}{216--229}.
\bibitem[{Vacchi and Cazzola(2015)}]{Cazzola15c}
\bibinfo{author}{Vacchi, E.}, \bibinfo{author}{Cazzola, W.},
  \bibinfo{year}{2015}.
\newblock \bibinfo{title}{{Neverlang: A Framework for Feature-Oriented Language
  Development}}.
\newblock \bibinfo{journal}{{Computer Languages, Systems \& Structures}}
  \bibinfo{volume}{43}, \bibinfo{pages}{1--40}.
\newblock \DOIprefix\doi{10.1016/j.cl.2015.02.001}.
\bibitem[{Vacchi et~al.(2014a)Vacchi, Cazzola, Combemale and
  Acher}]{Cazzola14e}
\bibinfo{author}{Vacchi, E.}, \bibinfo{author}{Cazzola, W.},
  \bibinfo{author}{Combemale, B.}, \bibinfo{author}{Acher, M.},
  \bibinfo{year}{2014}a.
\newblock \bibinfo{title}{{Automating Variability Model Inference for
  Component-Based Language Implementations}}, in: \bibinfo{editor}{Heymans,
  P.}, \bibinfo{editor}{Rubin, J.} (Eds.), \bibinfo{booktitle}{{Proceedings of
  the 18th International Software Product Line Conference (SPLC'14)}},
  \bibinfo{publisher}{ACM}, \bibinfo{address}{Florence, Italy}. pp.
  \bibinfo{pages}{167--176}.
\bibitem[{Vacchi et~al.(2013)Vacchi, Cazzola, Pillay and
  Combemale}]{Cazzola13g}
\bibinfo{author}{Vacchi, E.}, \bibinfo{author}{Cazzola, W.},
  \bibinfo{author}{Pillay, S.}, \bibinfo{author}{Combemale, B.},
  \bibinfo{year}{2013}.
\newblock \bibinfo{title}{{Variability Support in Domain-Specific Language
  Development}}, in: \bibinfo{editor}{Erwig, M.}, \bibinfo{editor}{Paige,
  R.F.}, \bibinfo{editor}{Van~Wyk, E.} (Eds.), \bibinfo{booktitle}{{Proceedings
  of 6\textsuperscript{th} International Conference on Software Language
  Engineering (SLE'13)}}, \bibinfo{publisher}{Springer},
  \bibinfo{address}{Indianapolis, USA}. pp. \bibinfo{pages}{76--95}.
\bibitem[{Vacchi et~al.(2014b)Vacchi, Olivares, Shaqiri and
  Cazzola}]{Cazzola14c}
\bibinfo{author}{Vacchi, E.}, \bibinfo{author}{Olivares, D.M.},
  \bibinfo{author}{Shaqiri, A.}, \bibinfo{author}{Cazzola, W.},
  \bibinfo{year}{2014}b.
\newblock \bibinfo{title}{{Neverlang 2: A Framework for Modular Language
  Implementation}}, in: \bibinfo{booktitle}{{Proceedings of the 13th
  International Conference on Modularity (Modularity'14)}},
  \bibinfo{publisher}{ACM}, \bibinfo{address}{Lugano, Switzerland}. pp.
  \bibinfo{pages}{23--26}.
\bibitem[{Vanbrabant(2008)}]{Vanbrabant08}
\bibinfo{author}{Vanbrabant, R.}, \bibinfo{year}{2008}.
\newblock \bibinfo{title}{{Google Guice: Agile Lightweight Dependency Injection
  Framework}}.
\newblock \bibinfo{publisher}{Apress}.
\bibitem[{V\"olter(2011)}]{Volter11}
\bibinfo{author}{V\"olter, M.}, \bibinfo{year}{2011}.
\newblock \bibinfo{title}{{Language and IDE Modularization and Composition with
  MPS}}, in: \bibinfo{editor}{L\"ammel, R.}, \bibinfo{editor}{Saraiva, J.a.},
  \bibinfo{editor}{Visser, J.} (Eds.), \bibinfo{booktitle}{{Proceedings of the
  4th International Summer School on Generative and Transformational Techniques
  in Software Engineering (GTTSE'11)}}, \bibinfo{publisher}{Springer},
  \bibinfo{address}{Braga, Portugal}. pp. \bibinfo{pages}{383--430}.
\bibitem[{V{\"o}lter and Pech(2012)}]{Voelter12}
\bibinfo{author}{V{\"o}lter, M.}, \bibinfo{author}{Pech, V.},
  \bibinfo{year}{2012}.
\newblock \bibinfo{title}{{Language Modularity with the MPS Language
  Workbench}}, in: \bibinfo{booktitle}{{Proceedings of the 34th International
  Conference on Software Engineering (ICSE'12)}}, \bibinfo{publisher}{IEEE},
  \bibinfo{address}{Z\"urich, Switzerland}. pp. \bibinfo{pages}{1449--1450}.
\bibitem[{Wachsmuth et~al.(2014)Wachsmuth, Konat and Visser}]{Wachsmuth14}
\bibinfo{author}{Wachsmuth, G.H.}, \bibinfo{author}{Konat, G.D.P.},
  \bibinfo{author}{Visser, E.}, \bibinfo{year}{2014}.
\newblock \bibinfo{title}{{Language Design with the Spoofax Language
  Workbench}}.
\newblock \bibinfo{journal}{{IEEE Software}} \bibinfo{volume}{31},
  \bibinfo{pages}{35--43}.
\bibitem[{Wende et~al.(2009)Wende, Thieme and Zschaler}]{Wende09}
\bibinfo{author}{Wende, C.}, \bibinfo{author}{Thieme, N.},
  \bibinfo{author}{Zschaler, S.}, \bibinfo{year}{2009}.
\newblock \bibinfo{title}{{A Role-Based Approach towards Modular Language
  Engineering}}, in: \bibinfo{editor}{van~den Brand, M.},
  \bibinfo{editor}{Ga\v{s}evi\'c, D.}, \bibinfo{editor}{Gray, J.} (Eds.),
  \bibinfo{booktitle}{{Proceedings of the 2nd International Conference on
  Software Language Engineering (SLE'09)}}, \bibinfo{publisher}{Springer},
  \bibinfo{address}{Denver, CO, USA}. pp. \bibinfo{pages}{254--273}.
\bibitem[{White et~al.(2009)White, Hill, Gray, Tambe, Gokhale and
  Schmidt}]{White09}
\bibinfo{author}{White, J.}, \bibinfo{author}{Hill, J.H.},
  \bibinfo{author}{Gray, J.}, \bibinfo{author}{Tambe, S.},
  \bibinfo{author}{Gokhale, A.}, \bibinfo{author}{Schmidt, D.C.},
  \bibinfo{year}{2009}.
\newblock \bibinfo{title}{{Improving Domain-specific Language Reuse with
  Software Product-Line Configuration Techniques}}.
\newblock \bibinfo{journal}{IEEE Software} \bibinfo{volume}{26},
  \bibinfo{pages}{47--53}.
\bibitem[{W\"urthinger et~al.(2013)W\"urthinger, Wimmer, Wo{\ss}, Stadler,
  Duboscq, Humer, Richards, Simon and Wolczko}]{Wurthinger13}
\bibinfo{author}{W\"urthinger, T.}, \bibinfo{author}{Wimmer, C.},
  \bibinfo{author}{Wo{\ss}, A.}, \bibinfo{author}{Stadler, L.},
  \bibinfo{author}{Duboscq, G.}, \bibinfo{author}{Humer, C.},
  \bibinfo{author}{Richards, G.}, \bibinfo{author}{Simon, D.},
  \bibinfo{author}{Wolczko, M.}, \bibinfo{year}{2013}.
\newblock \bibinfo{title}{{One VM to Rule Them All}}, in:
  \bibinfo{editor}{Hirschfeld, R.} (Ed.), \bibinfo{booktitle}{{Proceedings of
  the 2013 ACM International Symposium on New Ideas, New Paradigms, and
  Reflections on Programming \& Software (Onward!'13)}},
  \bibinfo{organization}{ACM}, \bibinfo{address}{Indianapolis, IN, USA}. pp.
  \bibinfo{pages}{187--204}.
\bibitem[{Zschaler et~al.(2009)Zschaler, S\'anchez, Santos, Alf\'erez, Rashid,
  Fuentes, Moreira, Ara\'ujo and Kulesza}]{Zschaler09}
\bibinfo{author}{Zschaler, S.}, \bibinfo{author}{S\'anchez, P.},
  \bibinfo{author}{Santos, J.}, \bibinfo{author}{Alf\'erez, M.},
  \bibinfo{author}{Rashid, A.}, \bibinfo{author}{Fuentes, L.},
  \bibinfo{author}{Moreira, A.}, \bibinfo{author}{Ara\'ujo, J.},
  \bibinfo{author}{Kulesza, U.}, \bibinfo{year}{2009}.
\newblock \bibinfo{title}{{VML*---A Family of Languages for Variability
  Management in Software Product Lines}}, in: \bibinfo{editor}{van~den Brand,
  M.}, \bibinfo{editor}{Ga\v{s}evi\'c, D.}, \bibinfo{editor}{Gray, J.} (Eds.),
  \bibinfo{booktitle}{{Proceedings of the 2nd International Conference on
  Software Language Engineering (SLE'09)}}, \bibinfo{publisher}{Springer},
  \bibinfo{address}{Denver, CO, USA}. pp. \bibinfo{pages}{82--102}.

\end{thebibliography}

\bigskip

\begin{wrapfigure}{l}{2cm}
    \includegraphics[width=2cm,keepaspectratio]{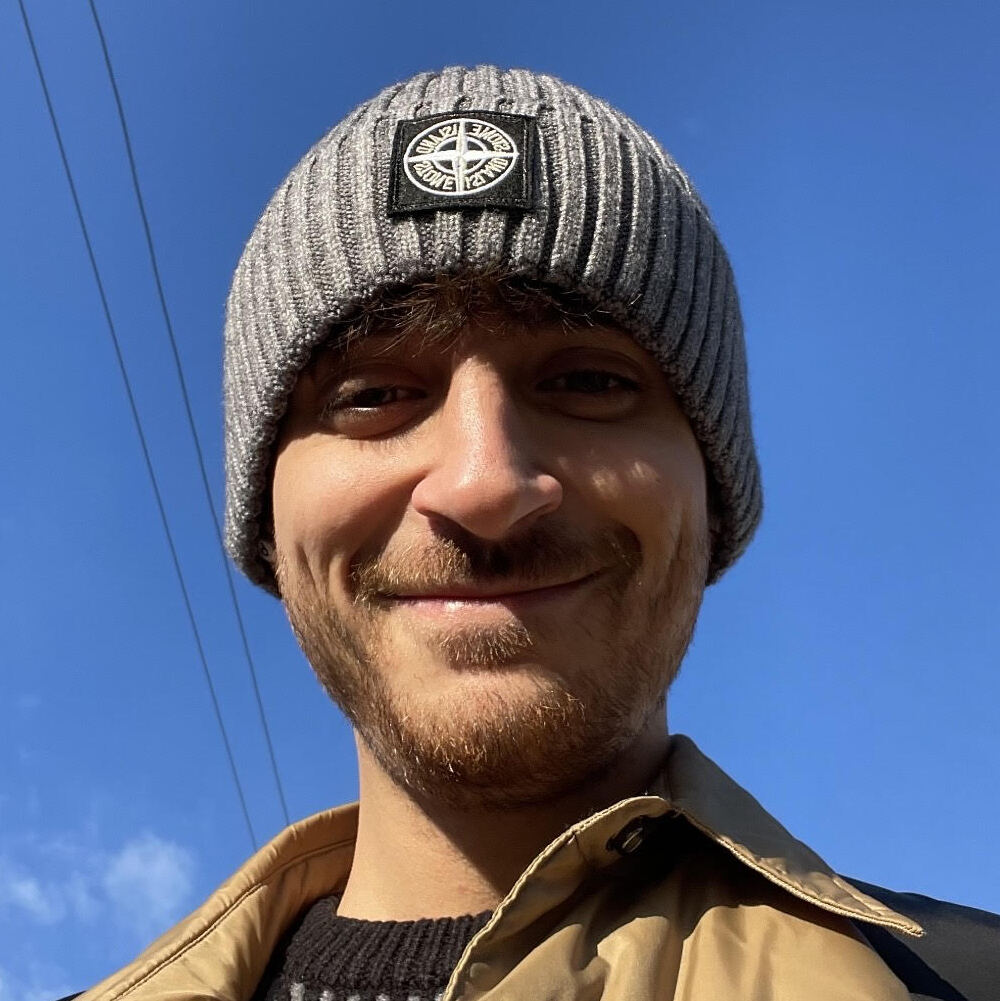}
\end{wrapfigure}\par
\noindent\textbf{Federico Bruzzone} is currently a Ph.D. student in Computer Science at Universit\`a degli Studi di Milano, Italy. He was born in 2000 and since he was a child he has been passionate about computer science and music. He got his bachelor degree in Musical Computer Science, the master degree in Computer Science and currently he is involved in the research activity of the ADAPT Lab. His main research interests are (but are not limited to) programming languages and compilers, software maintenance and evolution. For any question he can be contacted at \url{federico.bruzzone@unimi.it}.\smallskip

\begin{wrapfigure}{l}{2cm}
    \includegraphics[width=2cm,keepaspectratio]{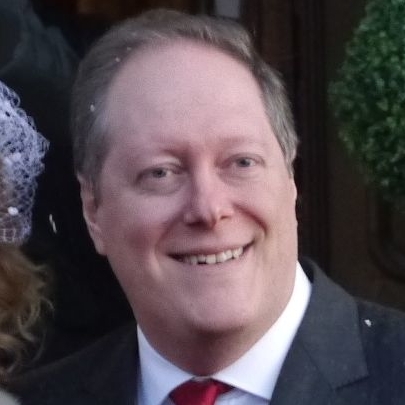}
\end{wrapfigure}\par
\noindent\textbf{Walter Cazzola} is currently a Full Professor in the Department of Computer Science of the Università degli Studi di Milano, Italy and the Chair of the ADAPT laboratory. Dr\@. Cazzola designed the mChaRM framework, @Java, [a]C\#, Blueprint programming languages and he is currently involved in the designing and development of the Neverlang language workbench. He also designed the JavAdaptor dynamic software updating framework and its front-end FiGA\@. He has written over 100 scientific papers. His research interests include (but are not limited to) software maintenance, evolution and comprehension, programming methodologies and languages. He served on the program committees or editorial boards of the most important conferences and journals about his research topics. He is associate editor for the Journal of Computer Languages published by Elsevier. More information about Dr\@. Cazzola and all his publications are available at \url{http://cazzola.di.unimi.it} and he can be contacted at \url{cazzola@di.unimi.it} for any question.\smallskip

\begin{wrapfigure}{l}{2cm}
    \includegraphics[width=2cm, keepaspectratio]{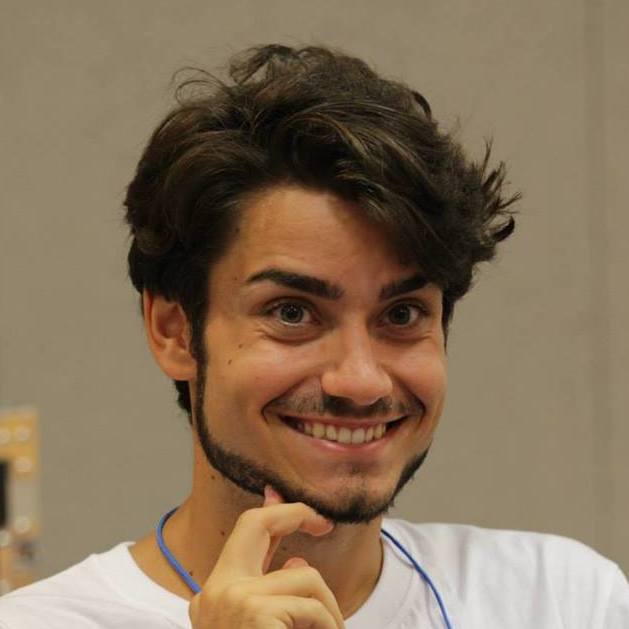}
\end{wrapfigure}\par
\noindent\textbf{Luca Favalli} is currently a Computer Science Postdoctoral Researcher at Università degli Studi di Milano. He got his PhD in computer science from the Università degli Studi di Milano. He is involved in the research activity of the ADAPT Lab and in the development of the Neverlang language workbench and of JavAdaptor. His main research interests are software design, software (and language) product lines and dynamic software updating with a focus on how they can be used to ease the learning of programming languages. He can be contacted at \url{favalli@di.unimi.it} for any question.\smallskip

\end{document}